\documentclass{emulateapj}
\usepackage{natbib}
\newcommand{\HI}{\ion{H}{1}}
\newcommand{\HII}{\ion{H}{2}}
\newcommand{\epot}{\ensuremath{\epsilon_{\rm pot}}}
\newcommand{\kms}{\mbox{km~s$^{-1}$}}
\newcommand{\mjybm}{\mbox{mJy~bm$^{-1}$}}
\newcommand{\jybm}{\mbox{Jy~bm$^{-1}$}}

\slugcomment{Astrophysical Journal, accepted 6 Dec 2003}

\shorttitle{Radial Gas Flows in Spiral Galaxies}

\begin{document}

\title{A Search for Kinematic Evidence of Radial Gas Flows
in Spiral Galaxies}

\author{Tony Wong\altaffilmark{1} and Leo Blitz}
\affil{Astronomy Department and Radio Astronomy Laboratory, 
University of California, Berkeley, CA 94720}
\email{Tony.Wong@csiro.au, blitz@gmc.berkeley.edu}
\and
\author{Albert Bosma}
\affil{Observatoire de Marseille, 2 Place Le Verrier, F-13248 
Marseille Cedex 4, France}
\email{bosma@oamp.fr}
\altaffiltext{1}{Present address: CSIRO Australia Telescope 
National Facility, PO Box 76, Epping NSW 1710, Australia}

\begin{abstract}

CO and \HI\ velocity fields of seven nearby spiral galaxies, derived
from radio-interferometric observations, are decomposed into Fourier
components whose radial variation is used to search for evidence of
radial gas flows.  Additional information provided by optical or
near-infrared isophotes is also considered, including the relationship
between the morphological and kinematic position angles.  To assist in
interpreting the data, we present detailed modeling that demonstrates
the effects of bar streaming, inflow, and a warp on the observed
Fourier components.  We find in all of the galaxies evidence for
either elliptical streaming or a warped disk over some range in
radius, with deviations from pure circular rotation at the level of
$\sim$20--60 \kms.  Evidence for kinematic warps is observed in
several cases well inside $R_{25}$.  No unambiguous evidence for
radial inflows is seen in any of the seven galaxies, and we are able
to place an upper limit of $\sim$5--10 \kms\ (3--5\% of the circular
speed) on the magnitude of any radial inflow in the inner regions of
NGC 4414, 5033 and 5055.  We conclude that the inherent
non-axisymmetry of spiral galaxies is the greatest limitation to the
direct detection of radial inflows.

\end{abstract}


\section{Introduction}\label{sec:intro}

The potential importance of radial gas flows for understanding the
evolution of galaxies has long been recognized.  Their effect on
galactic morphology has been discussed in the context of evolution
along the Hubble sequence \citep*[e.g.,][]{Norman:96} and the formation
of exponential stellar disks \citep{Lin:87,Ferguson:01}.  Their
possible role in fueling nuclear activity and star formation in the
inner disks of galaxies has been discussed by \citet*{Shlosman:89} and
\citet{Blitz:96} respectively.  Their ability to account for the
steeper than expected abundance gradients in spiral galaxies
has been explored by \citet{Lacey:85}, \citet{Chamcham:94}, and
\citet{Portinari:00}.  Finally, the susceptibility of galactic disks
to radial flows has been discussed by \citet{Struck:91} and
\citet{Struck:99} from a hydrodynamic standpoint.  In general such
flows are expected to be mildly subsonic, i.e.\ up to a few \kms\ in
typical galaxies.

Despite the strong motivations for expecting radial flows to occur,
published measurements of radial gas flows in galaxy disks are limited
to a few strongly barred galaxies (e.g., NGC 7479,
\citealt{Quillen:95}; NGC 1530, \citealt*{Regan:97}), and even in these
cases it is difficult to measure inflow rates unambiguously.  Part of
the problem is achieving adequate velocity resolution: even slow
radial flows can be important on an evolutionary timescale, since 1
\kms\ $\approx$ 1 kpc Gyr$^{-1}$.  More fundamental is the ambiguity
in deriving the full velocity structure from only the
line-of-sight component of motion.  For gas in pure circular rotation,
the direction of the velocity vector at any point in the disk is
uniquely determined (assuming the orientation of the disk is known)
and its magnitude depends only on the galactocentric radius.  If
non-circular motions exist, however, the direction of local motion is
generally not known, and recovering the true velocity structure from
Doppler velocities alone is impossible without further assumptions.


\begin{table*}[t]
\footnotesize
\begin{center}
\caption{Properties of the Sample Galaxies\label{tbl:props}}
\bigskip
\setlength{\tabcolsep}{8pt}
\begin{tabular}{ccccccccc}
\tableline\tableline
 &  &  & $R_{25}$\tablenotemark{a} & $V_{\odot}$\tablenotemark{b} 
& Nuc. & CO Resolution & HI Resolution \\
Name & R.A. (2000) & Decl.\ (2000) & (arcsec) & (\kms) & 
Type\tablenotemark{c} &
\multicolumn{2}{c}{(arcsec $\times$ arcsec $\times$ \kms)} \\[0.5ex]
\tableline
NGC 4321 & 12:22:54.9 & 15:49:21 & 220 & 1571 & T2 &
    8.85 $\times$ 5.62 $\times$ 10 & 20.0 $\times$ 20.0 $\times$ 20.6\\
NGC 4414 & 12:26:27.1 & 31:13:24 & 110 & 716 & T2: &
    6.53 $\times$ 4.88 $\times$ 10 & 17.5 $\times$ 15.3 $\times$ 10.3\\
NGC 4501 & 12:31:59.2 & 14:25:14 & 210 & 2281 & S2 &
    7.89 $\times$ 6.23 $\times$ 20 & 23.2 $\times$ 22.3 $\times$ 10.3\\
NGC 4736 & 12:50:53.1 & 41:07:14 & 340 & 308 & L2 &
    6.86 $\times$ 5.02 $\times$ 10 & 15.0 $\times$ 15.0 $\times$ 5.15\\
NGC 5033 & 13:13:27.5 & 36:35:38 & 320 & 875 & S1.5 &
    7.52 $\times$ 6.49 $\times$ 20 & 19.9 $\times$ 17.0 $\times$ 20.6\\
NGC 5055 & 13:15:49.3 & 42:01:45 & 380 & 504 & T2 &
    5.69 $\times$ 5.30 $\times$ 10 & 12.9 $\times$ 12.8 $\times$ 10.3\\
NGC 5457 & 14:03:12.5 & 54:20:55 & 870 & 241 & H &
    6.88 $\times$ 6.36 $\times$ 10 & 15.0 $\times$ 15.0 $\times$ 5.15\\
\tableline
\end{tabular}
\end{center}
\tablenotetext{a}{Semimajor axis at 25.0 mag arcsec$^{-2}$, from RC3.}
\tablenotetext{b}{Heliocentric velocity, from NED.}
\tablenotetext{c}{Nuclear spectral class and type, from
\citet{Ho:97cat}.  ``S'' refers to Seyfert, ``H'' to \HII, ``L'' to
LINER, and ``T'' to transition (\HII/LINER).  Uncertain
classifications are marked by a colon (:).} 
\end{table*}


One approach for measuring inflow speeds is to model the gravitational
potential based on the observed stellar distribution and then use the
model to predict the gas inflow rate towards the center.  While such
modeling is becoming less daunting computationally, the results still
depend quite heavily on the exact techniques employed and on
assumptions regarding the mass-to-light ($M/L$) ratio and the
deprojection of surface brightness into a volume mass density.
Determining the appropriate $M/L$ ratio is especially problematic when
observations of the stellar velocity dispersion are unavailable. An
upper limit to $M/L$ is provided by the ``maximum disk'' hypothesis,
which adopts the maximum $M/L$ ratio for the disk that is consistent
with the rotation curve and observed gas and stellar profiles.
However, the maximum disk hypothesis has been questioned, both in the
context of the Milky Way (\citealt{Merrifield:92}; but see
\citealt{Sackett:97}) as well as for other galaxies
\citep{Bottema:93}.  An alternative is to determine the $M/L$ ratio
from stellar population synthesis models.  For the strongly barred
galaxy NGC 7479, \citet{Quillen:95} found a consistent value of $M/L_K
\approx 1.35\, M_\odot/L_{K\odot}$ using both maximum disk and
population synthesis techniques.  By estimating the torque from the
resulting gravitational potential, they inferred a radial inflow speed
of 10--20 \kms\ for molecular gas along the bar.  However, they only
integrated the torque across a linear dust lane seen in CO emission;
integrating over the entire gas distribution would likely result in a
lower inflow speed, in better agreement with numerical simulations
\citep[e.g.,][]{Athan:92}.

In this study, we will instead attempt to detect radial inflows
directly, from the atomic and molecular gas kinematics of nearby,
moderately inclined spiral galaxies.  We consider three basic types of
deviations from a circularly rotating disk: uniform (axisymmetric)
inflow, elliptical streaming in a non-axisymmetric potential
(including the special case of spiral streaming), and circular orbits
that are inclined with respect to the main disk (an outer warp).
Since the latter two situations are not necessarily accompanied by net
inflow, we aim to distinguish them from the pure inflow case by
employing a Fourier analysis of the velocity field.  Optical and
near-infrared images are used as an independent indicator of possible
bars or oval distortions.  Although our models are simplistic, they
have very few free parameters and hence can be applied to a large
number of galaxies with relatively little fine-tuning.

This paper is organized as follows.  In \S\ref{sec:obs} we summarize
the observational data, most of which come from CO mapping as part of
the BIMA Survey of Nearby Galaxies \citep{Regan:01,Helfer:03} and from
previously published VLA \HI\ data.  In \S\ref{sec:anal} we present
velocity fields, tilted-ring modeling and isophote fits for the seven
galaxies in our sample.  In \S\ref{sec:method} we describe the
signatures of different types of non-circular motions, as revealed by
a harmonic decomposition of the velocity field.  In
\S\ref{sec:results} we analyze the velocity fields for evidence of
radial gas flows, making use of complementary information from the
optical and near-infrared isophote fits.  \S\ref{sec:disc} contains a
discussion and summary of this work.

\section{Observations}\label{sec:obs}

\subsection{Properties of the Sample}

The sample of seven galaxies used in this study is the same as in
\citet[][hereafter Paper I]{Wong:02}.  Basic properties of the
galaxies are summarized in Table~\ref{tbl:props}.  All exhibit strong
CO emission at a range of radii (out to at least $R \sim 50\arcsec$)
and have available \HI\ maps at a resolution of $\sim$20\arcsec\ or better.
They have also been selected to possess fairly regular velocity
fields---none are involved in strong tidal interactions (although NGC
4321 and 4501 are members of the Virgo cluster) and none are
classified as strongly barred (SB) in RC3 \citep{RC3}.  Only two of
the galaxies, NGC 4501 and 5033, contain Seyfert nuclei; the rest show
\HII, LINER, or composite nuclear spectra \citep{Ho:97cat}.  At a median
distance of 16 Mpc, adopted for the Virgo cluster using the Cepheid
measurement of \citet{Ferrarese:96}, $1\arcsec = 80$ pc, and 1 kpc =
13\arcsec.

The exclusion of strongly barred galaxies may seem puzzling given that
the gravitational torque exerted by a bar is considered the most
effective mechanism for funneling gas into the central regions
\citep[e.g.,][]{Combes:99}.  Many strongly barred galaxies exhibit
dust lanes (usually along the leading side of the bar) which are
thought to be associated with shocks, and hydrodynamical simulations
indicate that these shocks can lead to strong radial inflows
\citep{Athan:92}.  Disentangling radial inflow from strong bar
streaming, however, requires the type of detailed modeling for which
our method is unsuitable, since our elliptical streaming models assume
non-intersecting gas orbits, a condition which would not be satisfied
where shocks occur.  Hence we will focus on unbarred or weakly barred
galaxies, which might still exhibit detectable (albeit smaller) radial
flows.

\subsection{CO and HI Observations}

CO(1--0) observations were conducted with the BIMA\footnote{The
Berkeley-Illinois-Maryland Association is funded in part by the
National Science Foundation.} interferometer at Hat Creek, California
and the NRAO\footnote{The National Radio Astronomy Observatory is a
facility of the National Science Foundation, operated under
cooperative agreement by Associated Universities, Inc.} 12~m telescope
at Kitt Peak, Arizona, mostly as part of the BIMA Survey of Nearby
Galaxies (BIMA SONG) \citep{Regan:01,Helfer:03}.  Details of the
observations and reduction procedures are given in Paper I.\@ The BIMA
visibilities were used to generate data cubes with synthesized beams
of $\sim$6\arcsec (see Table~\ref{tbl:props}).  The velocity channel
increment was 10 \kms\ for most of the cubes, but 20 \kms\ for the NGC
4501 and 5033 cubes.  Although the instrumental velocity resolution
was $\sim$4 \kms, choosing a larger velocity channel width can
significantly improve the signal-to-noise of the map, allowing for
better deconvolution.  The 12~m data were folded into the BIMA cubes
using the MIRIAD task IMMERGE, resulting in combined datacubes that
are sensitive to emission on both large and small scales.  Without the
single-dish data, the velocity field would be biased more strongly to
the emission peaks.

The \HI\ observations were obtained at the NRAO Very Large Array (VLA).
Six of the galaxies have appeared previously in the
literature, in papers by \citet{Knapen:93} (NGC 4321),
\citet{Thornley:97b} (NGC 4414), \citet{Braun:95} (NGC 4736 and 5457),
\citet{Thean:97} (NGC 5033), and \citet{Thornley:97a} (NGC 5055).  Reduced
datacubes for these galaxies were kindly provided by these authors.  The
NGC 4501 data are described in detail below.  All
galaxies were observed in the C configuration, which gives a
synthesized beam of $\sim$13\arcsec\ for uniform weighting and
$\sim$20\arcsec\ for natural weighting; three of the galaxies
(NGC~4736, 5055, and 5457) were also observed at higher resolution in
the B configuration.  Data obtained in the D configuration are
included for all galaxies except NGC~4501 and 5033, so that the
resulting maps are sensitive to even very extended ($\sim$15\arcmin)
emission.  The datacubes were imported into MIRIAD and transformed to
the spatial and velocity coordinate frames of the BIMA cubes.

\subsection{NGC 4501 Data}

The single-dish CO map for NGC 4501 used in Paper I was only 2\arcmin\
$\times$ 2\arcmin\ in size and had an angular resolution of 55\arcsec,
making it less than ideal for combination with the BIMA data.  For
this study we have instead used a CO map of NGC~4501 taken during
several sessions in 1992 and 1993 with the IRAM 30~m telescope on Pico
Veleta, Spain.  The map consisted of spectra taken at 249 positions
separated by 12\arcsec, covering a field roughly 3\arcmin $\times$
2\arcmin\ elongated along the major axis.  At $\lambda$ = 2.6~mm the
half-power beamwidth (HPBW) of the 30~m telescope is 23\arcsec.  The
noise level in the resulting map was 27 mK per 10.4 \kms\ channel.
The data were converted from a brightness temperature to flux density
scale assuming a telescope gain of 6.5 Jy~(K[$T_A^*$])$^{-1}$.  The
resulting CO flux within the primary beam function of the BIMA mosaic
is 2200 Jy~\kms, comparable to the 2300 Jy~\kms\ measured in the 12~m
map.  We used the IRAM map to regenerate a combined (single-dish +
interferometer) CO datacube for this galaxy using the IMMERGE
technique as described in Paper I.  In performing the combination, the
flux of the BIMA map had to be rescaled by a factor of 0.8 to produce
consistency with the IRAM map.  This scaling factor is similar to
factors needed to reconcile the BIMA and 12-m flux scales for the
other galaxies (Paper I), and is consistent with the calibration
uncertainties typical of millimeter-wave telescopes.


\begin{table}
\begin{center}
\caption{Parameters For NGC 4501 \HI\ Observations\label{tbl:vlapars}}
\bigskip
\begin{tabular}{lr}
\tableline\tableline
R.A. of phase center (1950.0) &     12$^{\mathrm h}$29$^{\mathrm m}$27\fs0\\
Decl. of phase center (1950.0) &    14\arcdeg 41\arcmin 43\farcs0\\
Central velocity (\kms, heliocentric) & 2280\\ 
Velocity range (\kms) &           630\\
Time on source (hr) &             5.8\\
Bandwidth (MHz) &                 3.125\\
Number of channels &              63\\
Channel separation (\kms) &       10.3\\
Synthesized beam width &          23\farcs2 $\times$ 22\farcs9\\
Noise level (1$\sigma$) &         0.55 mJy beam$^{-1}$\\
\tableline
\end{tabular}
\end{center}
\end{table}


As the \HI\ data for NGC~4501 used in this study (as well as Paper I)
have not previously been published, we also briefly describe the
observations and data reduction here.  The observations were made on
1991 January 20 as part of a program (AG 318) designed by Gunn, Knapp,
van Gorkom, Athanassoula and Bosma to determine rotation properties of
spiral galaxies. The instrumental parameters are given in
Table~\ref{tbl:vlapars}. Standard VLA calibration procedures were
applied in AIPS. The continuum was subtracted by making a linear
visibility fit for a range of line free channels at either side of the
band. The line free channels were determined by visual inspection of
the data. The line images were made with the task IMAGR, using a
``robust'' weighting scheme intermediate between uniform and natural
weighting, as well as a tapering of the longer baselines. This resulted
in a synthesized beam of 23\farcs2 $\times$ 22\farcs9 (FWHM), close to
the beam size of the 30-m CO data.

\subsection{NIR and Optical Images}

Near-infrared (NIR) images were obtained from the Two Micron All
Sky Survey (2MASS) in the $K_s$ (2.2 $\mu$m) filter for all galaxies
except NGC 4414 and 5055, for which $K^\prime$ images from
\citet{Thornley:96} were used instead.  Astrometry and photometry from
2MASS and \citet{Thornley:96} were adopted; surface photometry from
the latter was found to be consistent with 2MASS for NGC 4414 to
within $<$0.5 mag arcsec$^{-2}$.  Astrometric coordinates
should be accurate to less than 1\arcsec, as indicated by a comparison
with Digitized Sky Survey (DSS) images using foreground stars.  In
some cases (NGC 4501, 4736, 5033, 5457), the galaxy fell near the edge
of a 2MASS Atlas Image, and several Atlas images were mosaiced
together using the IRAF task COMBINE, after subtracting a
constant background level from each.

We also obtained $I$-band images for all galaxies except NGC 4736.
For NGC 4321, 4414, 4501, 5033, and 5055, these images come from the
on-line catalog of \citet{Frei:96}.  For NGC 5457, we used an $I$
image taken with the 1.5~m Palomar telescope in 2000 April, taken by
M. Regan for the BIMA SONG project.  For NGC 4736, we used instead the
$\lambda$=665 nm continuum image corresponding to the H$\alpha$ image
of \citet{Martin:01}, approximating an $R$-band image.  As most of
these images were not photometrically calibrated, and some are
saturated near the galaxy's nucleus, they are only used here to
supplement the $K$-band data.  Astrometry for the \citet{Frei:96}
images proved difficult, since foreground stars have been removed, and
was accomplished by matching the location of giant \HII\ regions with
a DSS image (NGC 4321, 5033, 5055) or by a cross-correlation technique
(NGC 4414, 4501).  We expect the resulting coordinates to be accurate
to within 2\arcsec.


\begin{figure*}[p]
\begin{center}
\includegraphics{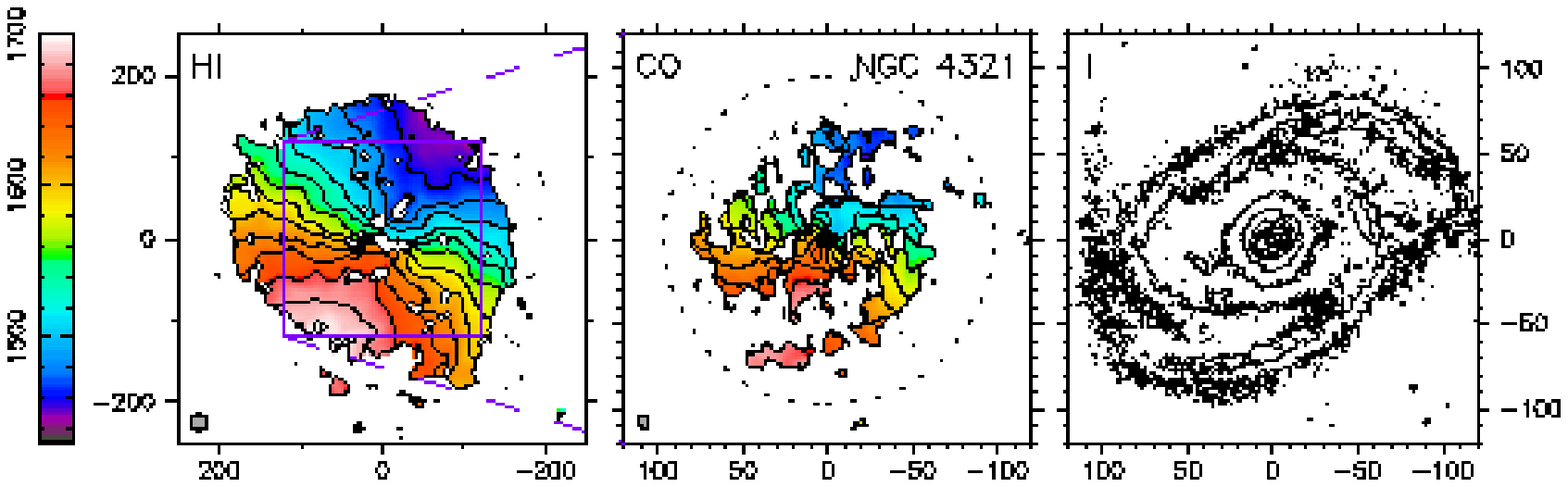}\\[0.5cm]
\includegraphics{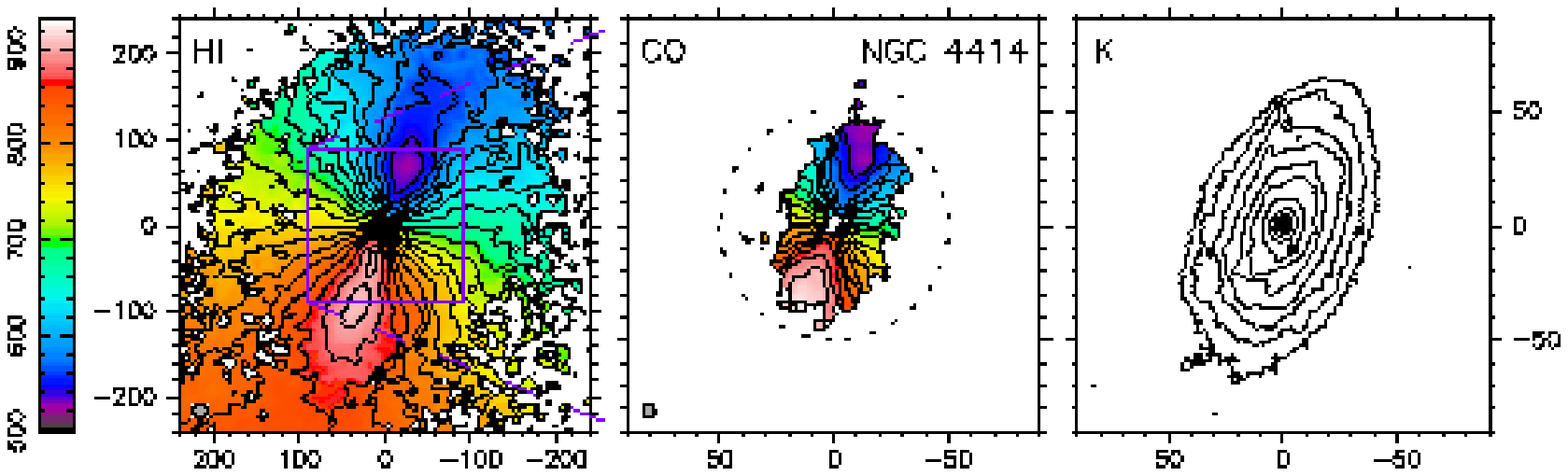}\\[0.5cm]
\includegraphics{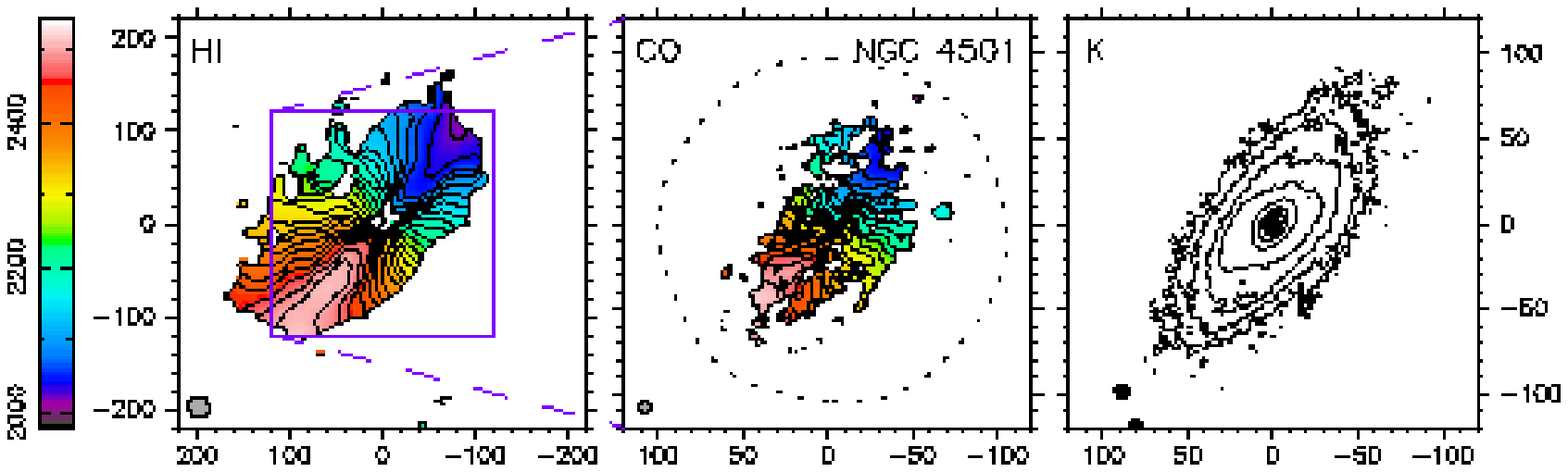}\\[0.5cm]
\caption{
Velocity fields for all seven galaxies derived from Gaussian fits as
described in the text.  For each galaxy, the left panel shows the \HI\
velocity field, the center panel shows the full-resolution CO velocity
field, and the right panel shows a contoured $I$ or $K$-band image
(note the presence of some foreground stars).  Units of the axes are
in arcseconds.  The velocity fields have been masked using a rotation
model as described in the text.  Isovelocity contours are spaced by 20
\kms.  The synthesized beam is shown on the lower left, and the CO
field of view (where the map sensitivity is at least half of its peak
value) is given by a dotted contour.
\label{fig:velfields}}
\end{center}
\end{figure*}

\begin{figure*}
\begin{center}
\includegraphics{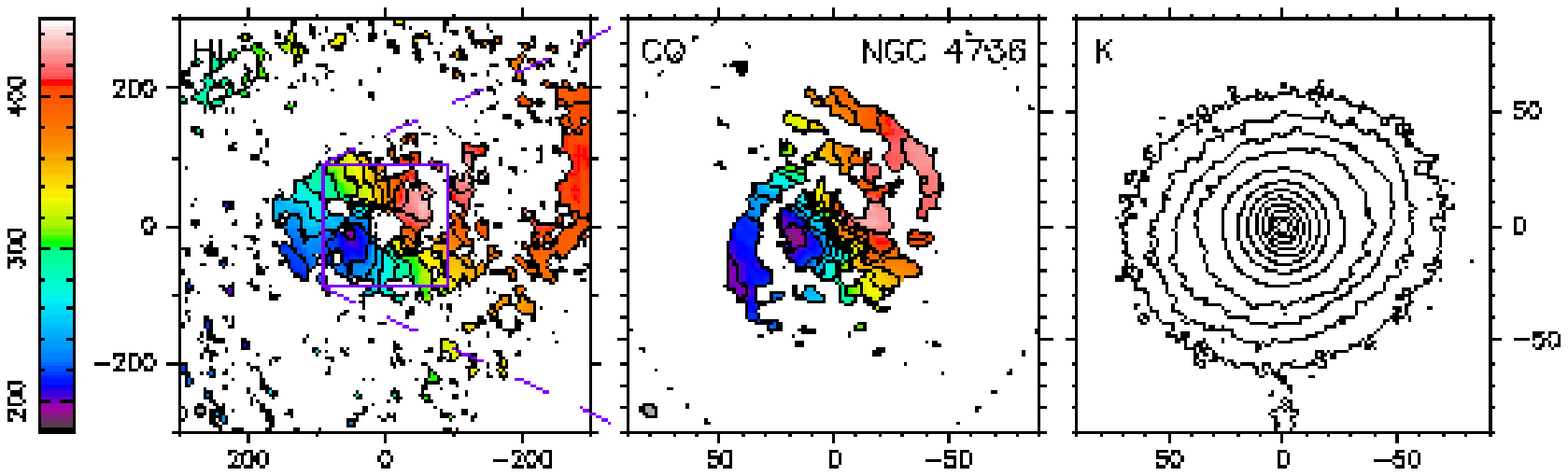}\\[0.5cm]
\includegraphics{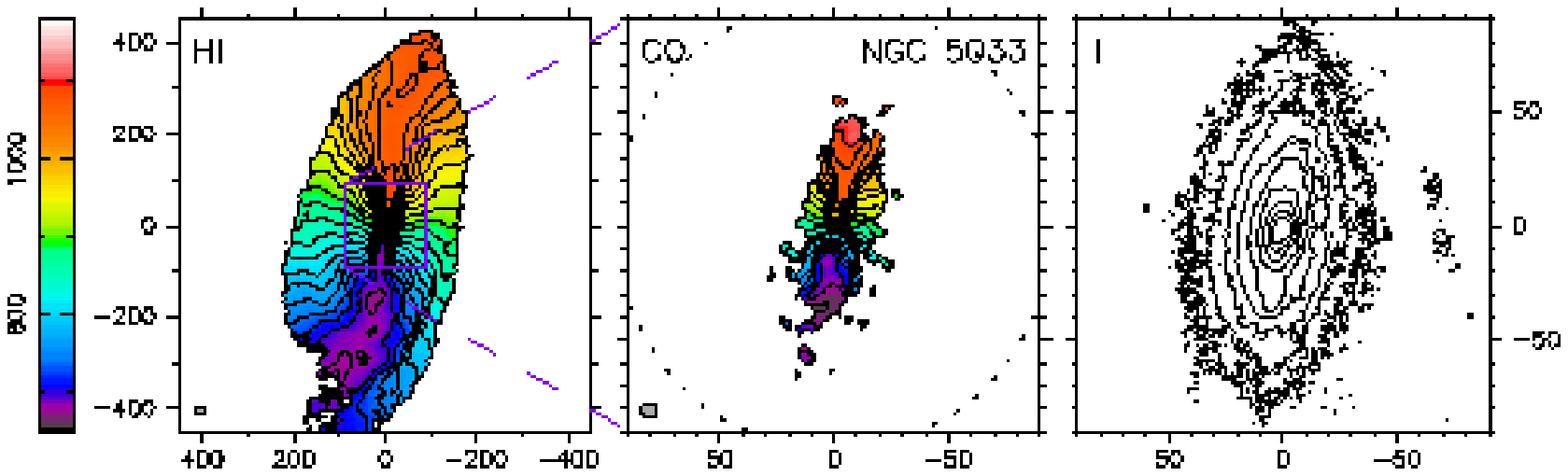}\\[0.5cm]
\includegraphics{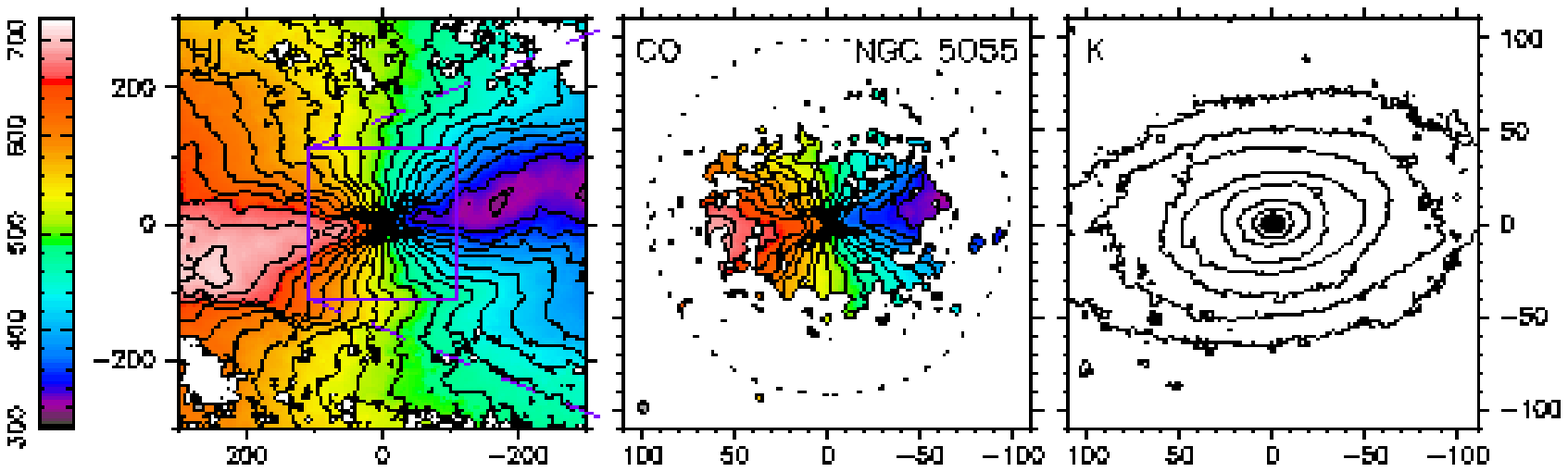}\\[0.5cm]
\includegraphics{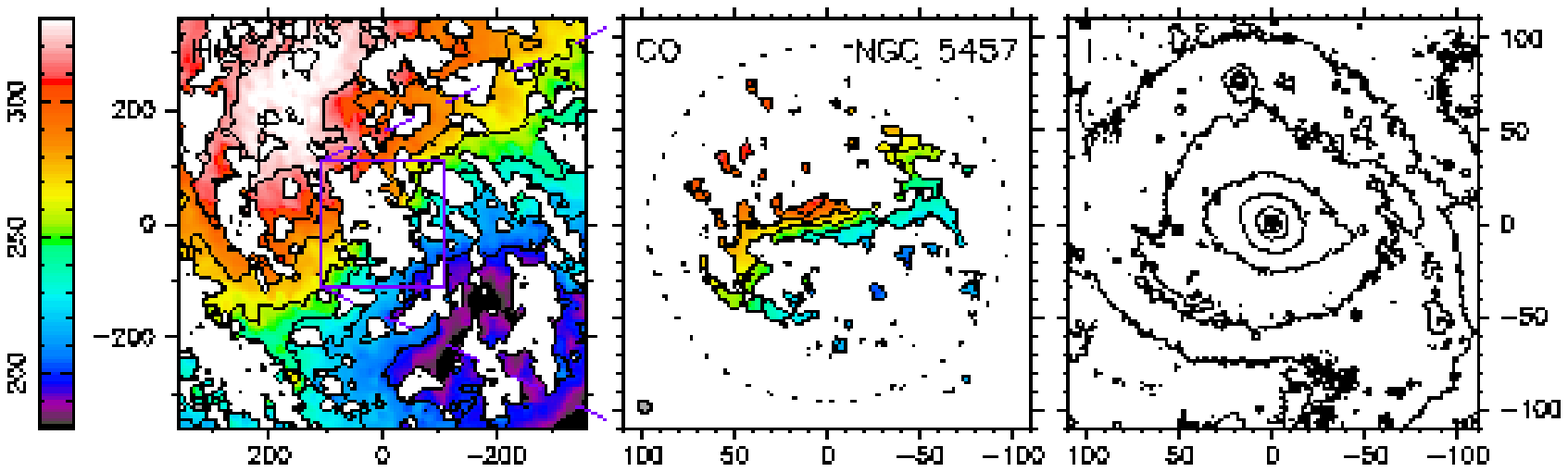}\\[0.5cm]
\end{center}
\end{figure*}

\section{Data Analysis}\label{sec:anal}

\subsection{Velocity Fields}

We derived CO and \HI\ velocity fields from the corresponding
datacubes by fitting a Gaussian to the spectrum at each pixel using a
customized version of the MIRIAD task GAUFIT.  Gaussian fits
were required to have an amplitude of $>$1.5$\sigma$, where $\sigma$
is the r.m.s.\ noise in a channel map.  Although this criterion allows
for the detection of broad lines which are relatively weak in a single
velocity channel, it has the drawback of also allowing for a large number
of spurious fits, as evidenced by velocities which are spatially
uncorrelated from pixel to pixel, producing a ``noisy'' appearance.
In contrast, the true velocity field should be dominated by galactic
rotation and smooth on scales smaller than a synthesized beam.
Additional thresholds were therefore imposed based on the integrated
flux of the Gaussian and the uncertainty in its mean velocity.
Although the exact threshold depended somewhat on the signal-to-noise
and flux level in the datacube, a Gaussian fit in a naturally weighted
CO datacube ($\sigma \sim 50$ \mjybm) would typically be required to
have a velocity-integrated intensity of $\gtrsim 3$ \jybm\ \kms\ (a
3$\sigma$ detection over 20 \kms) and a velocity uncertainty of
$\lesssim 20$ \kms\ to be included.  The assumption of Gaussian
profiles, of course, is only an approximation, but in most cases the
signal-to-noise could not justify the use of an asymmetric profile,
and other methods of deriving the velocity field (such as the first
moment) proved less robust and did not provide meaningful error
estimates.

Figure~\ref{fig:velfields} shows the CO and \HI\ velocity fields at
full resolution and a contoured $I$ or $K$-band image on the same
scale as the CO velocity field for comparison.  The BIMA field of view
(where the sensitivity drops to half of its peak value) is shown as a
dotted contour.  The velocity fields shown have been masked further
using a rotation model derived from the adopted rotation curve.  This
masking eliminates points which deviate by $\gtrsim$50 \kms\ from
circular rotation, as described in \S\ref{sec:rotcur}.  As noted in
Paper I, the CO emission in all galaxies except NGC 4736 and 5033
appears to extend to the edge of the observed field, suggesting that
our observations sample only the inner part of the molecular disk.

Note that while there is some subjectivity in determining the
appropriate thresholds for accepting the Gausssian fits, our velocity
fields do not automatically reject anomalous velocity gas that is
inconsistent with circular rotation, until the final masking using a
rotation model is performed.  Indeed an anomalous velocity component
is seen in the southeastern corner of the \HI\ disk in NGC 4501.  No
other clear examples of anomalous velocity gas were seen.


\begin{table}
\begin{center}
\caption{Galaxy Parameters From Kinematic Fits\label{tbl:rotpars}}
\bigskip
\begin{tabular}{ccccc}
\tableline\tableline
Galaxy & $(\Delta x_0,\Delta y_0)$\tablenotemark{a} 
	& $V_{\rm sys}$\tablenotemark{b}
	& $\Gamma_0$\tablenotemark{c} & $i$\\
& (\arcsec) & (\kms) & (\arcdeg) & (\arcdeg)\\[0.5ex]
\tableline
NGC 4321 & (0,0) & $1570\pm3$ & $153\pm3$ & $34\pm5$\\
NGC 4414 & (2,0) & $720\pm5$ & $159\pm2$ & $55\pm2$\\
NGC 4501 & (0,0) & $2270\pm1$ & $141\pm1$ & $64\pm2$\\
NGC 4736 & (1,1) & $317\pm4$ & $295\pm10$ & $32\pm8$\tablenotemark{d}\\
NGC 5033 & $(-1,0)$ & $880\pm8$ & $353\pm2$ & $68\pm2$\\
NGC 5055 & (0,0) & $506\pm3$ & $98\pm1$  & $63\pm1$\\
NGC 5457 & (0,0) & $257\pm1$ & $42\pm2$  & $21\pm3$\\
\tableline
\end{tabular}
\end{center}
\tablenotetext{a}{Position of kinematic center with respect to optical
center, in directions of RA and DEC.}
\tablenotetext{b}{Systemic LSR velocity.}
\tablenotetext{c}{Position angle of receding side of line of nodes,
measured E from N.}
\tablenotetext{d}{Inclination of 35\arcdeg\ was adopted based on the
photometric study of \citet{Mollen:95}.}
\end{table}



\begin{figure*}
\plottwo{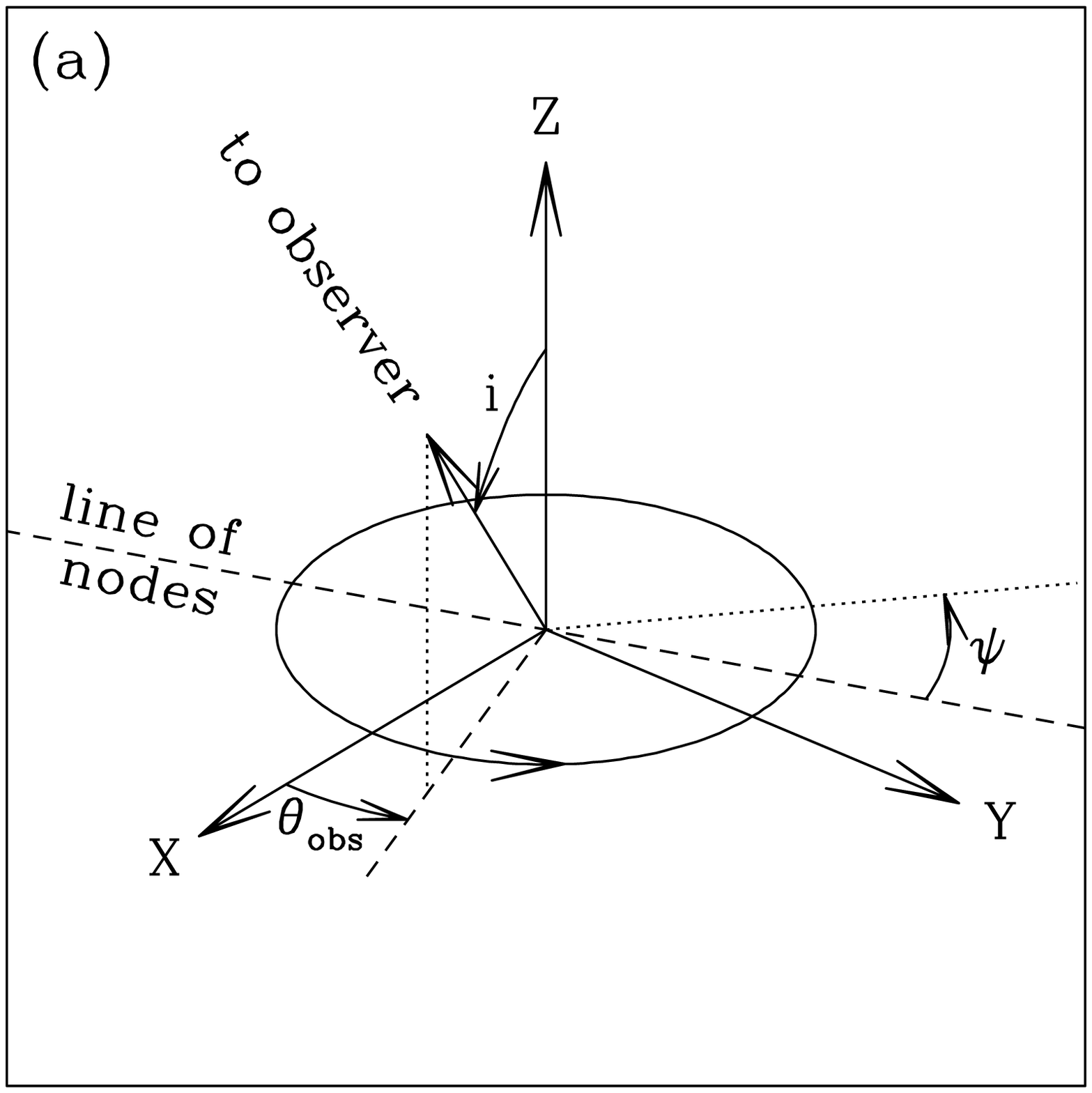}{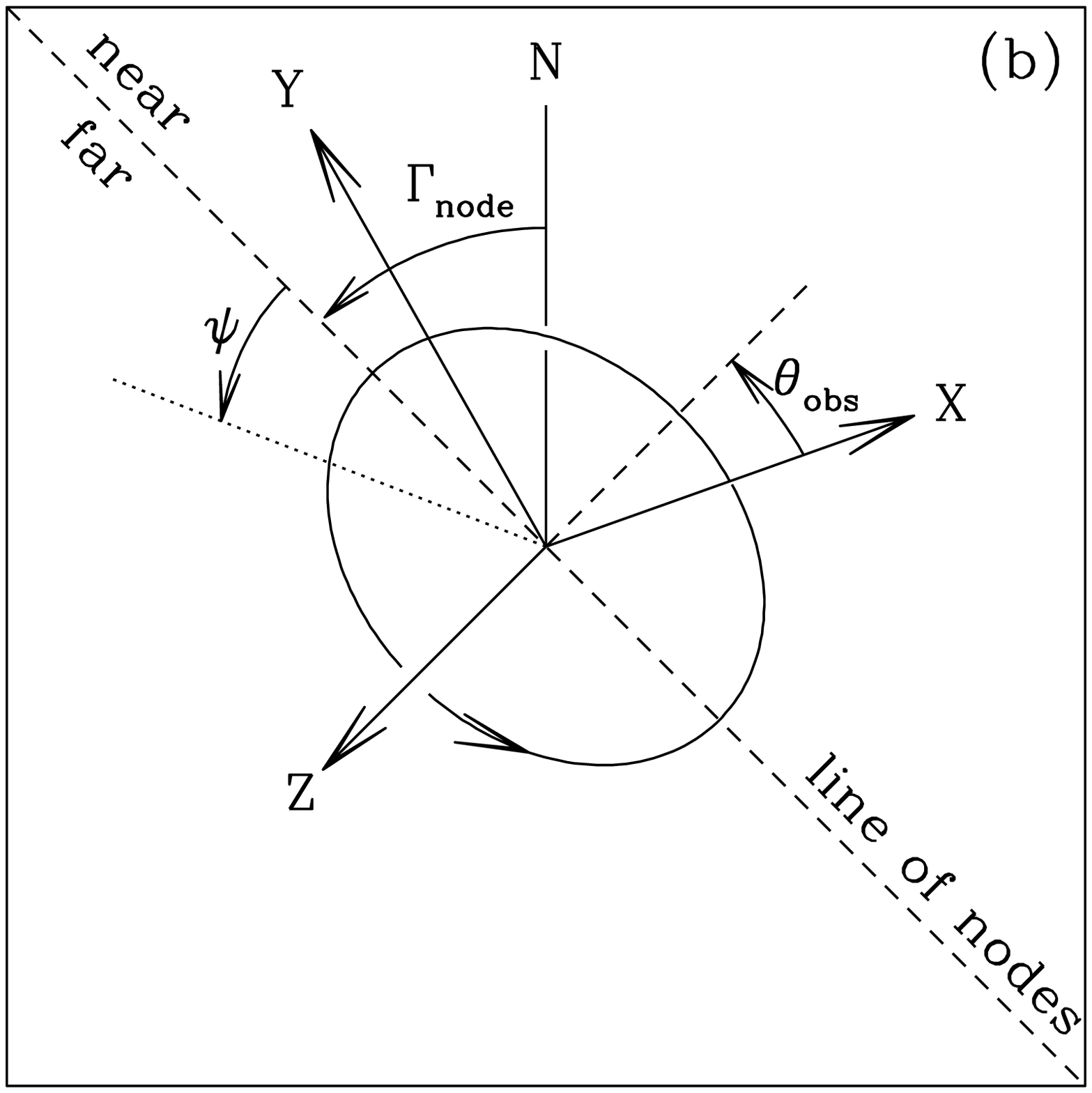}
\caption{
A circular orbit in the plane of the galaxy as viewed from two
different orientations: (a) oblique view, showing direction to the
observer; and (b) observer's view, with north up.  The angles shown
are as defined in the text.
\label{fig:geom}}
\end{figure*}

\subsection{Kinematic Disk Parameters}\label{sec:rotcur}

In preparation for the more detailed kinematic analysis needed to
search for radial flows, we estimated the basic disk orientation
parameters (mean velocity, rotation center, position angle, and
inclination) as well as the rotation curve by fitting tilted-ring
models to the CO and \HI\ velocity fields.  Four velocity fields were
analyzed for each galaxy: the CO and \HI\ at full resolution, the CO
data smoothed to the resolution of the \HI, and the \HI\ data smoothed
to 55\arcsec\ resolution.  We used the standard least-squares fitting
technique developed by \citet{Begeman:th}, as implemented in the
ROTCUR task within the NEMO software package \citep{Teuben:95}.  The
galactic disk is subdivided into rings, each of which is described by
6 parameters: the center coordinates $(x_0,y_0)$, position angle
($\Gamma_0$), inclination ($i$), offset velocity ($V_{\rm sys}$), and
circular velocity ($v_c$).  Starting with initial estimates for the
fitting parameters, the parameters are adjusted iteratively for each
ring until convergence is achieved.

In general it is not possible to fit all 6 parameters simultaneously,
due to kinematic disturbances and limited signal-to-noise in the
velocity field, compounded by correlations between fit parameters.  We
therefore adopted the following procedure for determining the fit
parameters:

\begin{enumerate}

\item Initial estimates for $i$ and $\Gamma_0$ were taken from the
isophotal fits (\S\ref{sec:isophot}), and the optical or radio
continuum center from the NASA Extragalactic Database (NED) was taken
as $(x_0,y_0)$.  Initial estimates for $V_{\rm sys}$ and $v_c$ were
taken from a position-velocity cut through the center at a position
angle of $\Gamma_0$, or from previously published studies.

\item An improved estimate of $\Gamma_0$ was determined by fixing
$(x_0,y_0)$, $V_{\rm sys}$, and $i$, and allowing the remaining
parameters ($\Gamma_0$ and $v_c$) to vary.

\item An improved estimate of $V_{\rm sys}$ was determined by fixing
$(x_0,y_0)$, $\Gamma_0$, and $i$.

\item An improved estimate of $(x_0,y_0)$ was determined by fixing
$V_{\rm sys}$, $\Gamma_0$, and $i$.

\item An attempt was made to fit for $i$ while $(x_0,y_0)$, $V_{\rm
sys}$, and $\Gamma_0$ were held fixed, although this could only be 
accomplished for some of the rings.

\end{enumerate}

At each step, an inner radius for the fit was chosen to exclude rings
within 2 beamwidths of the center, which were strongly affected by
finite resolution effects (beam smearing).  An outer radius was chosen
to ensure that the ring was well-sampled in azimuth and did not show
signs of a strong warp (a systematic drift in $\Gamma_0$ with radius).
Since the program fits each ring separately, a weighted average of the
values for all rings was used as the improved estimate for the next
step.  The full-resolution \HI\ maps were generally most useful for
constraining the fit parameters, since they contain a large number of
independent resolution elements in a region where $v_c$ is roughly
constant, but the CO maps provide better constraints on the kinematic
center since they have higher resolution (and often signal-to-noise)
in the central region.

Once an initial rotation curve and set of orientation parameters had
been determined, model velocity fields were generated (assuming a flat
rotation curve beyond the outer edge of the fitted disk) and compared
to the observed velocity fields.  Pixels in the velocity fields which
were discrepant by more than $\sim$50 \kms\ from the corresponding
model were flagged.  This threshold was increased when necessary so
that only points that were clearly noise were rejected: care was taken
to ensure that the threshold did not exclude gas that was continuous
in velocity with gas in the disk.  Although this process introduces
some noise bias (noise pixels with velocities within 50 \kms\ of the
model are not eliminated), it provides an excellent compromise between
sensitivity (which would be degraded by too strict a Gaussian fit
threshold) and noise rejection, as evidenced by the generally smooth
velocity contours in Fig.~\ref{fig:velfields}.

The cleaned velocity fields were then subjected to the ROTCUR analysis
again to derive ``final'' values for the fit parameters, which are
summarized in Table~\ref{tbl:rotpars}.  The uncertainties given
represent the range in values for the various velocity fields
analyzed, and are thus generally larger than the formal least-squares
errors for a single velocity field.  In \S\ref{sec:diskerr} we discuss
how errors in the adopted disk parameters would affect the resulting
harmonic decomposition.

Note that in most of the galaxies the kinematic center is displaced
$<1\arcsec$ from the center defined by the $K$-band or radio continuum
nucleus.  Only NGC~4414 and 4736 show a displacement of
$\sim$1\farcs5, which is of marginal significance given the
uncertainties in determining the optical and kinematic centers.  For
the subsequent analyses presented here, we have adopted the kinematic
rather than optical center.

\subsection{Isophote Fits}\label{sec:isophot}

In order to better constrain the influence of spiral structure and
bars on the observed kinematics, we performed ellipse fitting to the
optical and NIR isophotes using the IRAF task ELLIPSE.  The fitting
program outputs the center $(x,y)$ coordinates, semimajor axis $a$,
ellipticity $\epsilon = 1-b/a$, and position angle $\phi$ of each fit.
The ellipse centers were fixed during the fitting except in the case
of NGC~4321, for which a satisfactory fit could not be obtained
without allowing the center to wander by up to $\sim$10\arcsec.
However, the difficulty in fitting the isophotes does not reflect an
uncertainty in the center position: the nucleus of NGC~4321 is
well-determined photometrically in the $K$-band image (to within
$\lesssim$2\arcsec) and the kinematic center is also well-determined
(to within $\sim$1\arcsec) by the CO velocity field.  The results of
the isophotal analysis are discussed in \S\ref{sec:kinmorph}.

\section{Signatures of Non-Circular Motions in Disks}\label{sec:method}

In this section we present some general results for non-circular
motions in disks.  We adopt a right-handed $xyz$ coordinate system
with the main disk of the galaxy confined to the $xy$ plane.  As
discussed by \citet*{Franx:94}, the direction of our sight-line is
defined by two angles, $\theta_{\rm obs}$ and $i$, where $\theta_{\rm
obs}$ is measured in the $xy$ plane from the $x$-axis and $i$ is
measured from the $z$-axis [see Figure~\ref{fig:geom}(a)]. Note that
$i$ is restricted to the range $0<i<\pi/2$.  Figure~\ref{fig:geom}(b)
then shows the appearance of the galaxy as viewed in the sky.  Here
$\Gamma$ is the position angle measured in the sky plane, from north
to east, and $\Gamma_{\rm node}$ is the position angle of the line of
nodes (the intersection of the galaxy and sky planes).  Although in
principle we could define the $x$-axis so that $\theta_{\rm obs}$=0,
we choose to keep it independent of $\theta_{\rm obs}$ to more easily
handle the case of a bar.

Adopting the convention that positive line-of-sight (``radial'')
velocities correspond to recession, we therefore have:
\begin{equation}
V_{\rm los} = V_{\rm sys} + v_c \cos (\theta-\theta_{\rm obs}-\pi/2) \sin i
\end{equation}
for an axisymmetric model with rotation curve $v_c(R)$.  Here
$V_{\rm sys}$ is the mean (systemic) velocity of the galaxy.  For the
most general case of a two-dimensional velocity field,
\begin{equation}
V_{\rm los} (x,y) = V_{\rm sys} + v_\theta (x,y) \cos\psi\sin i + v_R (x,y)
\sin\psi\sin i\;,
\label{loseq}
\end{equation}
where $(v_R,v_\theta)$ is the velocity vector in the disk in polar
coordinates and $\psi\equiv\theta-\theta_{\rm obs}\mp\pi/2$ is the
azimuthal angle in the galaxy plane, measured from the receding side
of the line of nodes (the $-$ sign is taken for counterclockwise
rotation and the $+$ sign for clockwise).  With these definitions
$v_\theta$ is always positive, and $v_R>0$ indicates outflow for
counterclockwise rotation and inflow for clockwise rotation.

In each case below, we seek to express the model velocity field
$V_{\rm los}$ in terms of harmonic coefficients $c_j(R)$ and $s_j(R)$,
where
\begin{equation}
V_{\rm los} = c_0 + \sum_{j=1}^n \left[c_j \cos (j\psi) + s_j \sin (j\psi)
\right]
\label{eqn:decomp}
\end{equation}
For the simple models we consider, it is only necessary to expand out
to $n$=3.  For the case of pure circular rotation, $c_0$=$V_{\rm sys}$ and
$c_1$=$v_c \sin i$, with all other coefficients 0.
 
\subsection{Axisymmetric Radial Flow}

The simplest example of non-circular motion is one in which $v_R$ and
$v_\theta$ are functions of radius $R$ only:
\begin{equation}
V_{\rm los} = V_{\rm sys} + v_\theta(R)\cos\psi\sin i + v_R (R) \sin\psi\sin i
\end{equation}
This corresponds to circular motion superposed on axisymmetric inflow
or outflow.  The harmonic coefficients are simply $c_0$=$V_{\rm sys}$,
$c_1$=$v_\theta \sin i$, and $s_1$=$v_R \sin i$.  Of course, for the
inflow case a continuity problem arises unless some sink for the
inflowing gas exists: this might be star formation, a central black
hole, or an outflowing wind ejected out of the plane.

\subsection{Elliptical Streaming}\label{sec:ellstr}

A straightforward alternative to simple radial flows is the case in which
gas follows elliptic closed orbits, as might occur in a bar-like
potential.  For a general distortion of harmonic number $m$ rotating
at a fixed pattern speed $\Omega_p$, the potential is given by:
\begin{equation}
\Phi (R,\theta) = \Phi_0 (R) + \Phi_m(R) \cos [m(\theta-\Omega_pt)]
\label{potent}
\end{equation}
where $\Phi_0$ is the axisymmetric part of the potential.  For
simplicity we take $\Phi_0$ to be a spherical logarithmic potential,
giving a flat rotation curve with circular speed $v_c$:
\[\Phi_0 (R) = v_c^2 \ln R \;.\]
We assume a uniform $m$=2 perturbation to the potential given by:
\[\Phi_2 = -\frac{1}{2}\epot v_c^2\;,\]
so that the ellipticity of the potential at all radii is \epot.  The
sign of $\Phi_2$ is taken to be negative in order that the
equipotentials are elongated along the $x$-axis ($\theta$=0) at $t$=0.
Potentials of this type have been analyzed in a number of previous
studies \citep[e.g.,][]{Franx:94,Kuijken:94,Rix:95,Jog:00}.


\begin{figure*}
\begin{center}
\plottwo{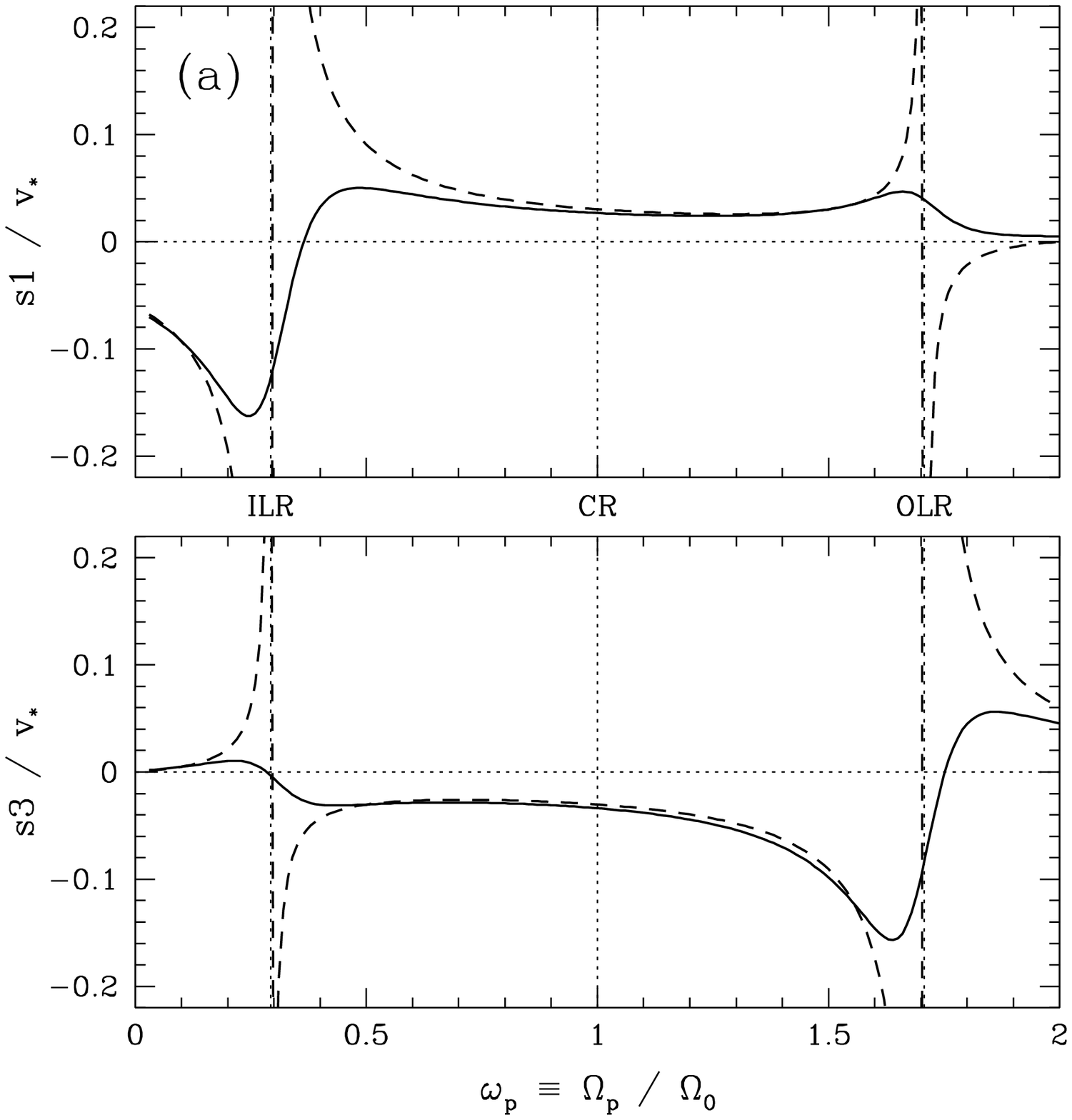}{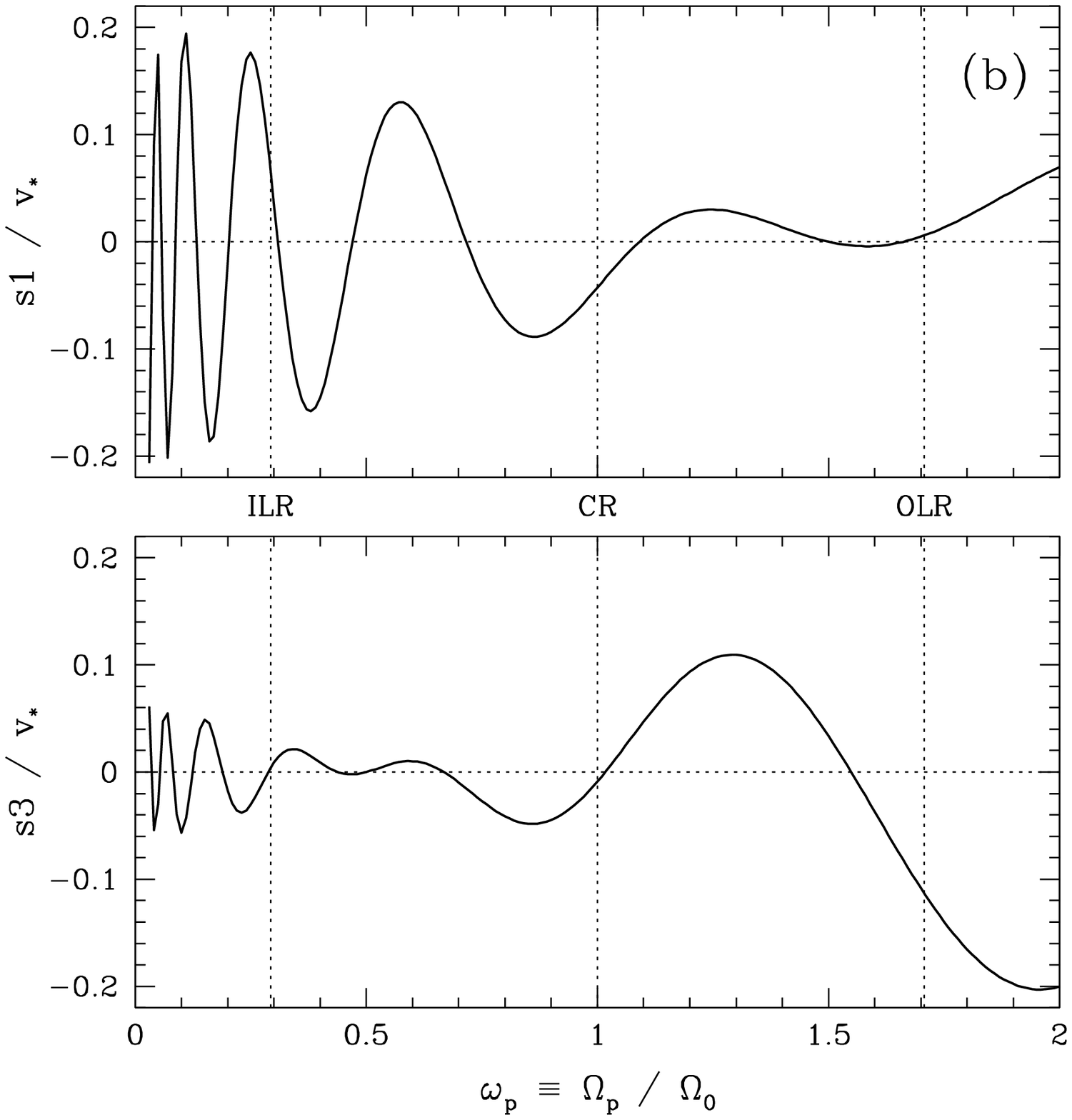}
\end{center}
\caption{
(a) Harmonic coefficients $s_1$ and $s_3$ normalized by $v_* = v_c\sin
i$, plotted as a function of $\omega_p \propto R$ for a collisionless
model ({\it dashed line}) and a dissipative model ({\it solid line})
with a flat rotation curve and a rotating, globally elongated
potential.  The locations of the ILR, CR, and OLR are marked as dotted
lines.  (b) Same but for a two-armed spiral model with a flat rotation
curve, pitch angle of 10\arcdeg, and velocity perturbation amplitude
$v_{\rm sp}=0.2v_c$.
\label{fig:s13models}}
\end{figure*}

First we consider collisionless models, which are strictly applicable
only to stellar orbits.  
Using the potential in Eq.~\ref{potent} and the equation of motion in
a frame rotating at angular velocity $\Omega_p$ \citep{Binney:87},
one can solve for the allowed closed, periodic orbits using
first-order epicyclic theory (this assumes the bar perturbation is
relatively weak).  \citet*{Schoen:97} showed that a potential
distortion of order $m$ introduces changes in the $(m-1)$st and $(m+1)$st
harmonic coefficients.  For the $m$=2 case, their expressions reduce to:
\begin{eqnarray}
c_0 & = & V_{\rm sys}\\
c_1 & = & v_* \left\{1-\frac{1}{4}[(2-3\omega_p)a_2 + (1-\omega_p)b_2]
	\cos 2\theta_{\rm obs}\right\} \nonumber\\
s_1 & = & v_* \left\{\frac{1}{4}[(2-3\omega_p)a_2 + (1-\omega_p)b_2]
	\sin 2\theta_{\rm obs}\right\} \nonumber\\
c_3 & = & v_* \left\{-\frac{1}{4}[(\omega_p-2)a_2 + (1-\omega_p)b_2]
	\cos 2\theta_{\rm obs}\right\} \nonumber\\
s_3 & = & v_* \left\{\frac{1}{4}[(\omega_p-2)a_2 + (1-\omega_p)b_2]
	\sin 2\theta_{\rm obs}\right\} \nonumber
\end{eqnarray}
where $v_* \equiv v_c \sin i$, $\omega_p \equiv \Omega_p/\Omega_0 =
R(\Omega_p/v_c)$, and the ellipticity coefficients
$a_2$ and $b_2$ are given by:
\begin{eqnarray}
a_2 & = & \frac{1}{(1-4\omega_p+2\omega_p^2)(1-\omega_p)}\,\epot\\
b_2 & = & \frac{2-3\omega_p+2\omega_p^2}{(1-4\omega_p+2\omega_p^2)
	(1-\omega_p)^2}\,\epot \nonumber
\end{eqnarray}
The coefficients $s_1$ and $s_3$ for this model are plotted with a
dashed line in Figure~\ref{fig:s13models}(a) as a function of
$\omega_p$, which is proportional to the radius $R$.  Note that
between the ILR (inner Lindblad resonance, at $\omega_p=1-\sqrt{2}/2$)
and OLR (outer Lindblad resonance, at $\omega_p=1+\sqrt{2}/2$), {\it
the $s_1$ and $s_3$ terms have opposite sign}.

Although gas is not expected to follow collisionless orbits,
\citet{Gerhard:86} and \citet{Franx:94} have argued that a
collisionless treatment should be applicable to gaseous orbits if
the potential is {\it non-rotating}, since gas will then be able to settle
into stable closed orbits.  The special case of a stationary
potential ($\omega_p$=0) was solved previously by \citet{Franx:94}:
\begin{eqnarray}
c_1 & = & v_*(1 - \epot \cos 2\theta_{\rm obs})\\
s_1 & = & v_* \epot \sin 2\theta_{\rm obs} \nonumber\\
c_3 & = & s_3 = 0  \nonumber
\end{eqnarray}
Here no resonances exist and all orbits are aligned along the
$y$-axis, perpendicular to the bar.  The fact that the $c_3$ and $s_3$
terms vanish has an important implication: {\it for a flat rotation
curve, the velocity field of a stationary bar potential resembles
axisymmetric inflow or outflow.}  Thus kinematics alone cannot
distinguish these two cases; in \S\ref{sec:phot} we will discuss how
photometric measurements might break this degeneracy.  Since
measurement of $c_1$ only provides the value of $v_*(1-\epot
\cos 2\theta_{\rm obs})$, and $v_*$ cannot be determined
independently, it is impossible to disentangle \epot\ from the viewing
angle $\theta_{\rm obs}$.  However, under the assumption that $c_1 \approx
v_*$, a measurement of the $s_1$ term provides an estimate of the
quantity $\epot \sin2\theta_{\rm obs}$, and hence a lower limit on \epot.

For the more general case of gas flow in a rotating potential, as in a
barred galaxy, it is possible to relax the collisionless assumption by
including a damping term in the equation of motion to simulate the
dissipative effect of gas viscosity.  This can be done as described in
Appendix~\ref{sec:damping}.  The corresponding harmonic coefficients
$s_1$ and $s_3$ are plotted with a solid line in
Figure~\ref{fig:s13models}(a).  Their behavior is qualitatively
similar to the collisionless model, aside from a smoother variation
near the resonances.  For the non-rotating case, the dissipative and
collisionless models yield the same coefficients.

To determine the effect of using a more realistic potential, we also
ran models using the analytic potential employed by \citet{Wada:94},
which gives a rotation curve that can be parametrized as $v_c \propto
R (R^2+a^2)^{-3/4}$.  Although the results differ in detail from the
flat rotation curve case, the $s_1$ and $s_3$ coefficients still have
opposite signs except near major resonances.  The region where the rotation
curve rises is associated with a second inner Lindblad resonance (the
inner ILR) across which the coefficients both change sign.

\subsection{Spiral Arm Streaming}

A two-armed spiral density wave can be thought of as an $m$=2
perturbation with a phase shift that varies with radius.  One would
therefore expect that the velocity perturbations due to spiral arms
would be in rough qualitative agreement with the results given in
\S\ref{sec:ellstr}.  For a more detailed analysis we adopt the
linearized equations given by \citet{Canzian:97} for the velocity
perturbations due to a spiral potential, which are based on the
original treatment given by \citet*{Lin:69}.  Substituting into their
equations, we find for a two-armed logarithmic spiral the harmonic
coefficients
\begin{eqnarray}
\frac{c_1-v_c}{\sin i} &=& \frac{v_{\rm sp}}{2}\left(\frac{\kappa}{2\Omega}-\nu
	\right)\sin \left(2\theta_{\rm sp} - \chi\right)\\
\frac{s_1}{\sin i} &=& \frac{v_{\rm sp}}{2}\left(\frac{\kappa}{2\Omega}-\nu
	\right)\cos \left(2\theta_{\rm sp} - \chi\right)\nonumber\\
\frac{c_3}{\sin i} &=& \frac{v_{\rm sp}}{2}\left(\frac{\kappa}{2\Omega}+\nu
	\right)\sin \left(2\theta_{\rm sp} + \chi\right)\nonumber\\
\frac{s_3}{\sin i} &=& \frac{v_{\rm sp}}{2}\left(\frac{\kappa}{2\Omega}+\nu
	\right)\cos \left(2\theta_{\rm sp} + \chi\right)\nonumber\;.
\end{eqnarray}
Here $\theta_{\rm sp}$ is the spiral phase, defined at a fiducial
radius $R_0$ by
\[\theta_{\rm sp} = \frac{\ln (R/R_0)}{\tan\chi}\;,\]
where $\chi$ is the {\it pitch angle}, the angle an arm makes with the
tangent to a circle.  The dimensionless frequency $\nu$ and epicyclic
frequency $\kappa$ are defined as
\begin{displaymath}
\nu \equiv \frac{m(\Omega_p-\Omega)}{\kappa}\;,\quad\quad
\kappa^2 \equiv  4\Omega^2 + R \frac{d\Omega^2}{dR}\;,
\end{displaymath}
and $v_{\rm sp}$ is a velocity amplitude that depends on the strength
of the spiral perturbation.

The $s_1$ and $s_3$ coefficients for a flat rotation curve are
plotted in Figure~\ref{fig:s13models}(b) as a function of $\omega_p
\propto R$.  A characteristic of the spiral wave is the sinusoidal
variation in the coefficients with radius---since the spiral has no
preferred orientation, the observer in effect samples different
orientations of the perturbed velocity field at different radii.

In a realistic treatment including the nonlinear effect of shocks, the
spiral velocity perturbations will not be purely sinusoidal
\citep[e.g.,][]{Roberts:87}, the peaks in Figure~\ref{fig:s13models}(b)
will be narrower, and additional Fourier terms will occur.  However,
two basic properties of the model are worth noting since they are
likely to be quite general.  First, there is the change in dominance
from the $s_1$ term to the $s_3$ term that occurs at corotation, which
\citet{Canzian:93} showed must occur when crossing a corotation
resonance.  The same phenomenon occurs in the bar model, although to a
lesser degree [Figure~\ref{fig:s13models}(a)].  Second, the signs of $s_1$
and $s_3$ are the same near corotation (where $\nu=0$), but tend to
take on opposite signs as one moves away from the corotation resonance
(CR).  For the bar models, on the other hand, the two coefficients
have opposite signs at nearly all radii.  As shown by
\citet{Schoen:97}, this difference can be attributed to additional
contributions to $s_1$ and $s_3$ that result from the radial variation
in the phase of the potential.

\subsection{Warped Disk}

Although not directly related to radial gas flows, warps are a common
feature in the outer \HI\ disks of galaxies, and are well-known to
have a significant effect on the observed velocity field
\citep[e.g.,][]{Bosma:78}.  A warp is characterized by gas moving in
circular orbits, but with the angular momentum vectors of the orbits
not aligned as they would be in the case of a flat disk.  Most
galaxies are not strongly warped within their optical disks \citep[for
a review of observed properties of warps see][]{Briggs:90}, so the
plane of the inner disk defines a natural coordinate system for the
galaxy.

The kinematics of warped galaxies can usually be well-modeled with a
tilted-ring fit \citep*{Rogstad:74,Bosma:81b}, in which the inclination
$i$ and the major axis position angle $\Gamma_{\rm maj}$ are allowed
to vary with radius.  In contrast to the cases of radial flows or
elliptical streaming, the kinematic axes $\Gamma_{\rm maj}$ and
$\Gamma_{\rm min}$ (see \S\ref{sec:phot}) remain perpendicular, even
as they drift with radius \citep{Bosma:81b}, since the gas flow is
still circular.  A harmonic decomposition of the velocity field using
fixed disk parameters will display the characteristic signatures due
to incorrect values of the position angle and inclination discussed
below (\S\ref{sec:diskerr}).  The coefficients $c_1$, $s_1$,
$c_3$, and $s_3$ will all change with radius, with the $c_1$ and $s_1$
terms being most affected.  Again this contrasts with the case of 
elliptical or spiral streaming, where changes in the $c_3$ and $s_3$
terms will dominate beyond the corotation radius.

\subsection{Errors in the Parameters and Beam Smearing}\label{sec:diskerr}

We have assumed up to this point that the disk parameters used in
making the harmonic expansion (namely, the kinematic center,
inclination, and position angle) are correct.  It is clear that this
will not generally be the case, since the disk parameters will likely
be derived from the velocity field under the assumption of circular
rotation.  \citet{Schoen:97} analyzed the changes in the harmonic
coefficients that result from using incorrect disk parameters, and
found that the Fourier components due to non-circular motions will mix
with those due to incorrect disk parameters.  To first order, and
assuming the errors in the position of the kinematic center are small
compared to the radii of interest (satisfied for our galaxies, see
Table~\ref{tbl:rotpars}), the following rules apply:
\begin{enumerate}
\item An error in the position angle $\Gamma_0$ leads to offsets
in the $s_1$ and $s_3$ terms proportional to $v_* \equiv v_c \sin i$.
The offset in $s_1$ is generally larger than that in $s_3$.
\item An error in the inclination $i$ leads to offsets in the
$c_1$ and $c_3$ terms which are proportional to $v_*$.  Because
the $c_1$ term is large, the effect will only be detectable in the
$c_3$ term.
\item An error in the kinematic center leads to offsets in the
$c_0$, $c_2$, and $s_2$ terms proportional to $v_*/R$.
\end{enumerate}

What is the effect of finite resolution (beam smearing) on the
coefficients?  Smoothing a datacube to lower resolution affects
primarily the $c_1$ term (since the rotation curve rise becomes more
gradual) and the $c_3$ term (since the isovels near the galaxy center
become more parallel, leading to a lower fitted inclination).  On the
other hand, the $s_1$ and $s_3$ terms, which are sensitive to the
kinematic position angle, should be relatively unaffected by beam
smearing outside of the central few resolution elements, since
smoothing an image should not lead to any net rotation of the velocity
contours.  (Of course, any small-scale variations in these terms will
be smoothed out, but there should be no {\it systematic} offsets.)  We
have confirmed these effects using simulations of axisymmetric
velocity fields convolved with both symmetric and asymmetric beams
\citep[for details see][]{Wong:th}.  We note, however, that our
simulations have assumed an axisymmetric filled disk; if the gas
distribution is very inhomogeneous, or confined to a small range in
azimuth, it is conceivable that beam smearing could affect the
measured $s_1$ and $s_3$ terms as well.


\begin{figure*}
\begin{center}
\plottwo{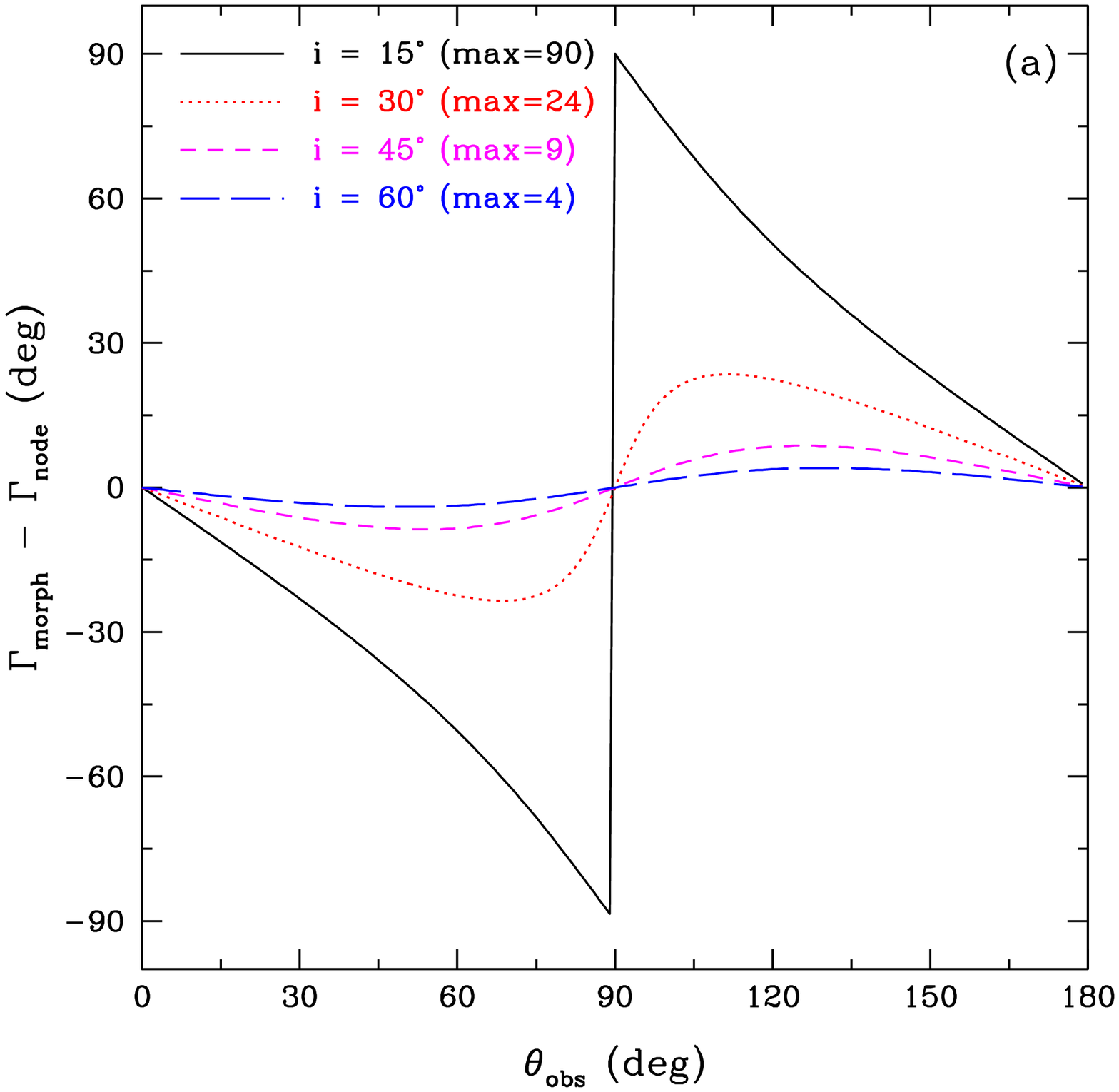}{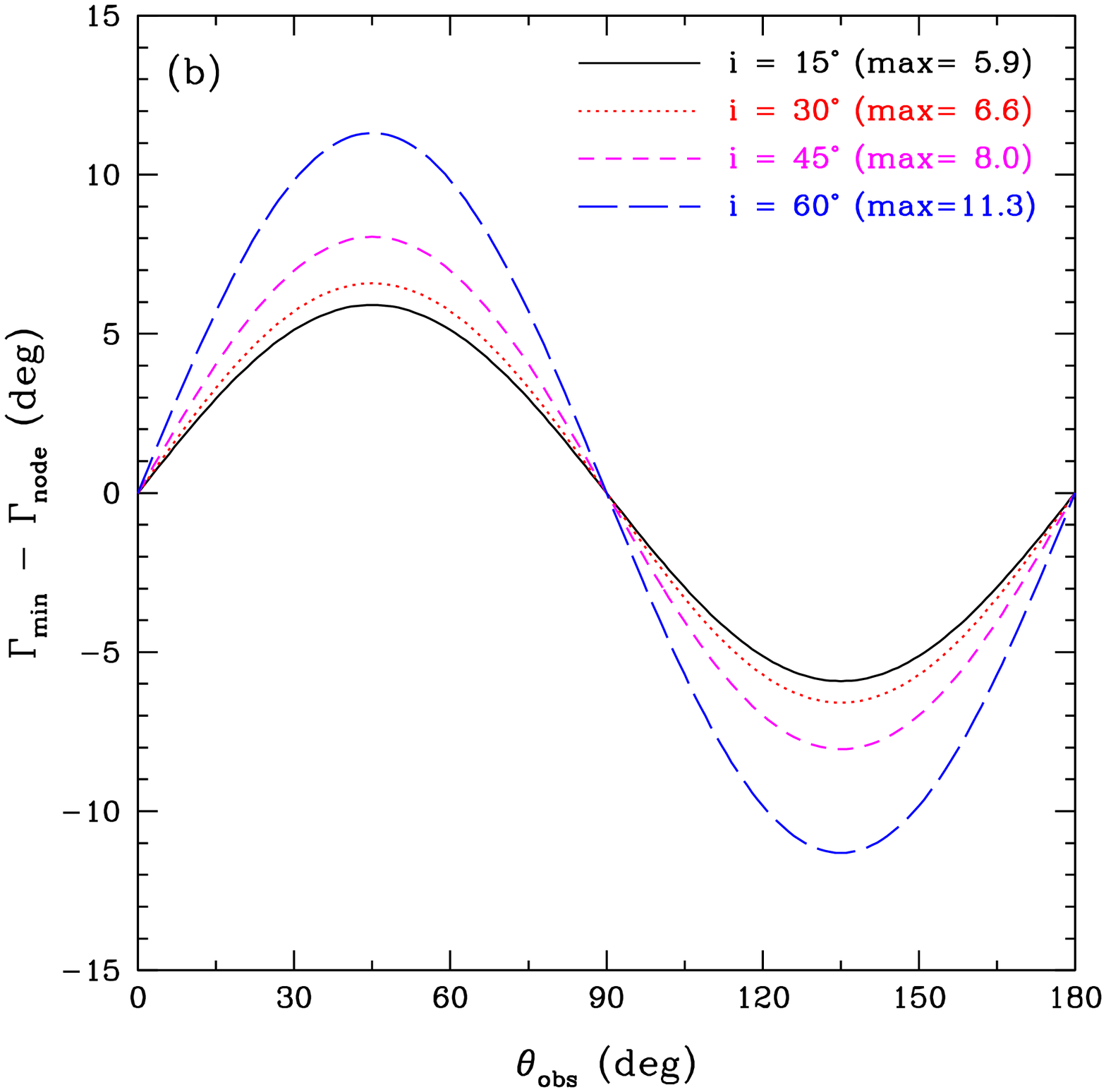}
\end{center}
\caption{
(a) Offset of morphological position angle from the line of
nodes, as a function of $\theta_{\rm obs}$, for isophotes of ellipticity
0.1.  (b) Offset of kinematic minor axis position angle from
the line of nodes, for an elongated potential with \epot=0.1.  In both
panels, a range of inclinations $i$ is shown.  The numbers in
parentheses in the legend give the maximum value (in degrees) for the
corresponding curve.
\label{fig:kinmorphpa}}
\end{figure*}

\subsection{Inclusion of Isophotal Data}\label{sec:phot}

Given that it is nearly impossible to kinematically distinguish a radial
inflow model from an elliptical streaming model when $\omega_p \equiv
\Omega_p/\Omega_0 \ll 1$, i.e.\ when the potential is rotating slowly
compared to the local circular speed, we must consider whether
additional constraints can be provided by optical or near-infrared
surface photometry.  It is worth noting at the outset that since the
isophotes (contours of constant surface brightness) of an axisymmetric
disk are concentric aligned ellipses, {\it any} change in the
isophotes over a region where the kinematics suggest pure radial flows
would raise the possibility that streaming motions are occurring.
A clearer prediction can only be made if one restricts
consideration to an elliptical streaming model in which the gas orbits
are elongated in the {\it same} direction as the isophotes.  Then, as
shown below, the morphological and kinematic position angles will
deviate in opposite directions with respect to the line of nodes.

First, however, we need to identify conditions under which the
elongation of the gas orbits will be parallel to the elongation of the
isophotes.  We consider three possibilities, two of which would 
satisfy this requirement:
\begin{enumerate}
\item For a stationary elongated potential, the closed loop orbits are
perpendicular to the elongation of the potential, so if the stars
follow loop-type orbits like the gas, the isophotes will be aligned
with the gas orbits.  In this case, however, the stars obviously
cannot support the potential---in fact they will tend to counteract
its asymmetry \citep{Jog:00}.  Rather, this case would be appropriate
for a disk galaxy embedded in a massive triaxial halo.
\item Most of the stars could follow box-type orbits that are aligned
with the potential, so that the isophotes are perpendicular to the gas
orbits.  This would allow a self-consistent solution between the stellar 
density and the potential it generates \citep*{deZeeuw:87},
and no triaxial halo would be required.
\item The potential of a {\it rotating} bar will be aligned with the
stable loop orbits between ILR and CR (these are often referred to as $x_1$
orbits), as shown in Figure~\ref{fig:orbits} in the Appendix, so again
the isophotes will be parallel to the gas orbits as in Case 1.
\end{enumerate}
Thus, by assuming that the gas orbits are elongated parallel to the
isophotes, we are assuming that Case 1 or 3 applies.  Note that Case
1, which implies a strongly sub-maximal disk, conflicts with most
observational studies of high surface brightness galaxies
\citep{Bell:01b,Bottema:93}.  On the other hand, Case 3 is not ruled
out by such considerations and should be readily apparent in the
isophotes, since bars are strong between the ILR and CR and weaker
outside this region.  We assume throughout our discussion that the
isophotes trace a thin disk, so we can neglect the complications of
projecting an inherently 3-dimensional structure onto the plane of the
sky.

If an orbit (or isophote) has an intrinsic axis ratio $Q = b/a$ and
position angle $\psi_0$ with respect to the line of nodes, how will it
appear when projected on the sky at some inclination angle $i$?  The
appropriate geometrical transformations can be found in
\citet{Teuben:91}, who finds that the new axis ratio $q$ and
position angle $\gamma_0$ (still measured with respect to the line of
nodes) are given by
\begin{equation}
q^2 = \frac{(A\cos^2 i + C)\cos 2\psi_0 + (A\cos^2 i - C)}
	{(A\cos^2 i + C)\cos 2\psi_0 - (A\cos^2 i - C)}\\
\end{equation}
\begin{equation}
\tan 2\gamma_0 = - \frac{2B\cos i}{A\cos^2 i - C}
\end{equation}
where the coefficients are defined as
\begin{eqnarray}
A & \equiv & Q^2 \cos^2 \psi_0 + \frac{\sin^2 \psi_0}{Q^2}\\
B & \equiv & \frac{(1-Q^2)}{Q^2} \sin\psi_0 \cos\psi_0\nonumber\\
C & \equiv & Q^2 \sin^2 \psi_0 + \frac{\cos^2 \psi_0}{Q^2}\;.\nonumber
\end{eqnarray}
Note that we have corrected some typesetting errors in the
original paper by \citet{Teuben:91}.


\begin{figure*}
\includegraphics[width=5.2cm]{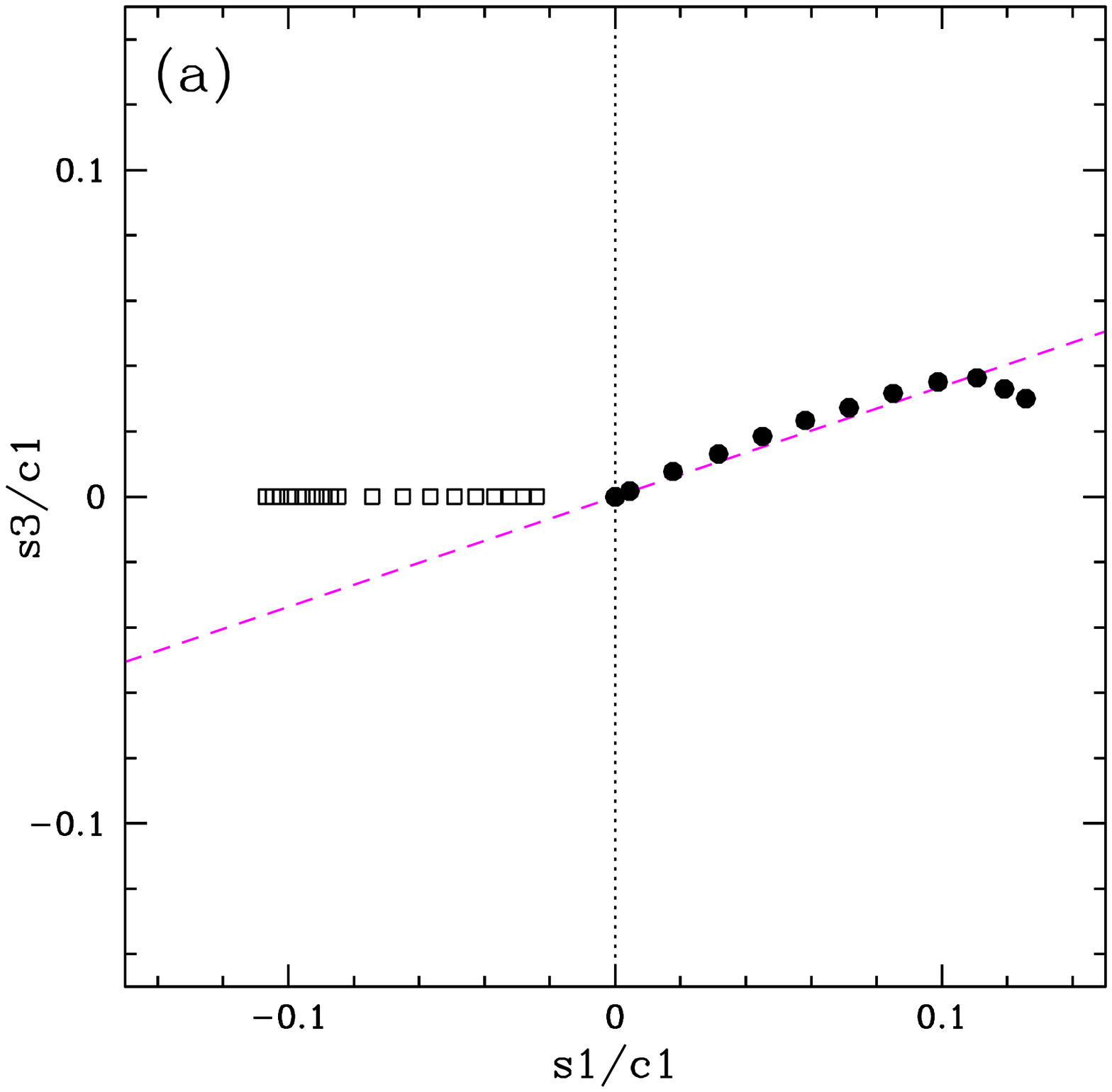}\hfill
\includegraphics[width=5.2cm]{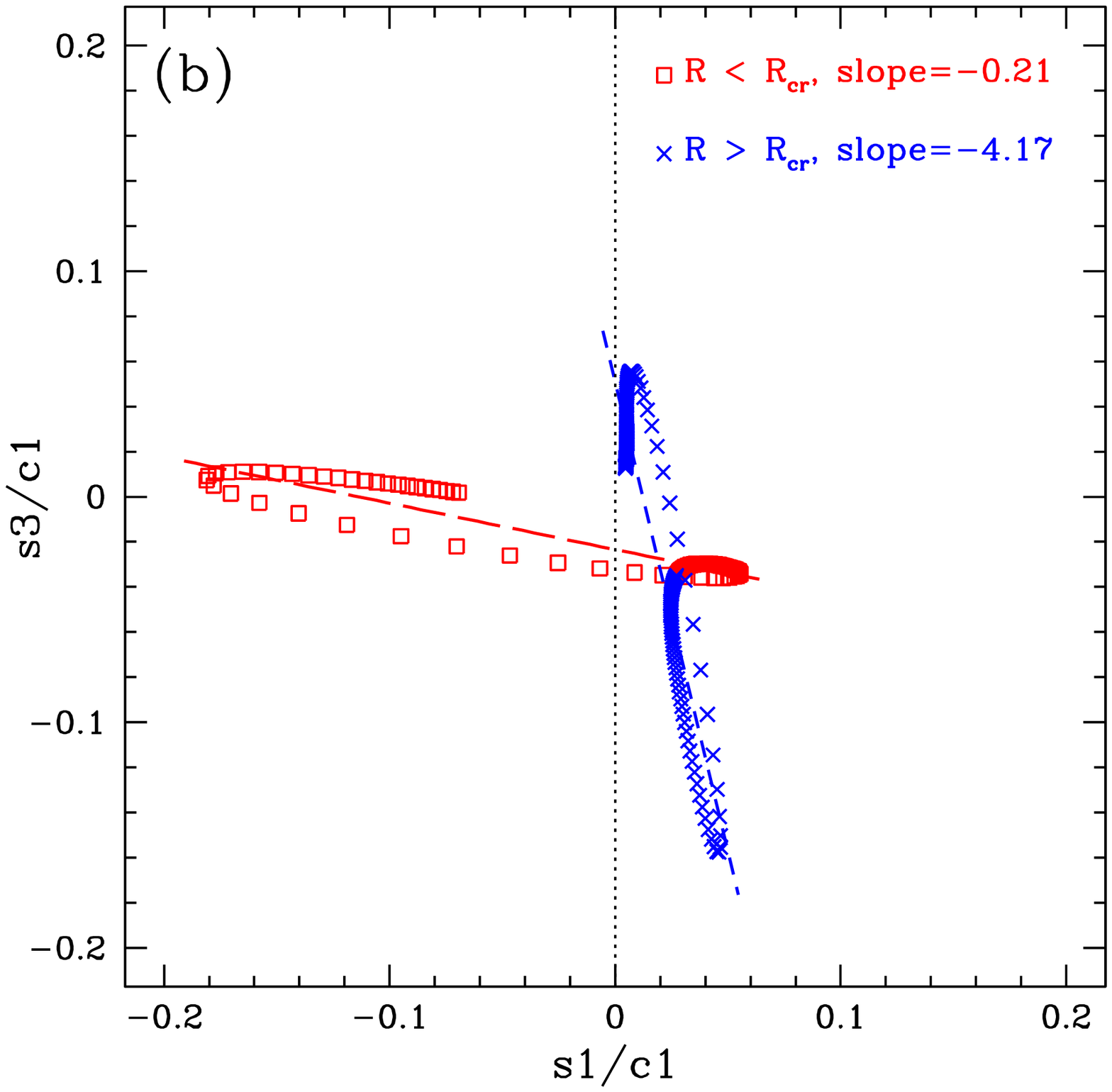}\hfill
\includegraphics[width=5.2cm]{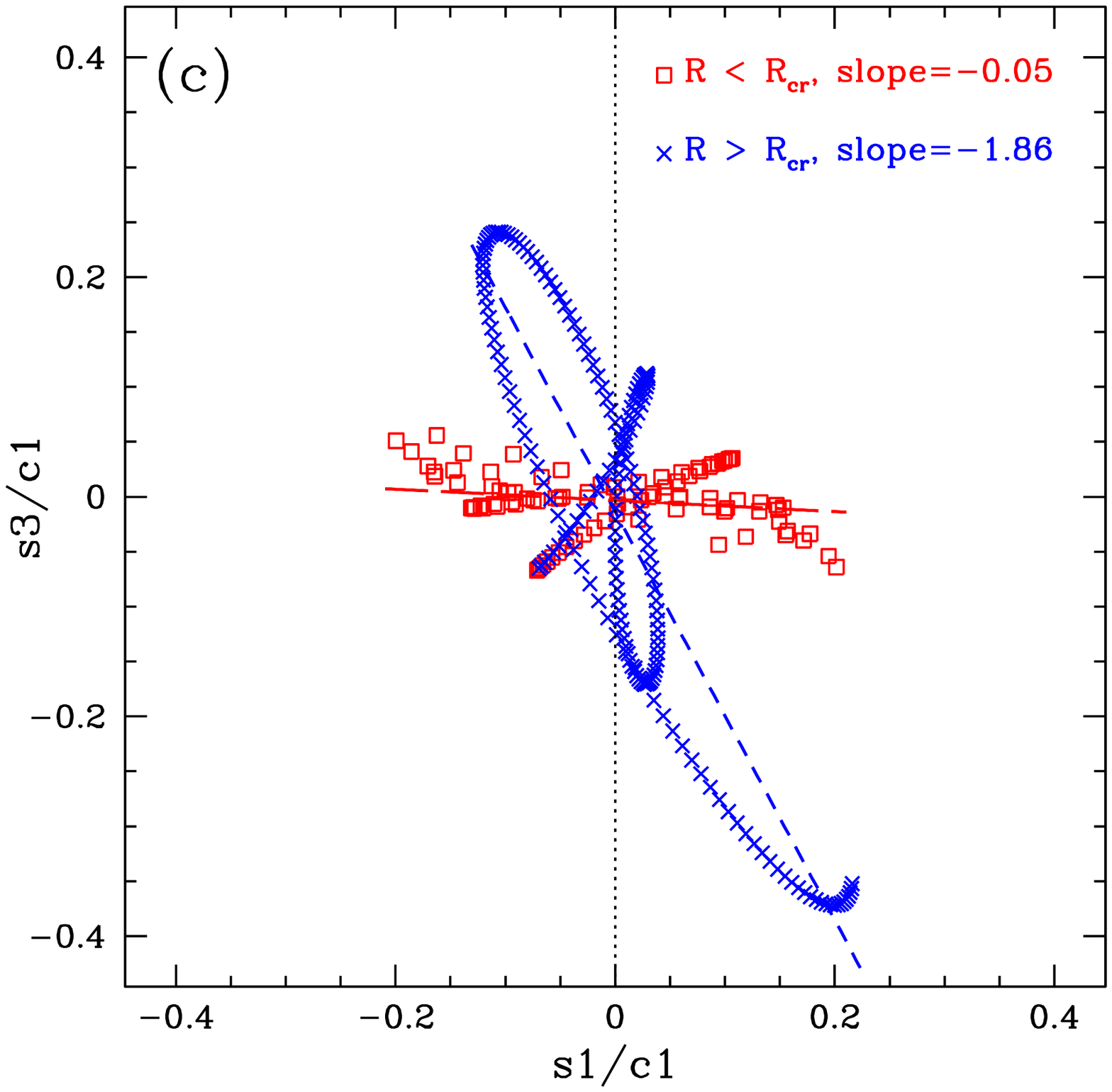}
\caption{
Comparison of $s_1$ and $s_3$ coefficients for (a) inflow ({\it open
squares}) and warp ({\it solid circles}) models, (b) bar streaming
model including dissipation, and (c) spiral streaming model with a
pitch angle of 10\arcdeg.  The dashed line in panel (a) represents the
``warp line'' where points should fall if there is an error in the
P.A., for the assumed inclination of 55\arcdeg.  Long-dashed and
short-dashed lines in panels (b) and (c) represent fits to points
inside and outside of corotation respectively (shown with different
plot symbols).
\label{fig:s1s3mod}}
\end{figure*}


Using these equations and the sky coordinate system described by
Figure~\ref{fig:geom}, we can calculate the offset $\gamma_0$ between
the apparent or {\it morphological} position angle of an elliptical
isophote ($\Gamma_{\rm morph}$) and the position angle of the true
line of nodes ($\Gamma_{\rm node}$).  For the model discussed in
\S\ref{sec:ellstr}, with equipotentials elongated along the $x$-axis, we
assume that the isophotes are elongated along the $y$-axis.  The
resulting offset angle is shown in Figure~\ref{fig:kinmorphpa}(a) as a
function of the viewing angle $\theta_{\rm obs}$ (measured from the
$+x$-axis), for an ellipticity of 0.1 ($Q$=0.9) and a range of
inclination angles.  Note that for low inclinations ($\cos i > Q$),
the offset angle can reach very large values if the line of sight is
close to the long axis of the ellipse.

For comparison, the offset between the kinematic minor axis (the locus
in the sky plane where $V_{\rm los}=V_{\rm sys}$) and the line of
nodes, for an orbit in an elongated potential, is plotted in
Figure~\ref{fig:kinmorphpa}(b) for the same set of inclination angles
(the kinematic major axis shows a similar but weaker effect).  This
angle can be expressed in terms of the harmonic coefficients as
\citep{Franx:94}:
\begin{equation}
\Gamma_{\rm min} - \Gamma_{\rm node} = \frac{\pi}{2} + \frac{s_1-s_3}
	{c_1-3c_3}\,\frac{1}{\cos i}\;.
\label{eqn:gammin}
\end{equation}
For all non-degenerate observing angles, the offset in $\Gamma_{\rm
min}$ is in the {\it opposite} direction from the offset in
$\Gamma_{\rm morph}$.  This is a robust result, depending purely on
geometry, as can be seen by drawing on a sheet of paper an ellipse
that is centered on, but has some angle to, the vertical and
horizontal axes.  The quadrants through which the apparent minor axis
passes will be different from the quadrants where the maximum
$y$-value is achieved, which is where the kinematic minor axis occurs
if the $x$-axis is the line of nodes.  Note also that a larger
inclination reduces $\Gamma_{\rm morph}$ but increases $\Gamma_{\rm
min}$, so {\it bars in highly inclined galaxies are best identified
kinematically rather than morphologically}.

\section{Results}\label{sec:results}

Before returning to the analysis of the seven galaxies in our sample,
let us summarize our basic approach to diagnosing various types of
non-circular motions.

Based on the results of \S\ref{sec:method}, the effects of elliptical
or spiral streaming are most easily distinguished from radial inflow
by examining the first and third order sine coefficients ($s_1$ and
$s_3$).  Variations in these coefficients as a function of radius can
be placed into three broad categories, corresponding roughly to warp,
elliptical streaming, and inflow models:
\begin{enumerate}
\item The $s_1$ and $s_3$ terms are correlated, with $ds_3/ds_1 > 0$.
\item The $s_1$ and $s_3$ terms are anti-correlated, with $ds_3/ds_1 <
0$.
\item The $s_1$ term is significant in some region but the $s_3$ term
is negligible ($s_3 \approx 0$).
\end{enumerate}
We reiterate that these criteria do not always permit a clear
distinction between physical models.  As noted in \S\ref{sec:ellstr},
a stationary bar potential (usually in the second category) can mimic
the kinematic signature of radial inflow (third category) inward of
the ILR.  A spiral streaming model (essentially equivalent to
elliptical streaming, but with the orbits misaligned) can produce all
three types of behavior, depending on location relative to the CR, but
will generally correspond to case 2.

Examples of how four different kinematic models appear in the
$s_1$-$s_3$ plane are given in Figure~\ref{fig:s1s3mod}.  For a pure
warp model, the $(s_1,s_3)$ points are predicted to lie along a line
we refer to as the ``warp line,'' defined by
\begin{equation}
\frac{\delta s_1}{\delta s_3} = \frac{3q^2+1}{1-q^2}\;,
\end{equation}
where $q \equiv \cos i$.  This is the predicted relation between $s_1$
and $s_3$ for an incorrect disk position angle (\S\ref{sec:diskerr}).
For elliptical streaming, the results depend somewhat on the choice of
model parameters, but in general one finds a shallow slope inside of
the corotation radius ($R_{\rm CR}$) and a steep slope outside.  This
is a reflection of the switch in dominance from the $s_1$ to the $s_3$
term as one crosses $R_{\rm CR}$ \citep{Canzian:97}, and also occurs
for the spiral streaming model.  As is clear from
Figure~\ref{fig:s1s3mod}, the greatest ambiguity arises when trying to
distinguish inflow from streaming motions in the region $R \ll R_{\rm
CR}$.  For a large sample of galaxies viewed from random orientations,
streaming motions should produce apparent outflow in roughly half of
the galaxies, but our sample is too small for such statistical
arguments to be useful.


\begin{figure*}[p]
\begin{center}
\includegraphics[scale=0.45]{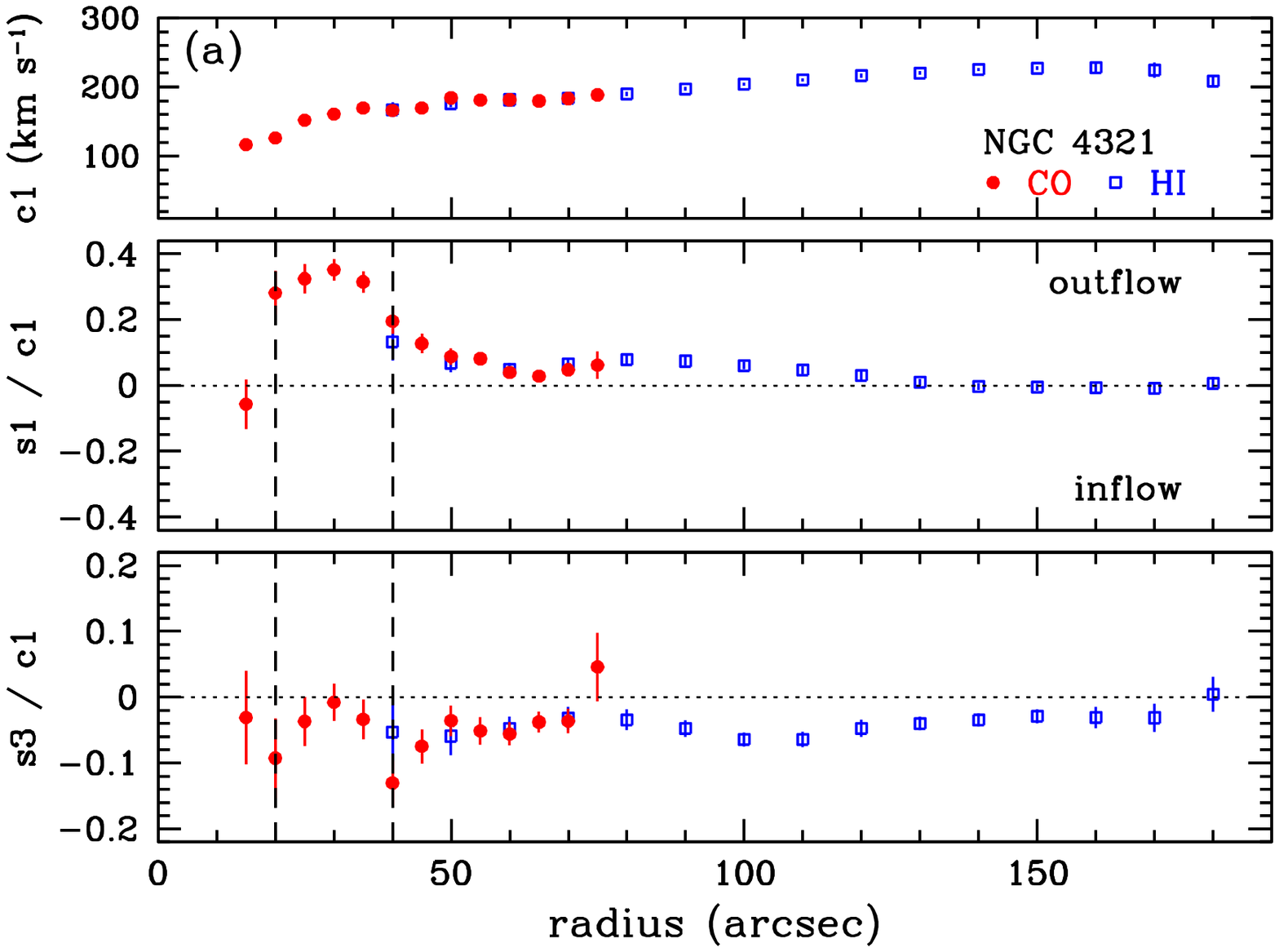}\hspace{1cm}
\includegraphics[scale=0.45]{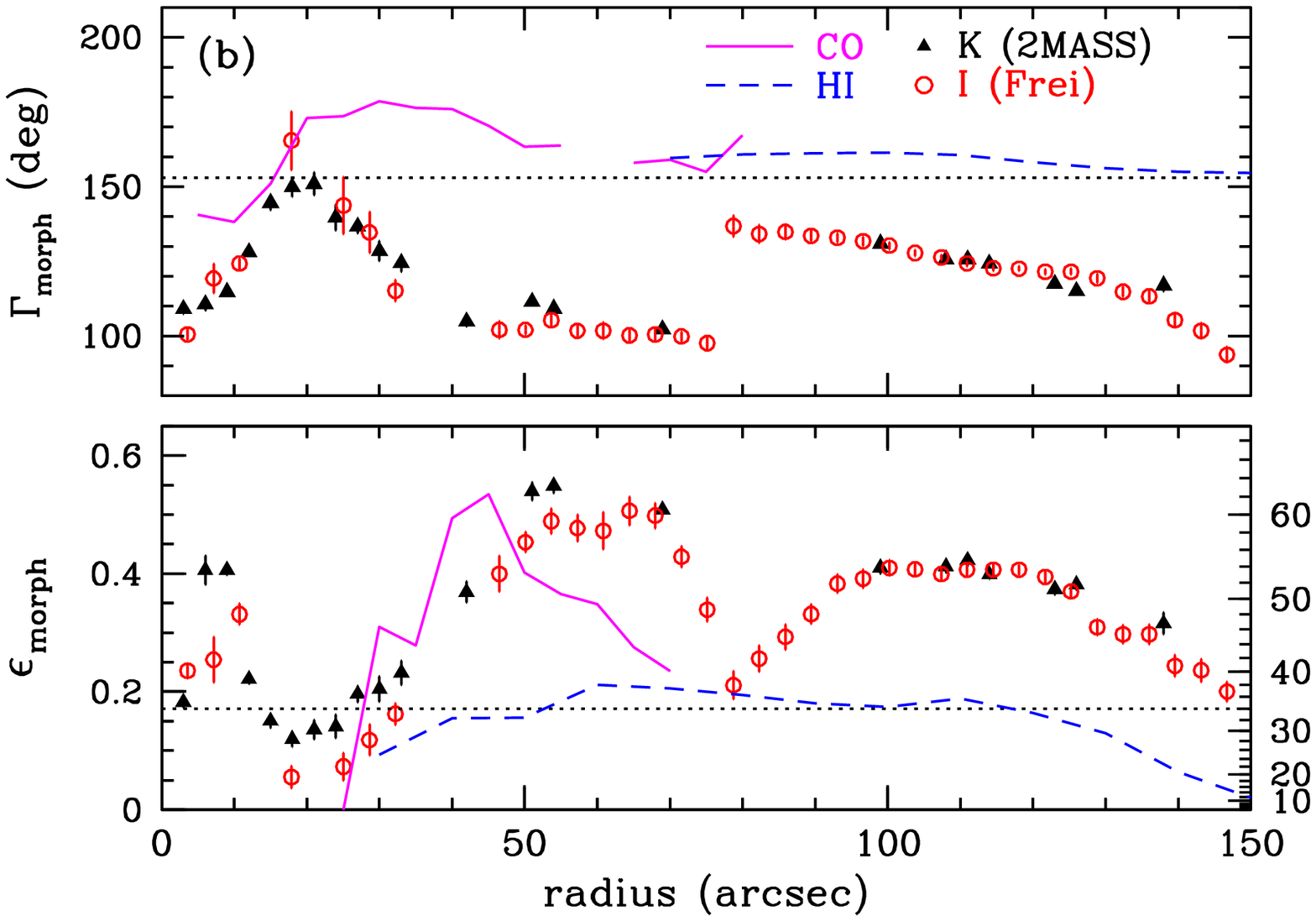}\\[5ex]
\includegraphics[scale=0.4]{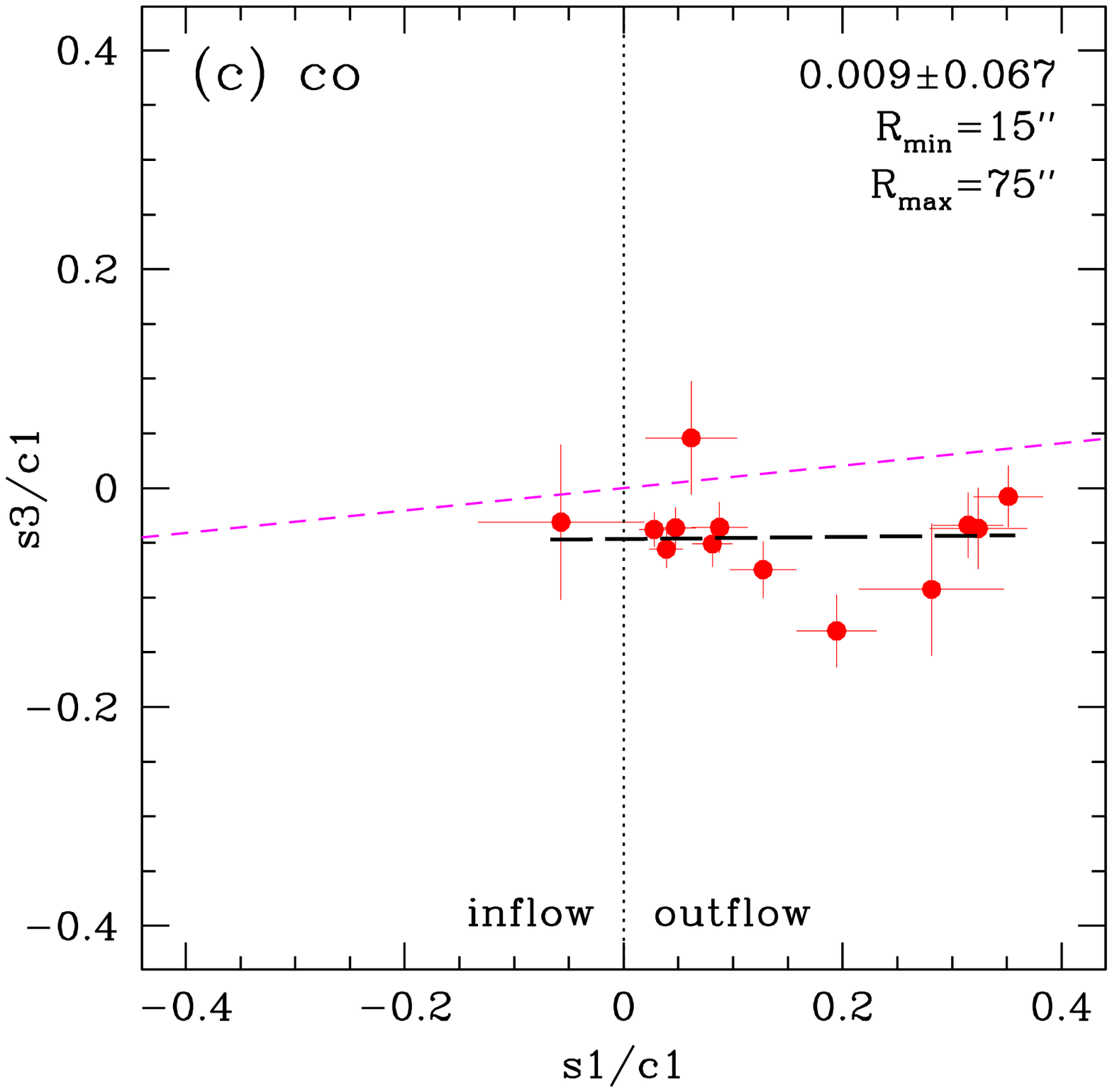}\hspace{1cm}
\includegraphics[scale=0.4]{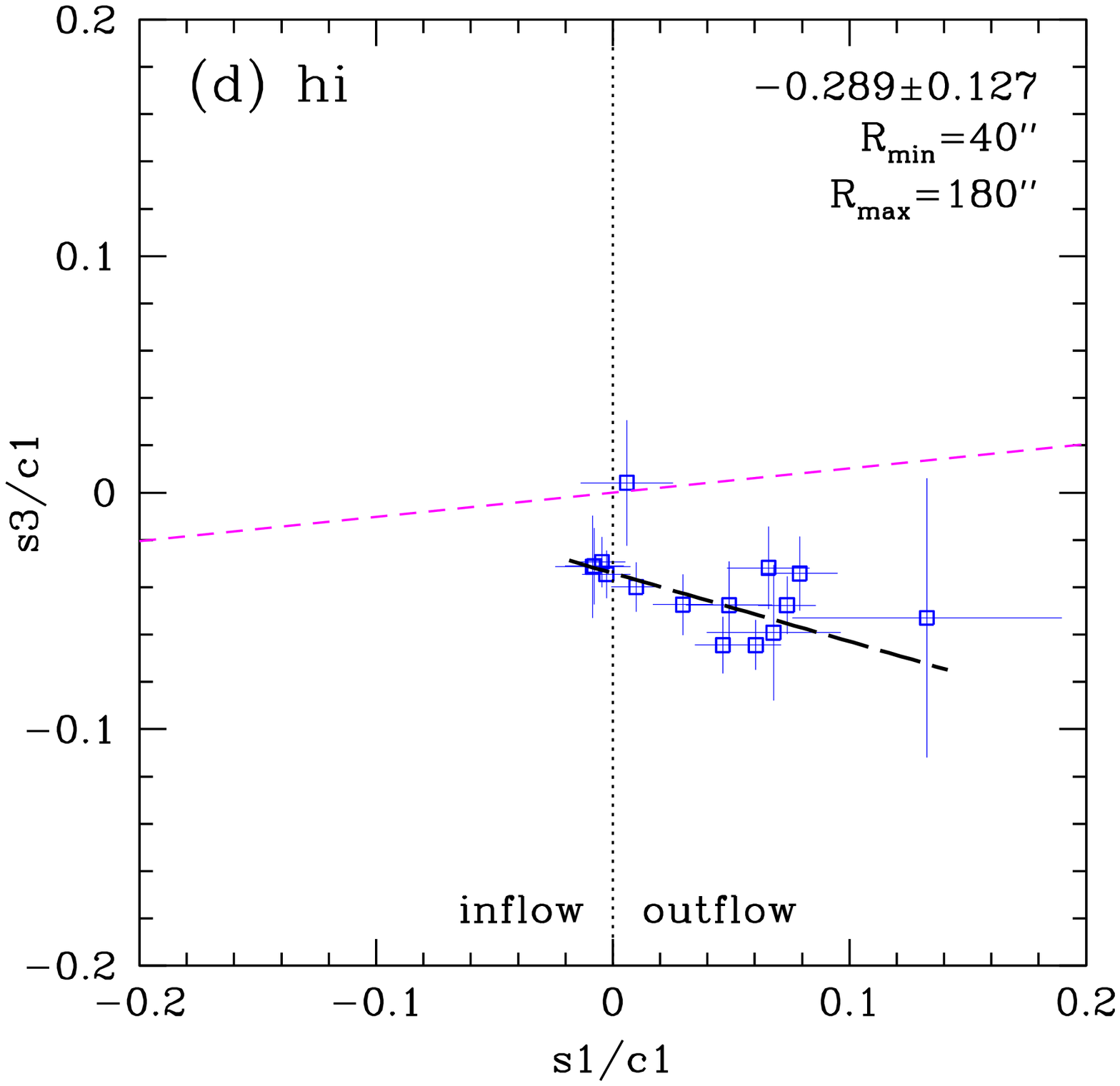}\\[5ex]
\end{center}
\caption{
(a) $c_1$, $s_1$ and $s_3$ coefficients plotted as functions of radius
for NGC 4321.  The $s_1$ and $s_3$ coefficients have been normalized
by $c_1$.  The CO data is indicated by filled circles and the \HI\
data by open squares.  Text labels indicate whether positive values of
$s_1$ correspond to radial inflow or outflow.  Vertical dashed lines
delineate candidate inflow regions given in Table~\ref{tbl:infcan}.
(b) Morphological position angle and ellipticity as a function of
radius from the isophote fits.  In the upper panel, the kinematic
minor axis position angle (shifted by $\pi/2$) is overplotted as a
solid line for the CO and a dashed line for the HI, for all points
with estimated error $<5\arcdeg$.  In the lower panel, the kinematic
inclination, expressed as an ellipticity using $\epsilon=1 - \cos i$,
is similarly plotted for points with error $< 0.05$.  Corresponding
values of $i$ in degrees are given on the right axis.  The dotted
horizontal lines represent the adopted values for the disk position
angle and inclination.  (c) Comparison of $s_1$ and $s_3$ coefficients
normalized by $c_1$, for CO and (d) HI.  The heavy long-dashed line
represents a linear least-squares fit to the data points (with slope
given in the upper right corner), while the short-dashed line (``warp
line'') represents the predicted relation between $s_1$ and $s_3$ for
an error in the position angle $\Gamma_0$.  The sign of $s_1$
corresponds to inflow or outflow as indicated at the bottom of each
plot.
\label{fig:s13:4321}}
\end{figure*}


\begin{figure}
\begin{center}
\includegraphics[scale=0.45]{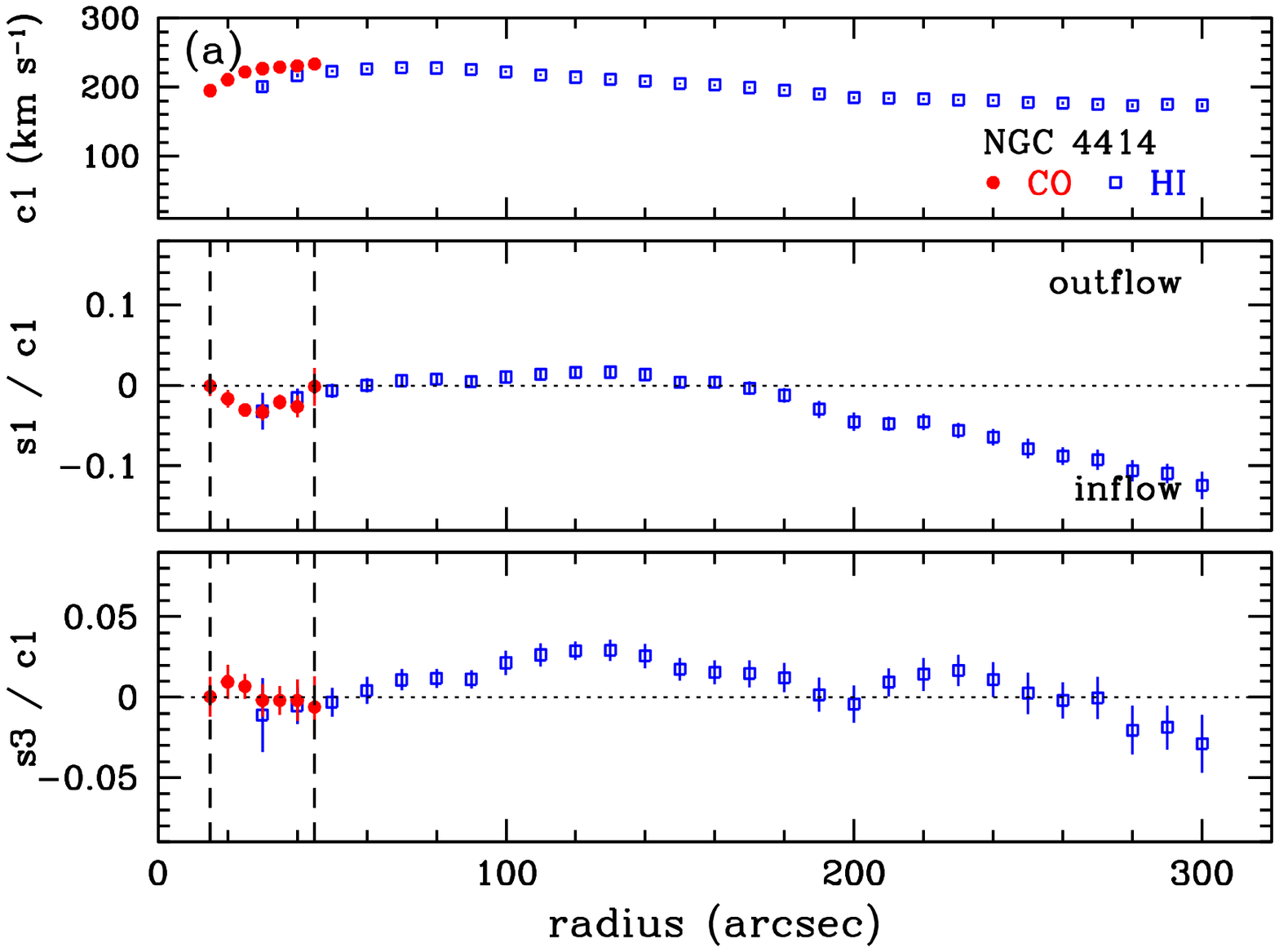}\\[2ex]
\includegraphics[scale=0.45]{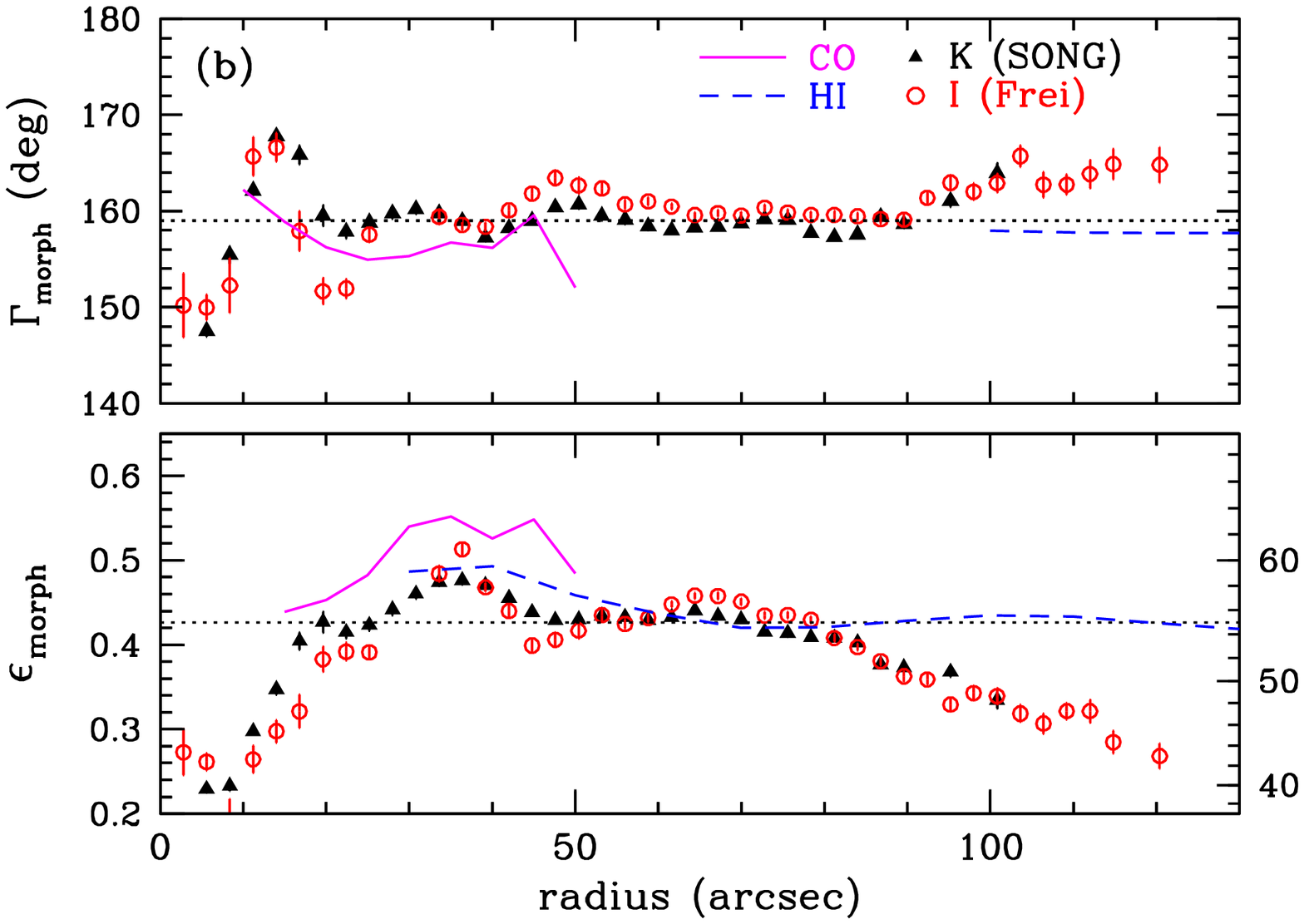}\\[2ex]
\includegraphics[scale=0.3]{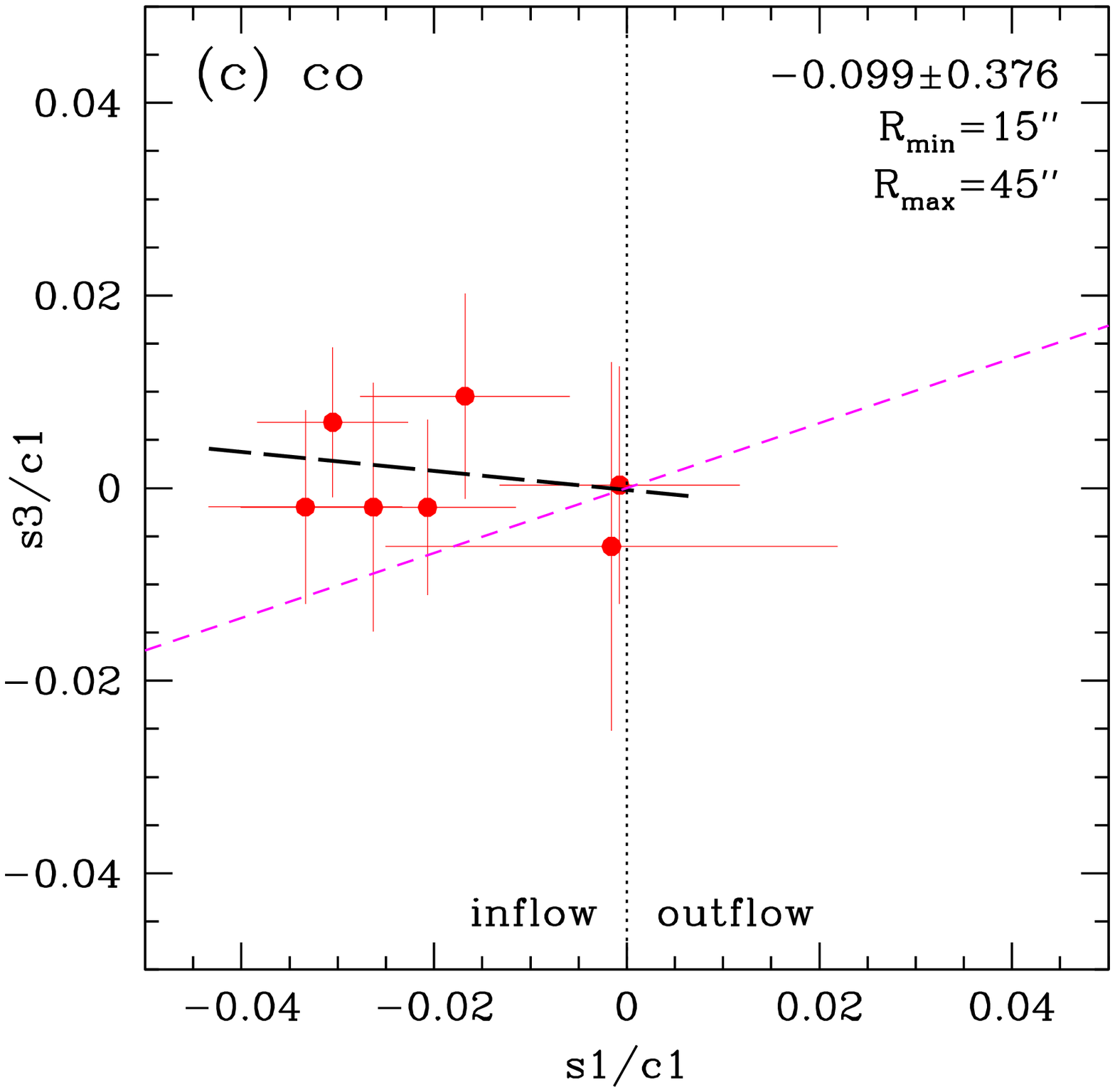}\\[3ex]
\includegraphics[scale=0.3]{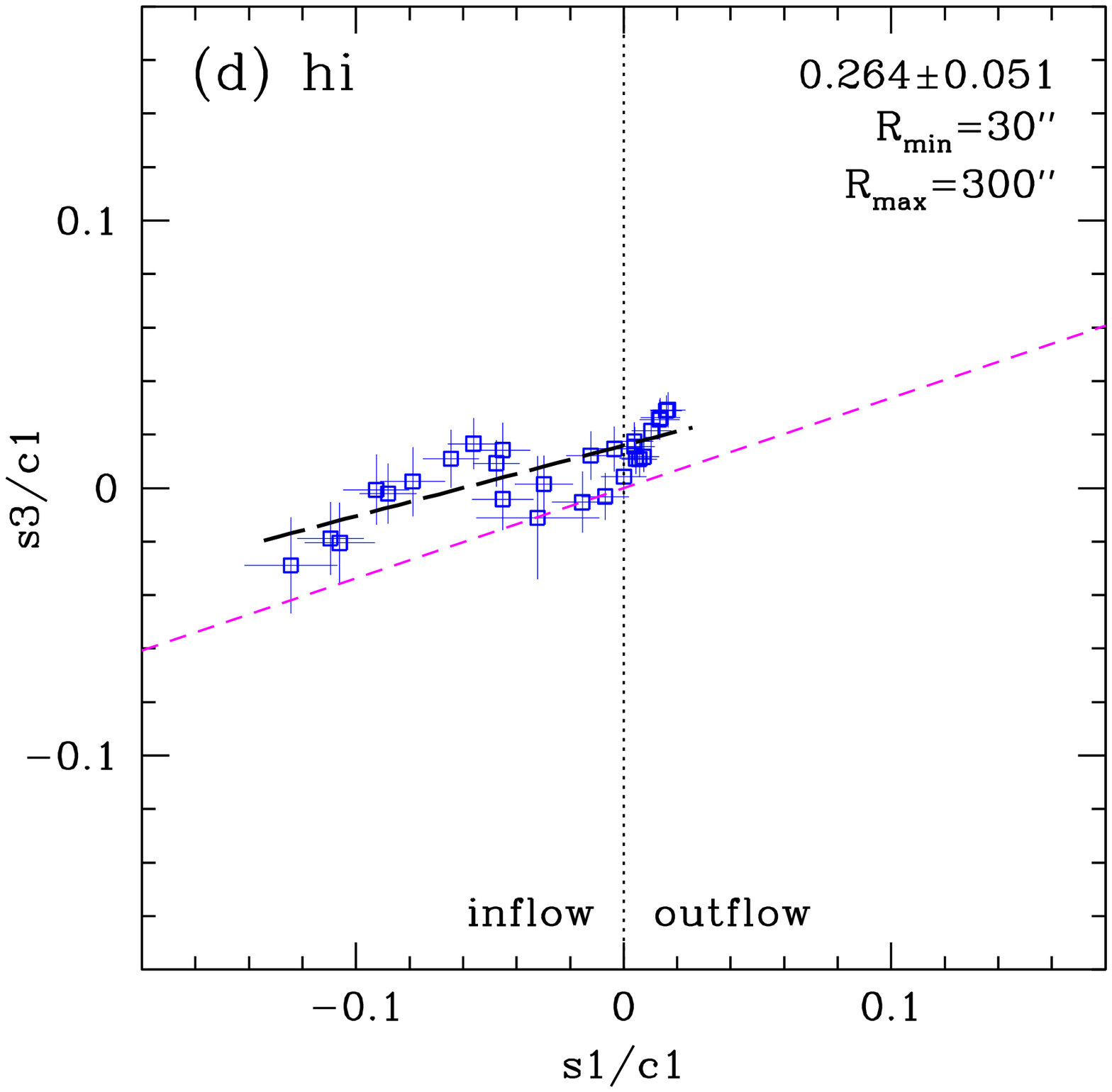}
\end{center}
\caption{
Same as Fig.~\ref{fig:s13:4321}, but for NGC 4414.
\label{fig:s13:4414}}
\end{figure}


\begin{figure}
\begin{center}
\includegraphics[scale=0.45]{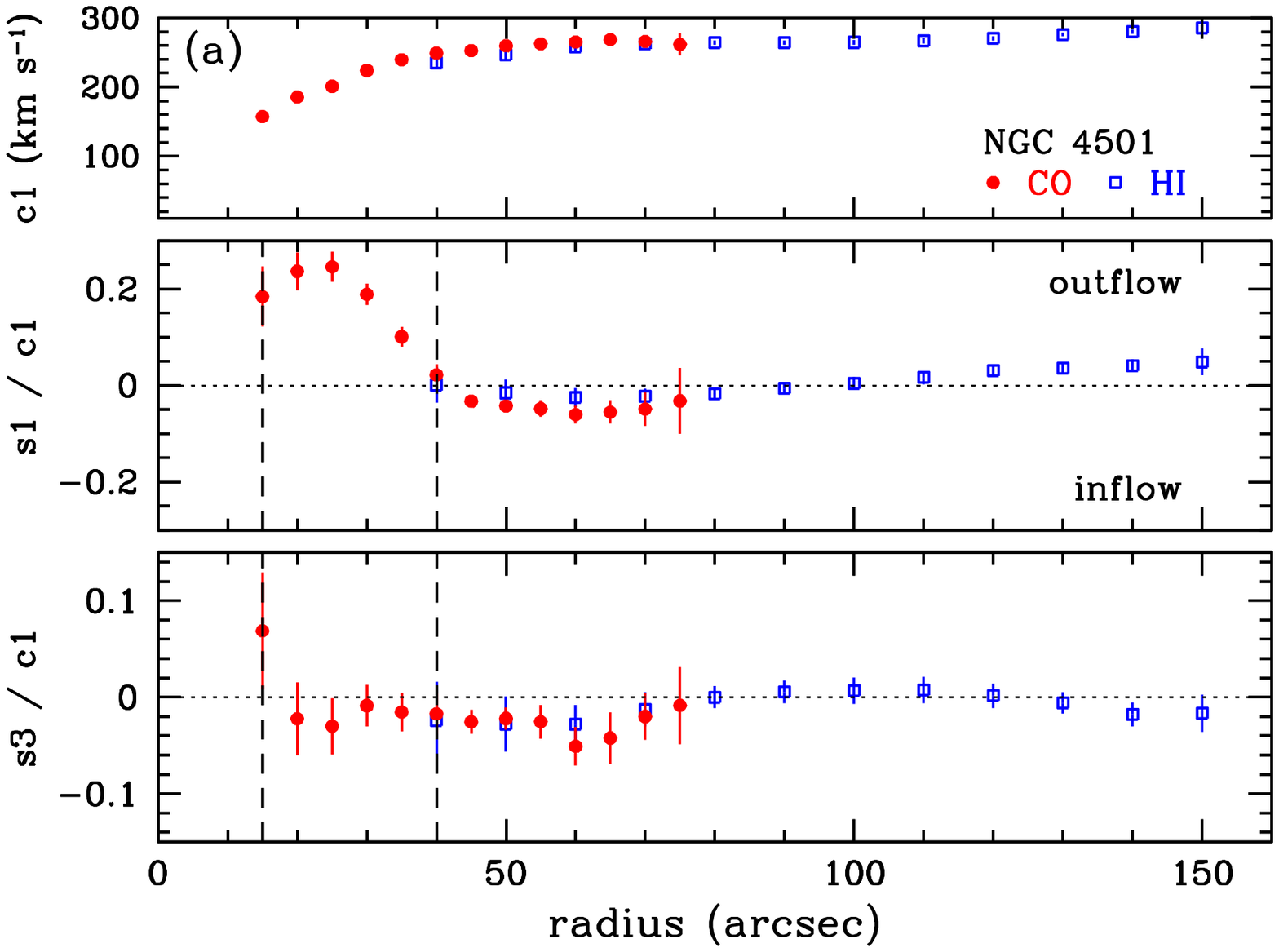}\\[2ex]
\includegraphics[scale=0.45]{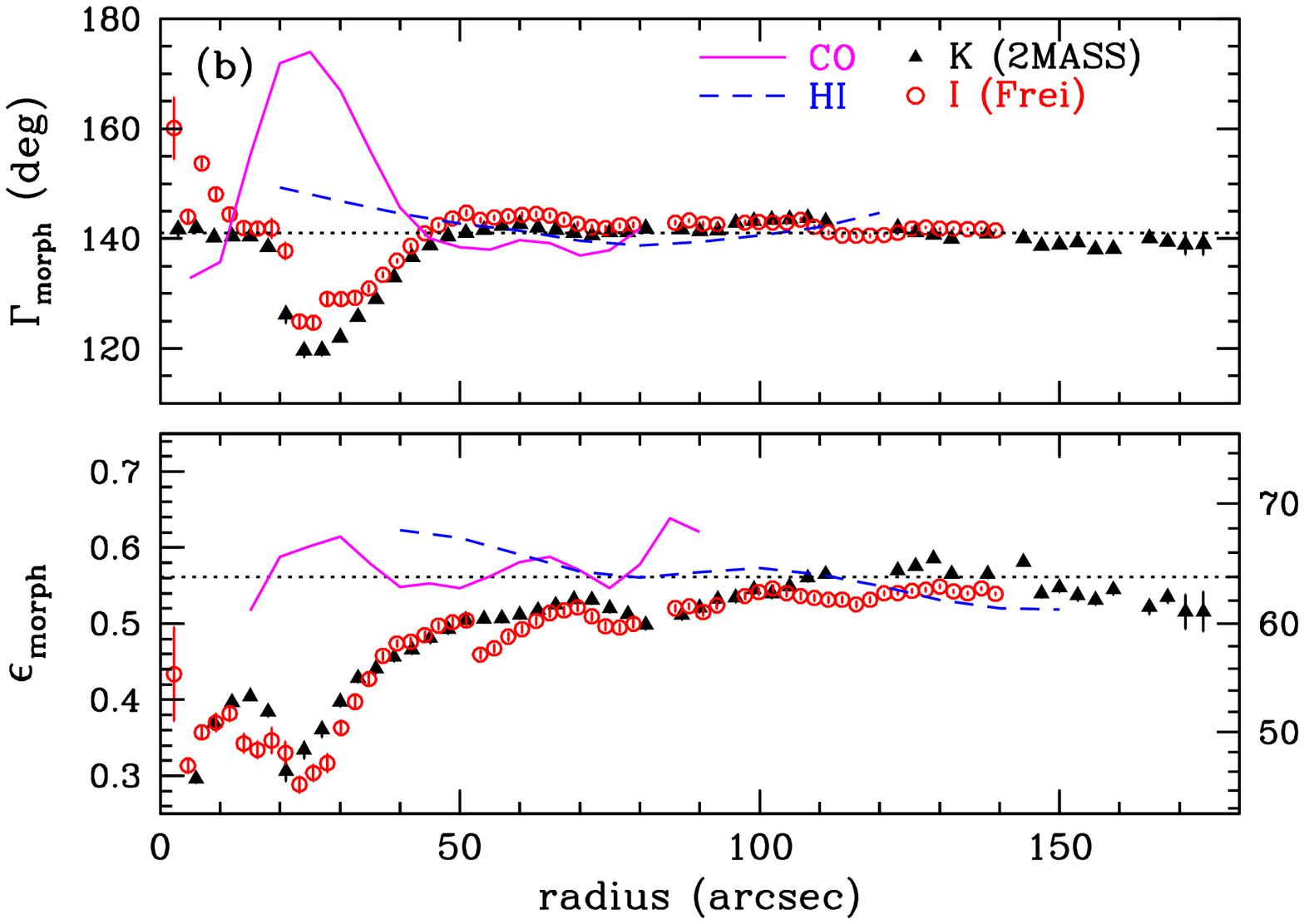}\\[2ex]
\includegraphics[scale=0.3]{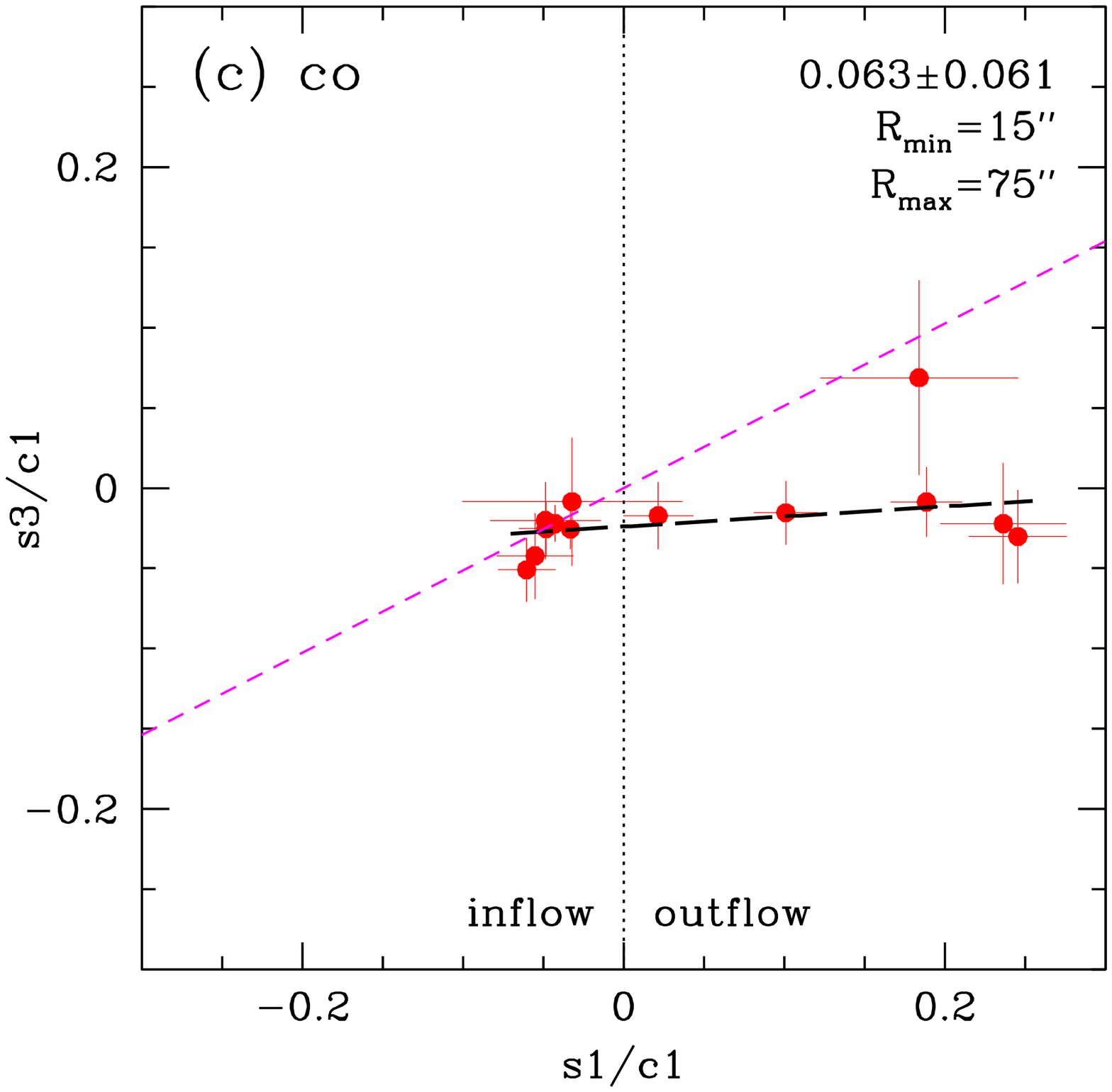}\\[3ex]
\includegraphics[scale=0.3]{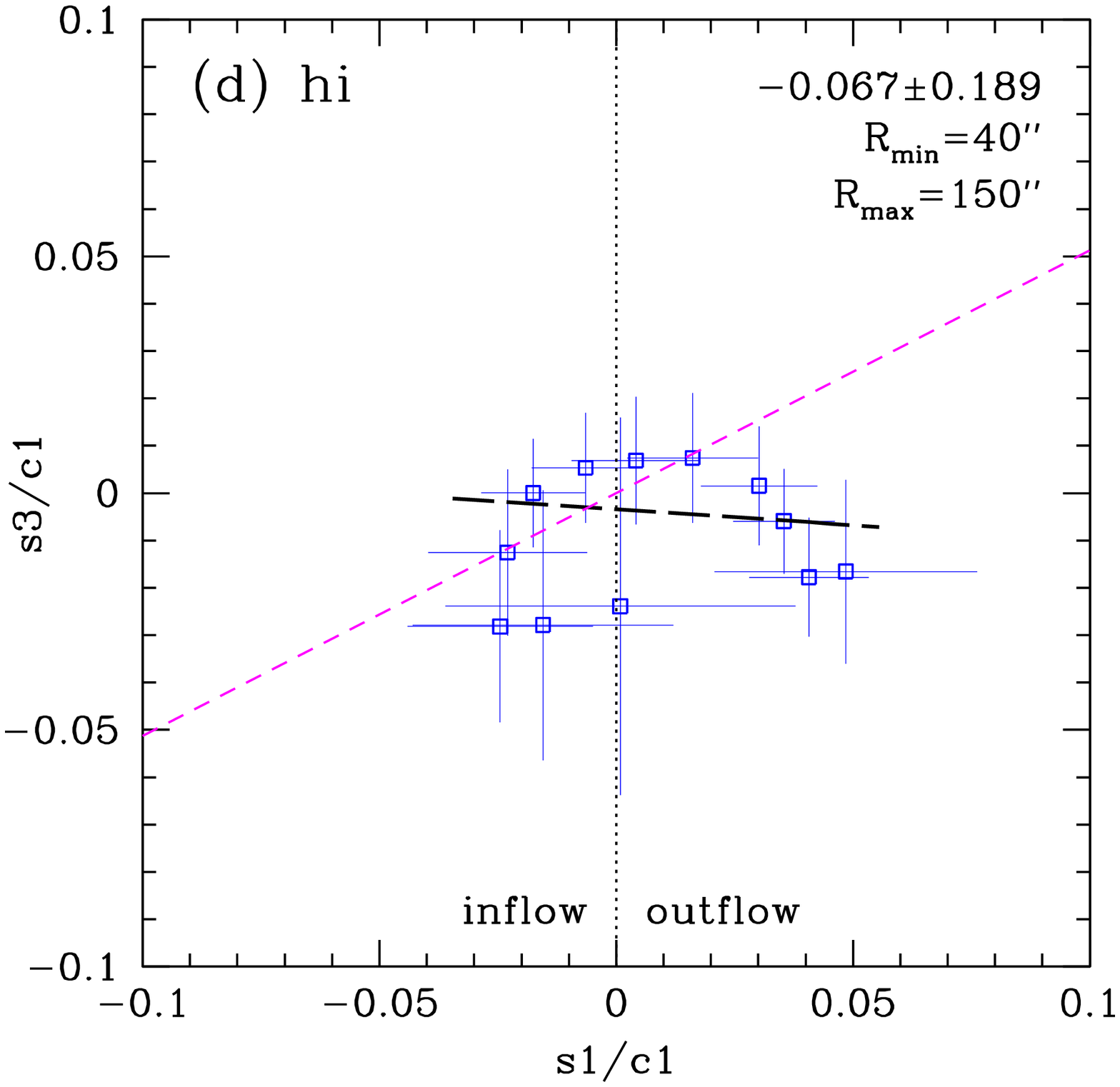}
\end{center}
\caption{
Same as Fig.~\ref{fig:s13:4321}, but for NGC 4501.
\label{fig:s13:4501}}
\end{figure}


\begin{figure}
\begin{center}
\includegraphics[scale=0.45]{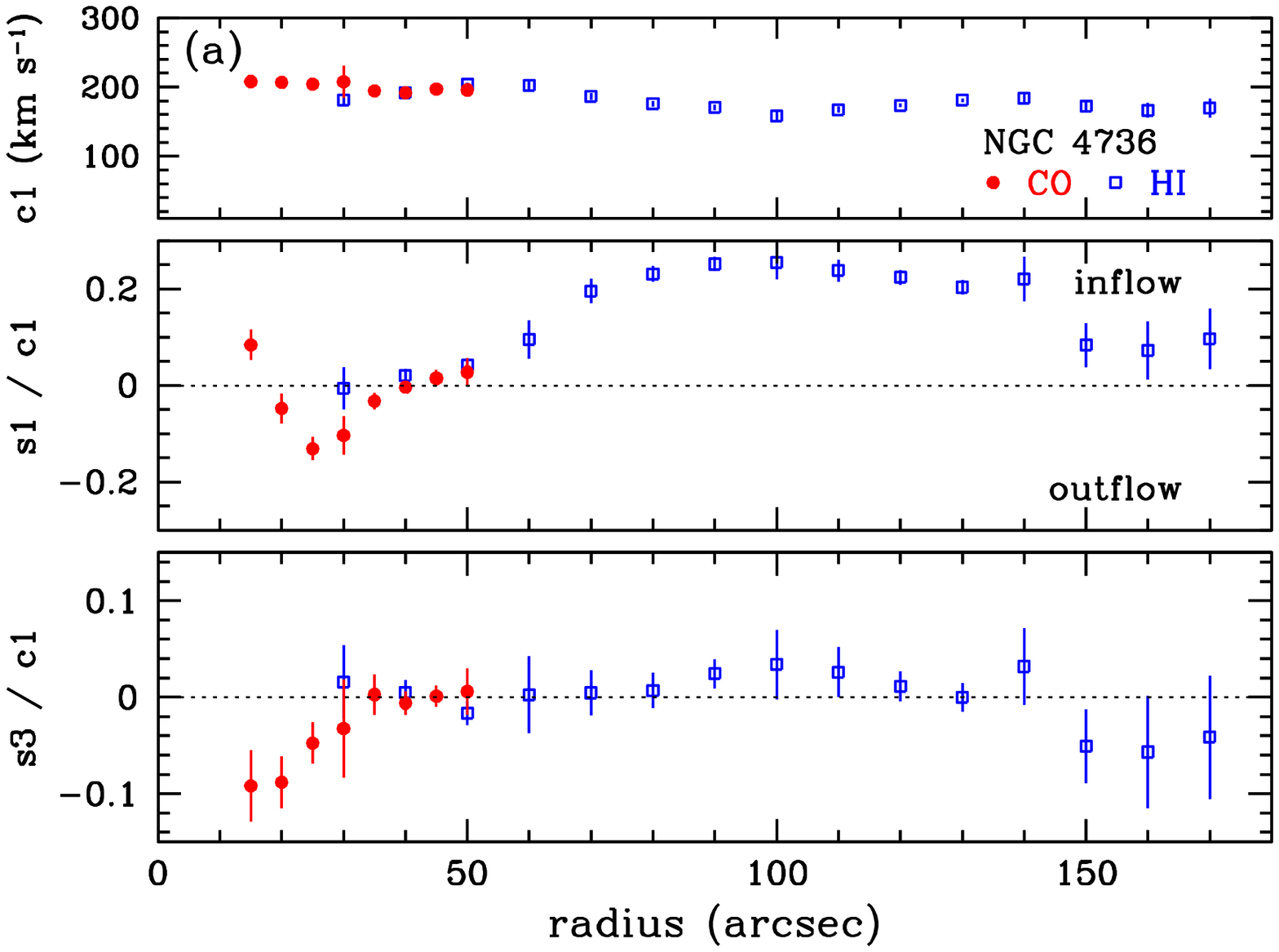}\\[2ex]
\includegraphics[scale=0.45]{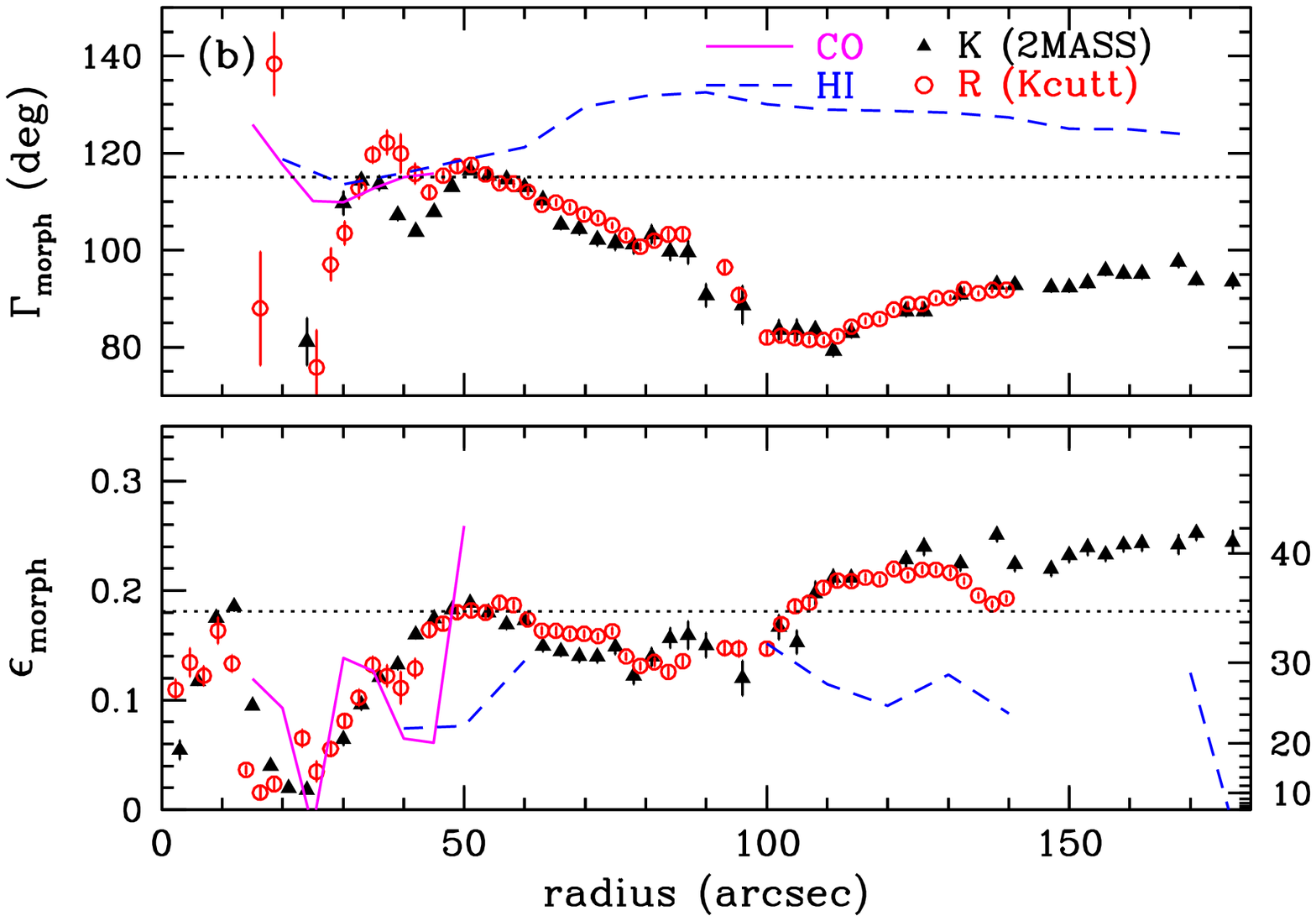}\\[2ex]
\includegraphics[scale=0.3]{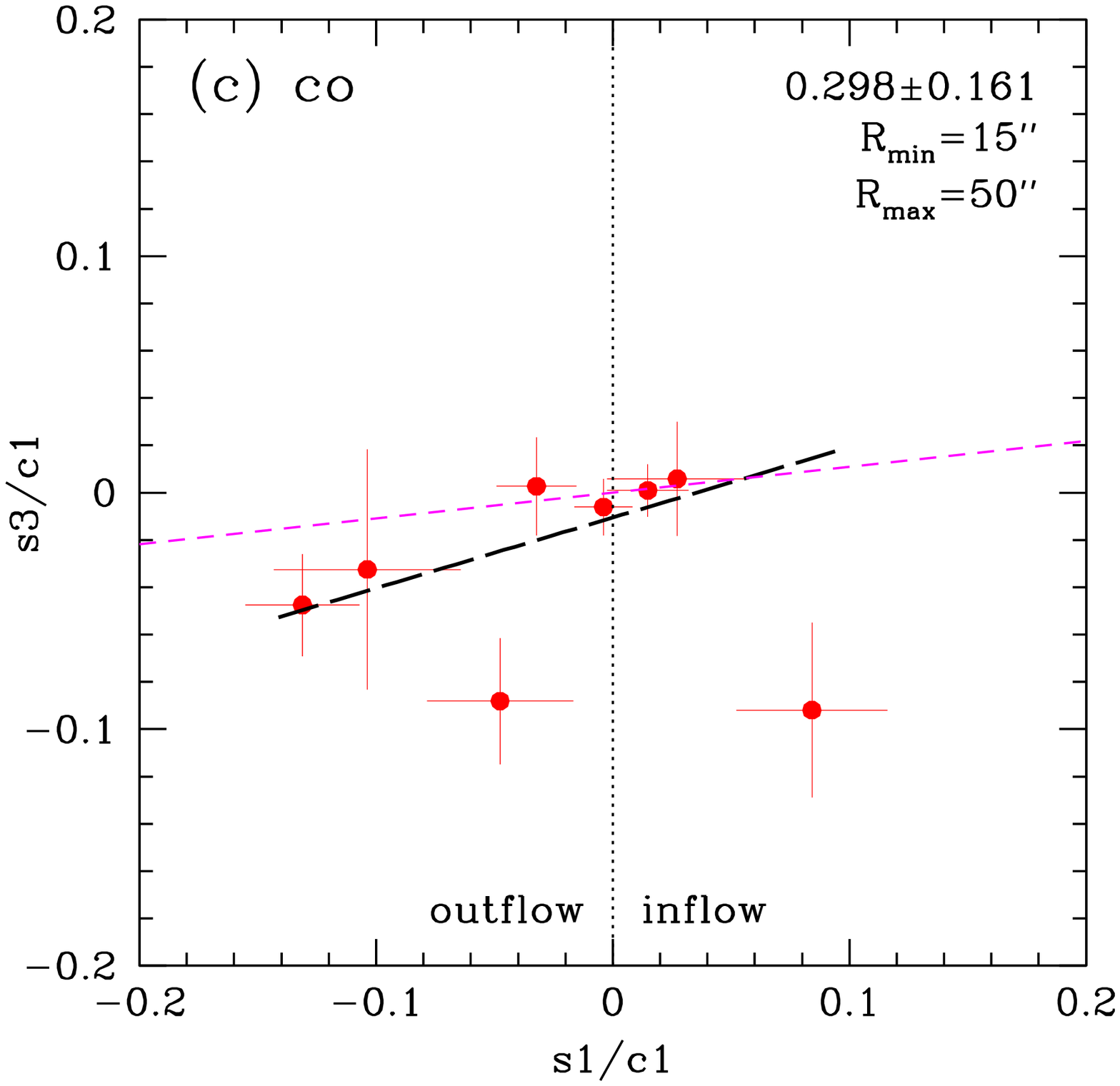}\\[3ex]
\includegraphics[scale=0.3]{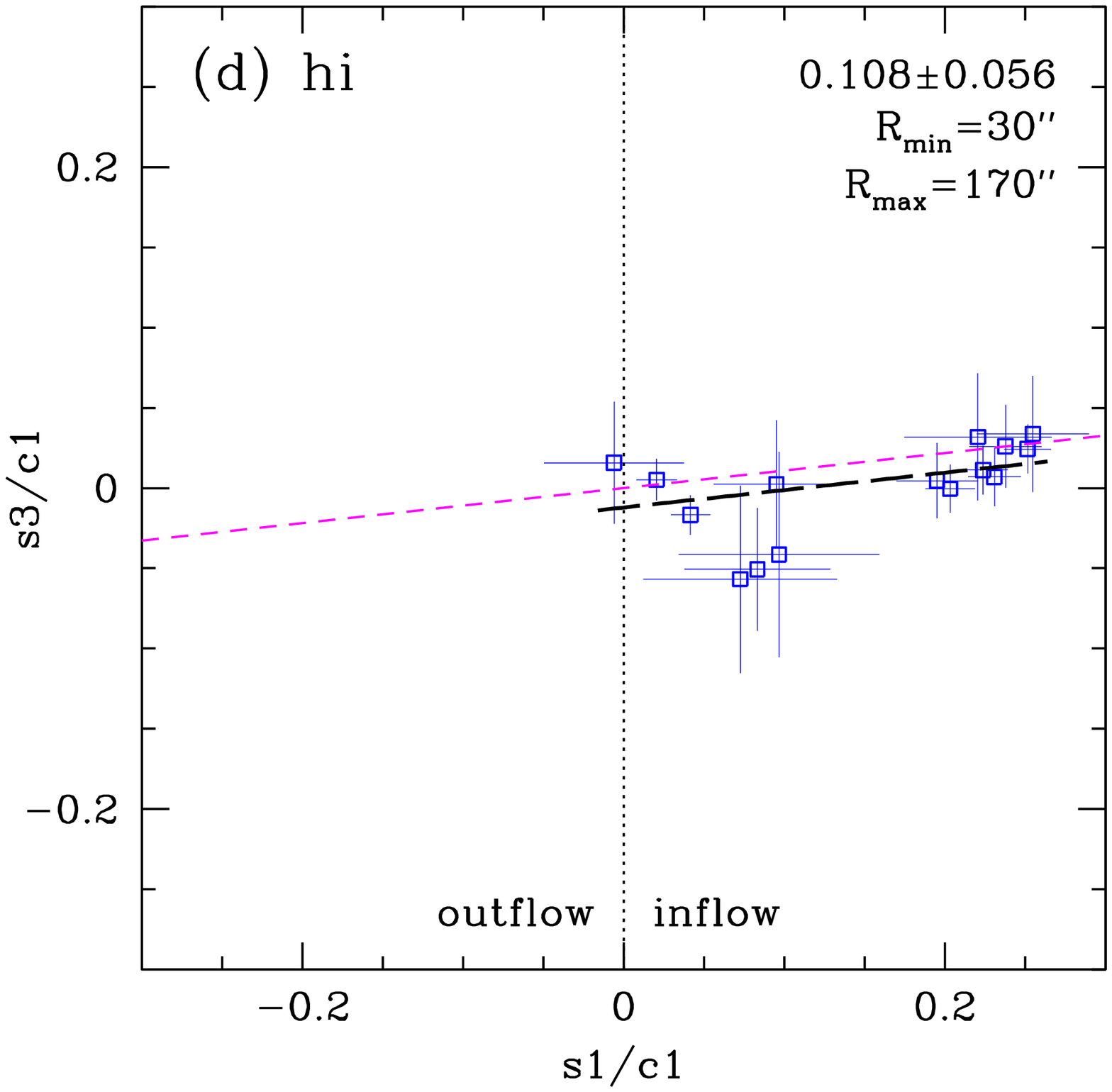}
\end{center}
\caption{
Same as Fig.~\ref{fig:s13:4321}, but for NGC 4736.
\label{fig:s13:4736}}
\end{figure}


\begin{figure}
\begin{center}
\includegraphics[scale=0.45]{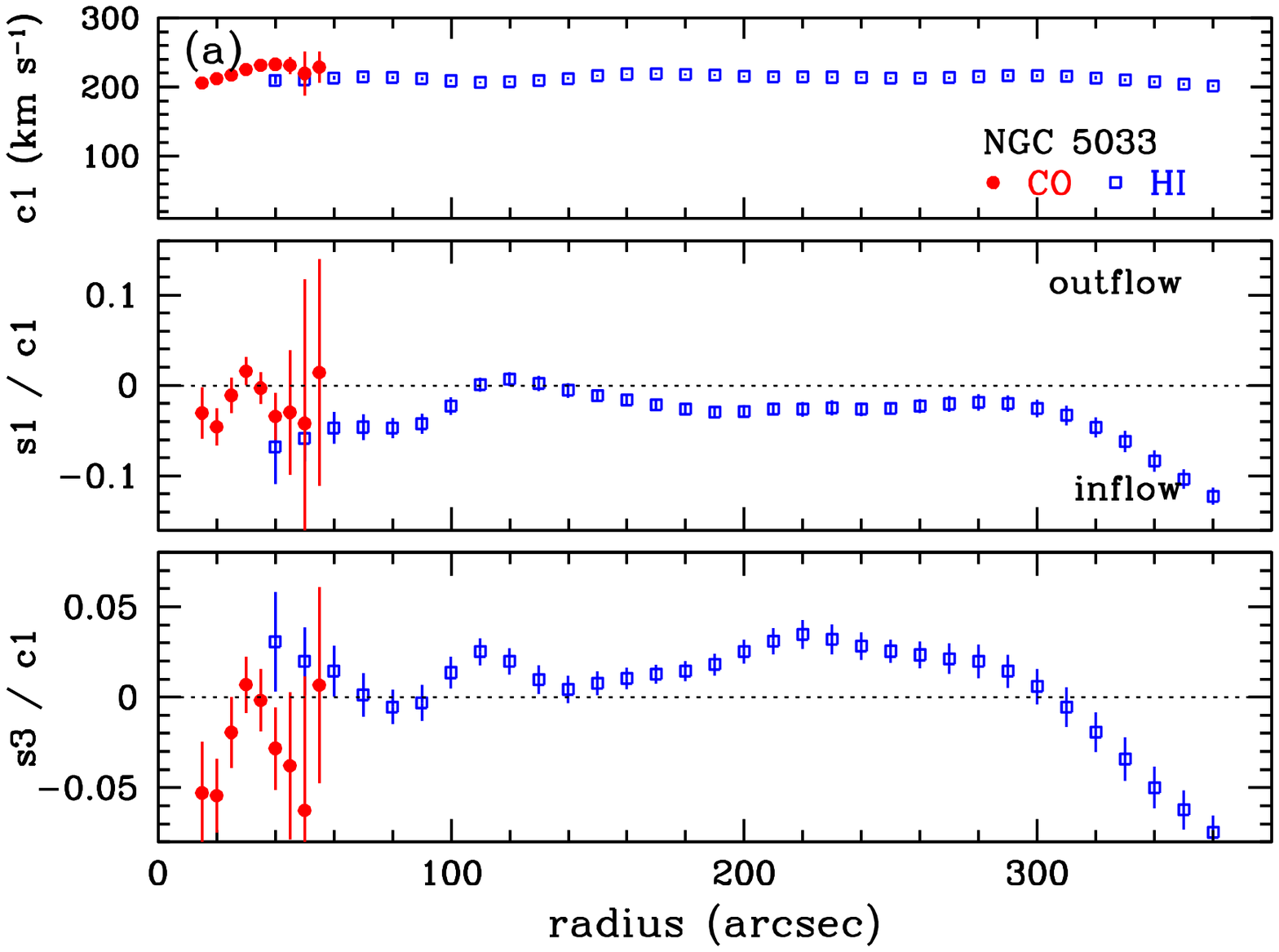}\\[2ex]
\includegraphics[scale=0.45]{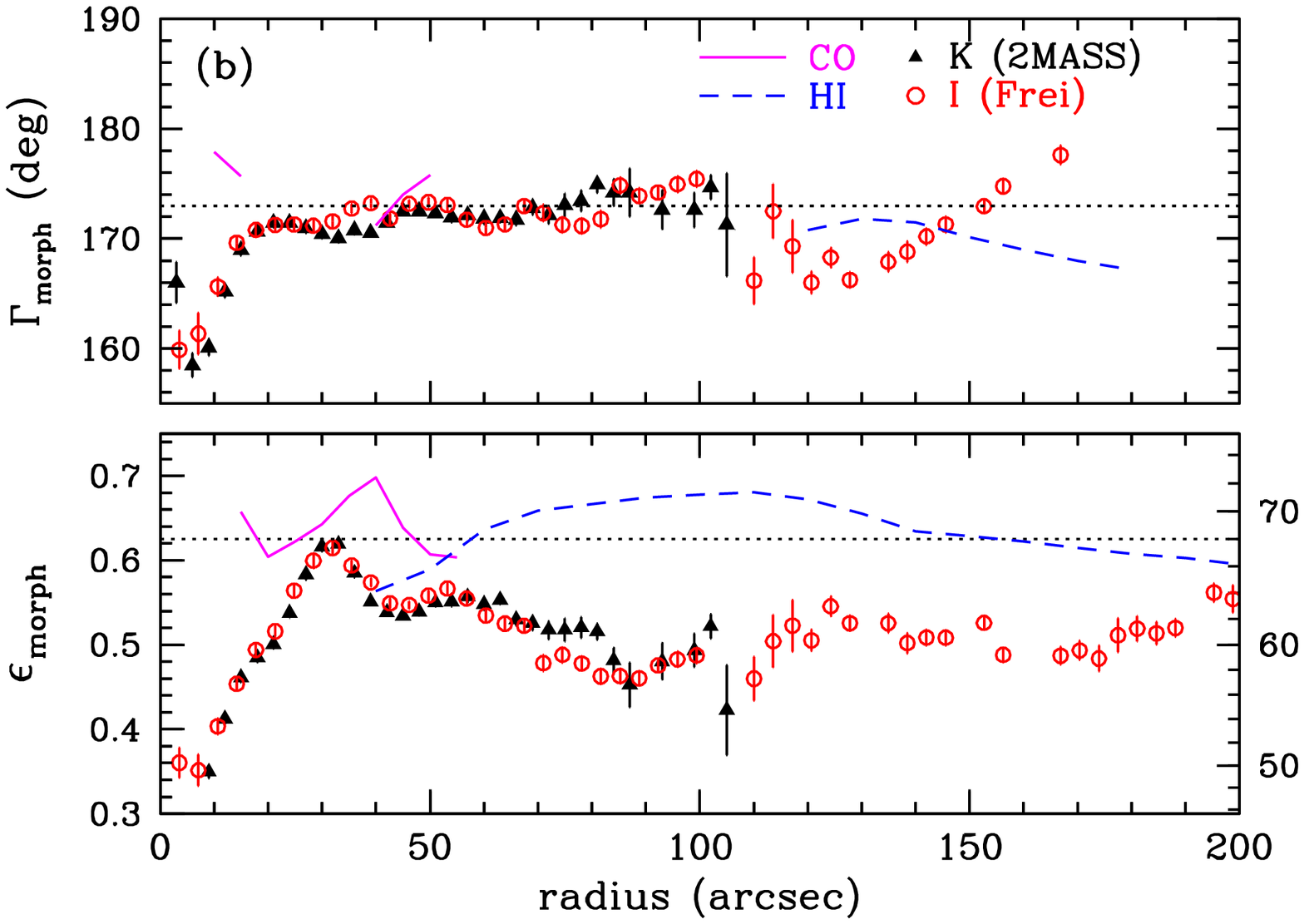}\\[2ex]
\includegraphics[scale=0.3]{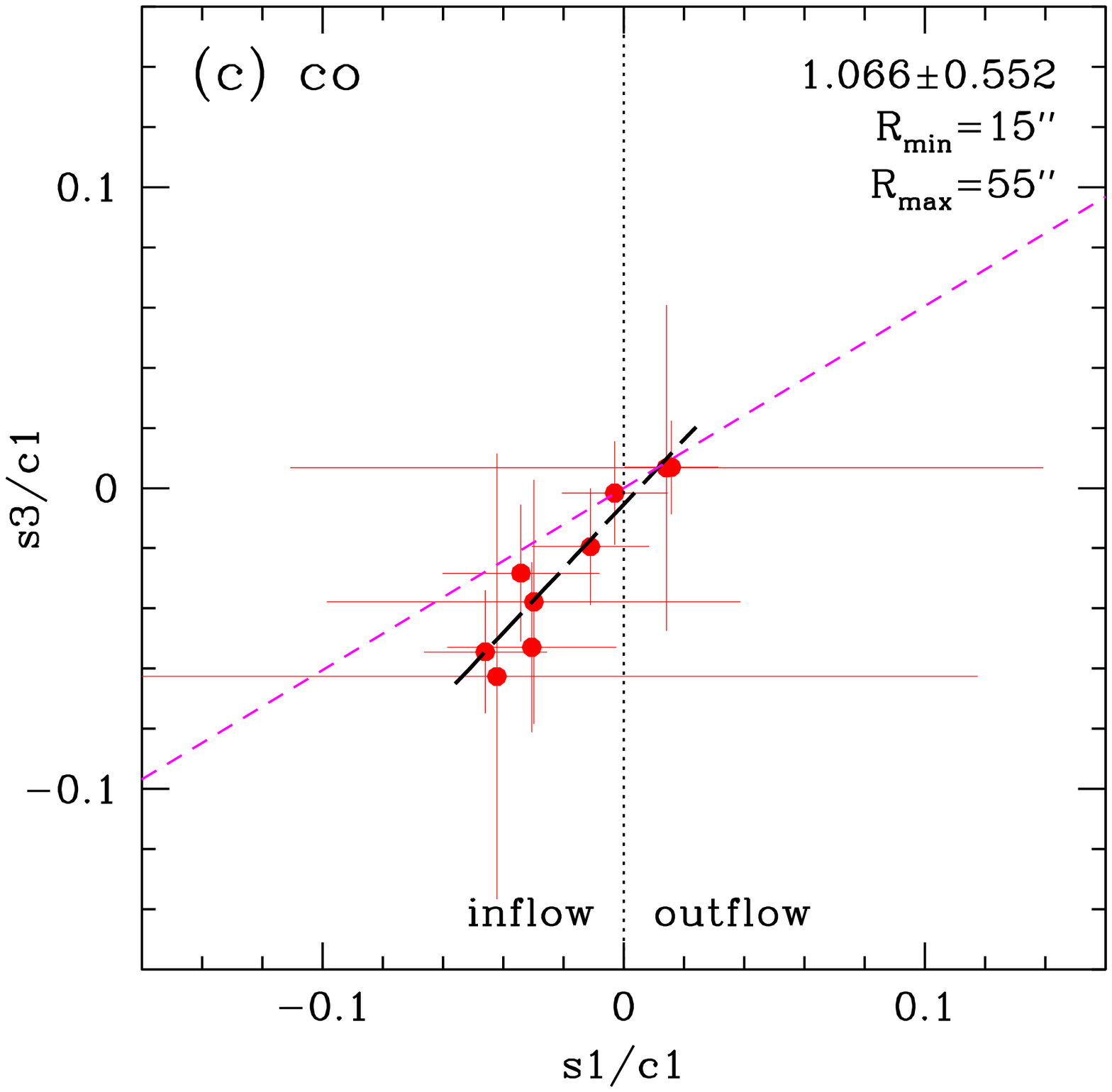}\\[3ex]
\includegraphics[scale=0.3]{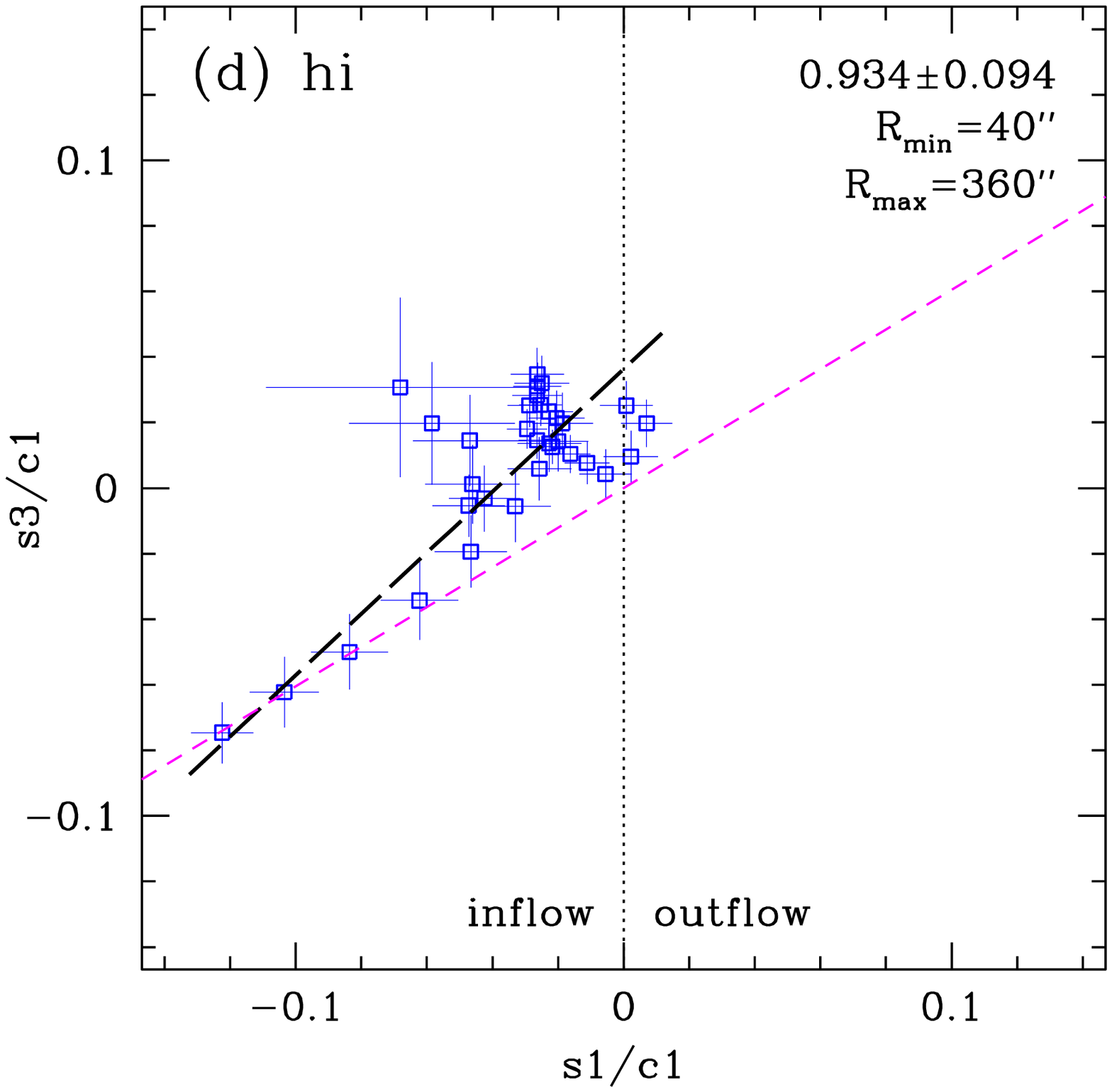}
\end{center}
\caption{
Same as Fig.~\ref{fig:s13:4321}, but for NGC 5033.
\label{fig:s13:5033}}
\end{figure}


\begin{figure}
\begin{center}
\includegraphics[scale=0.45]{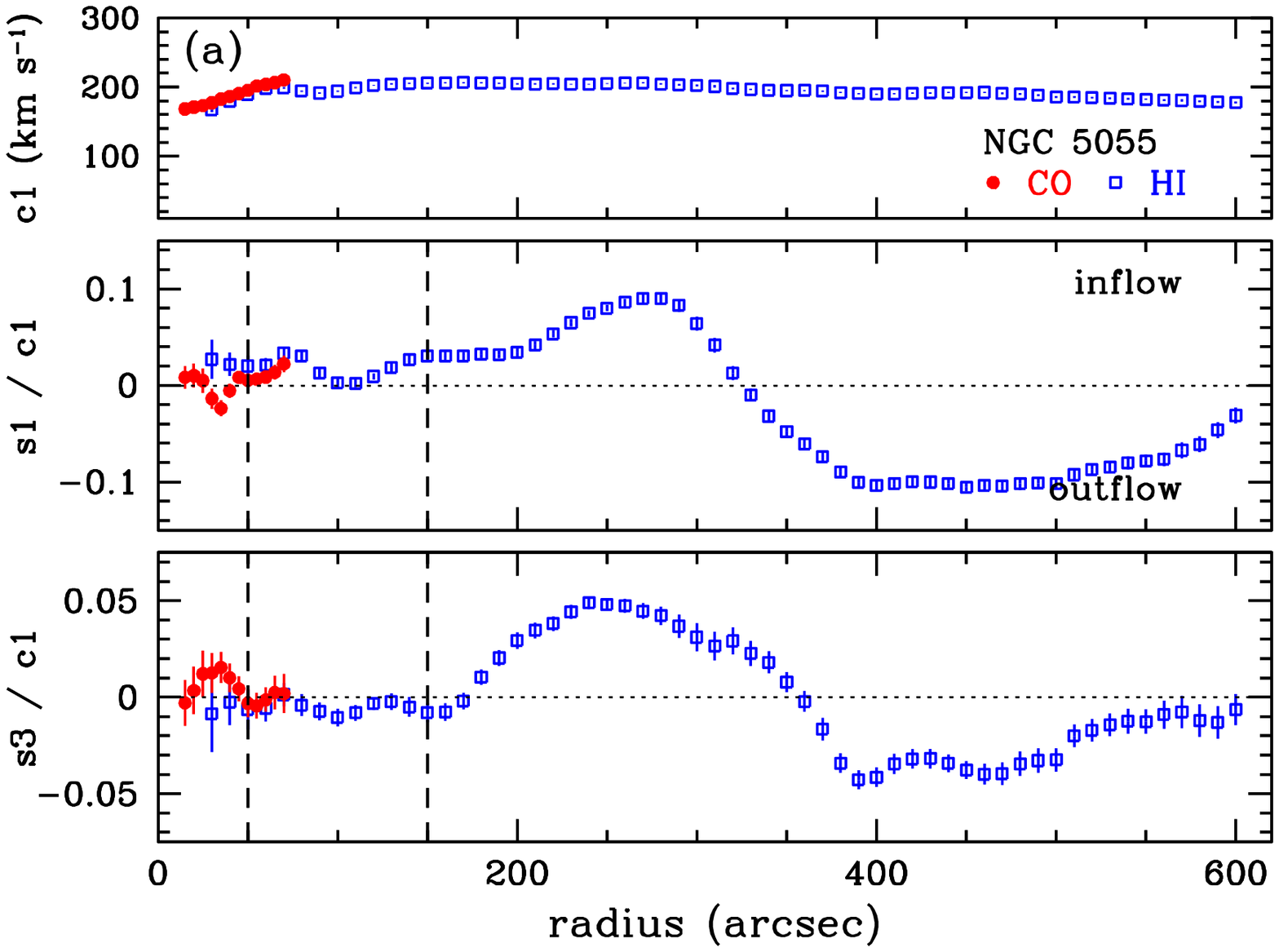}\\[2ex]
\includegraphics[scale=0.45]{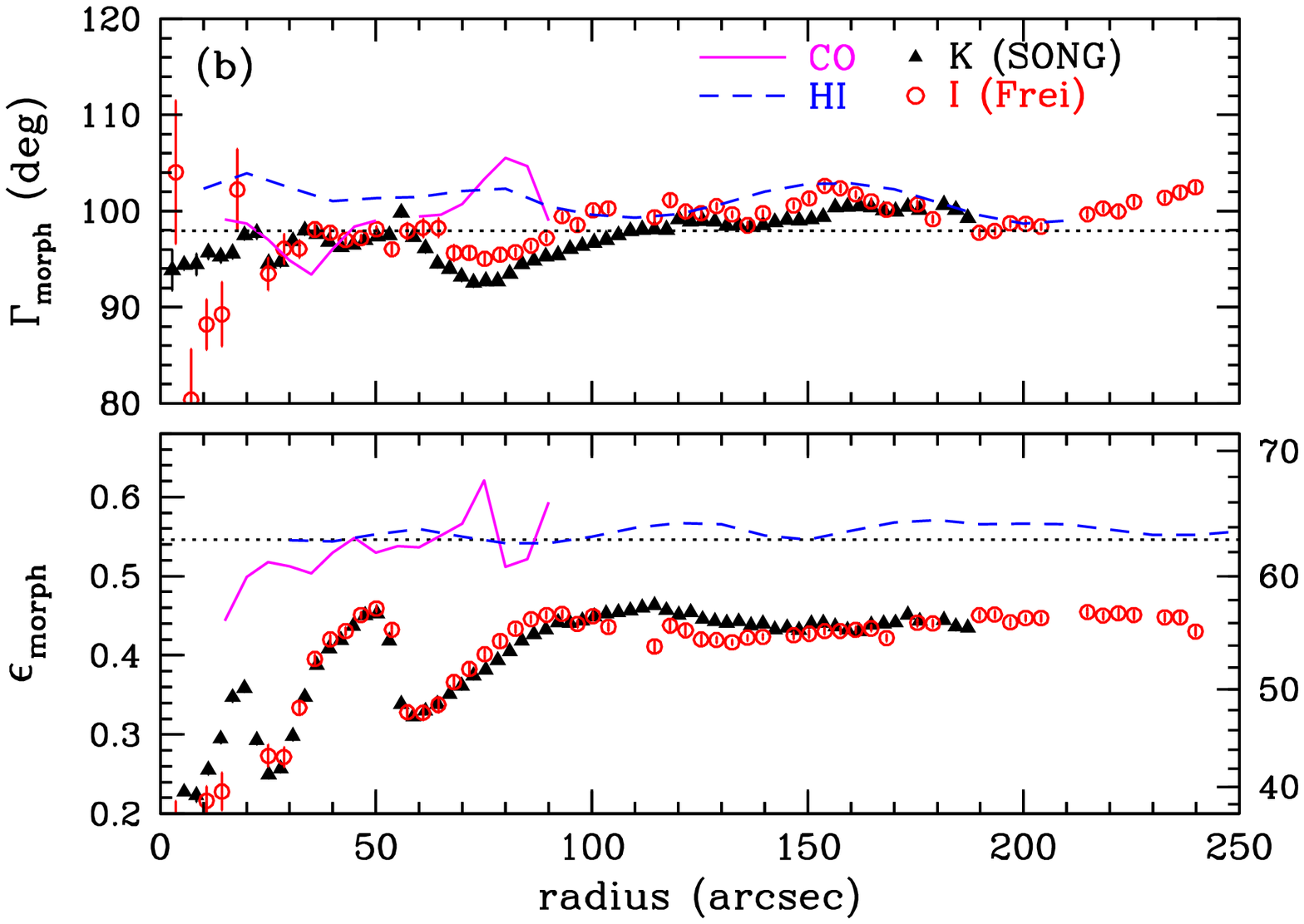}\\[2ex]
\includegraphics[scale=0.3]{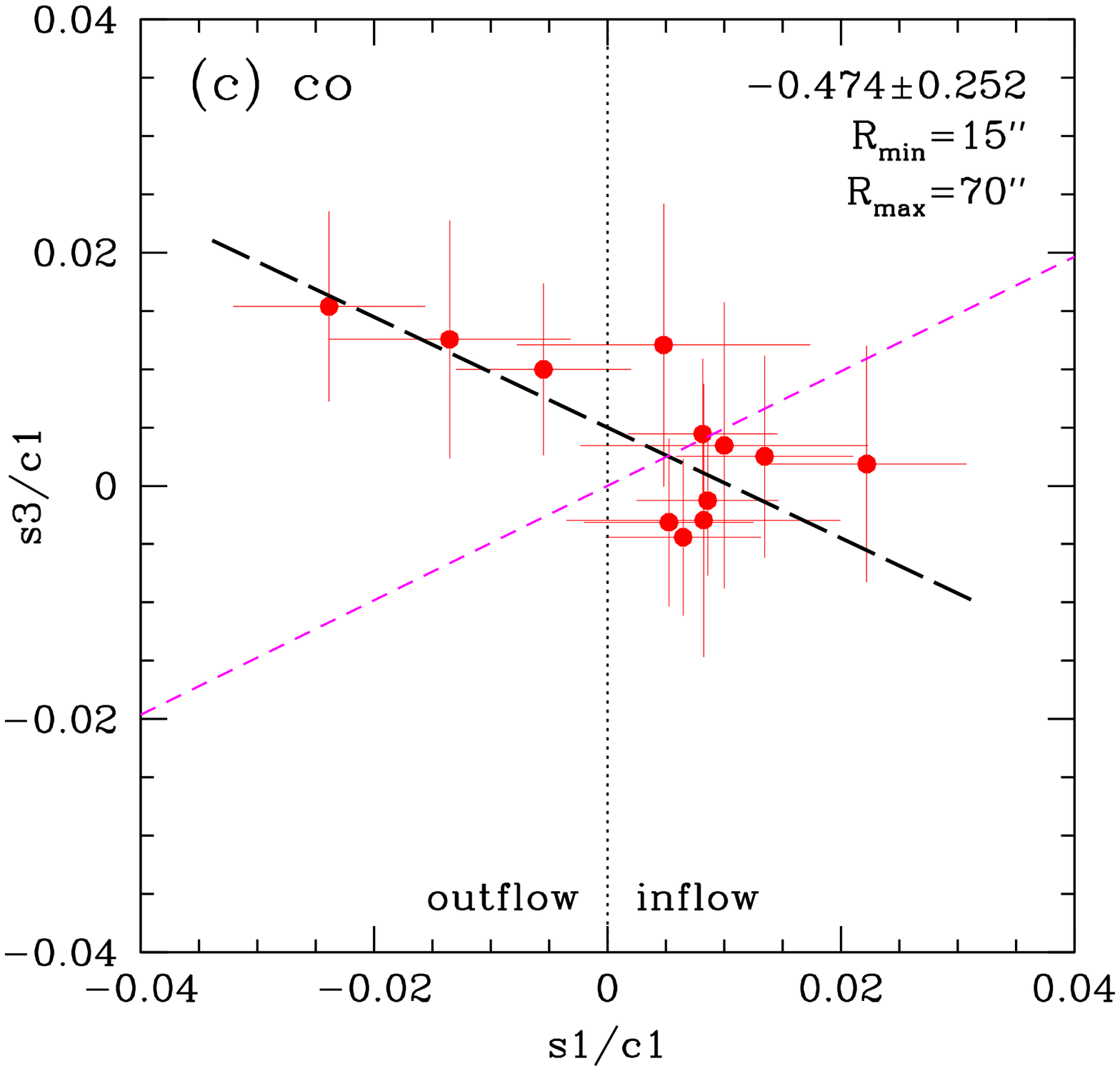}\\[3ex]
\includegraphics[scale=0.3]{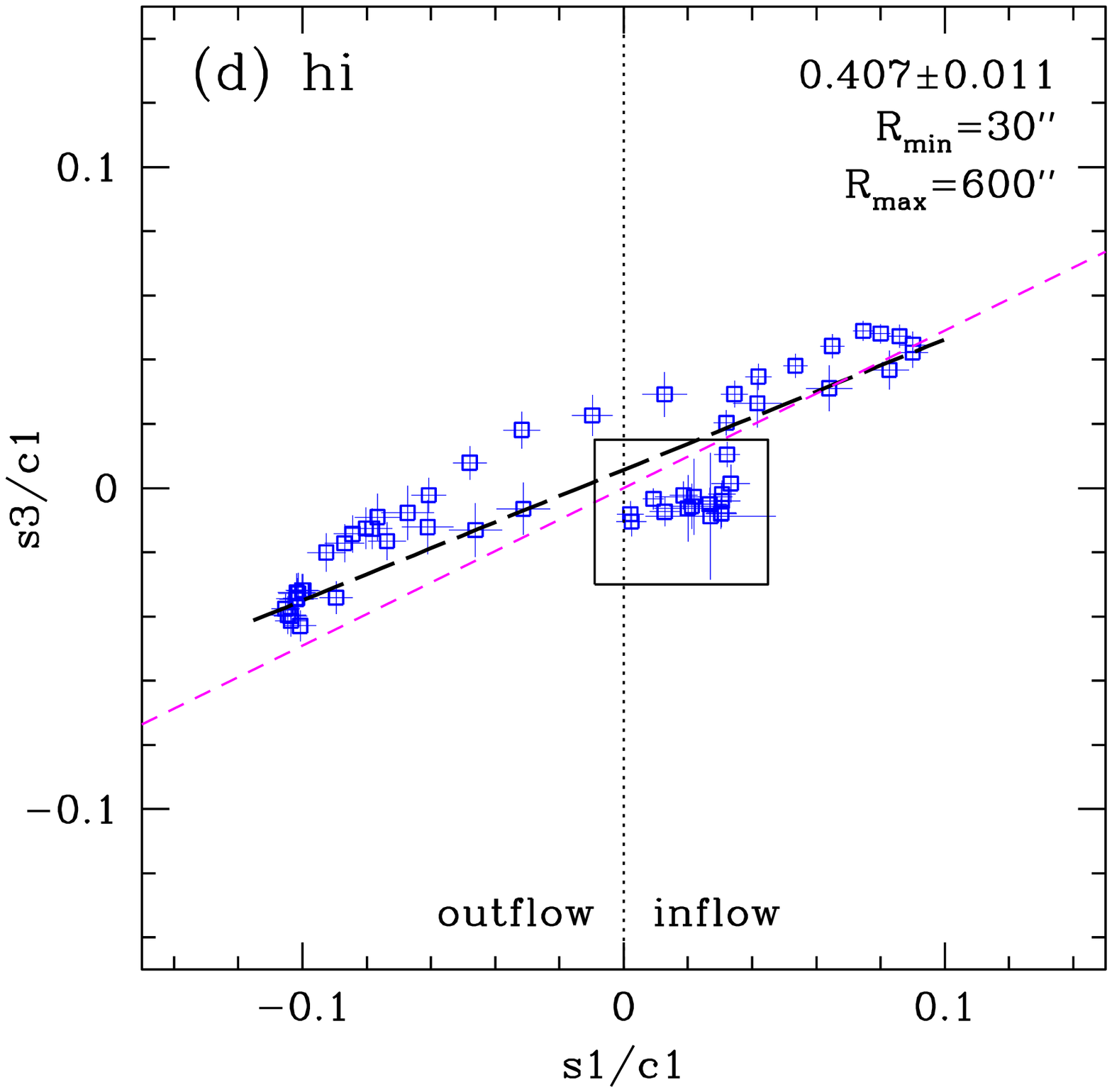}
\end{center}
\caption{
Same as Fig.~\ref{fig:s13:4321}, but for NGC 5055.  The box in
panel (d) indicates the region of possible inflow between
$R$=50\arcsec--150\arcsec.
\label{fig:s13:5055}}
\end{figure}


\begin{figure}
\begin{center}
\includegraphics[scale=0.45]{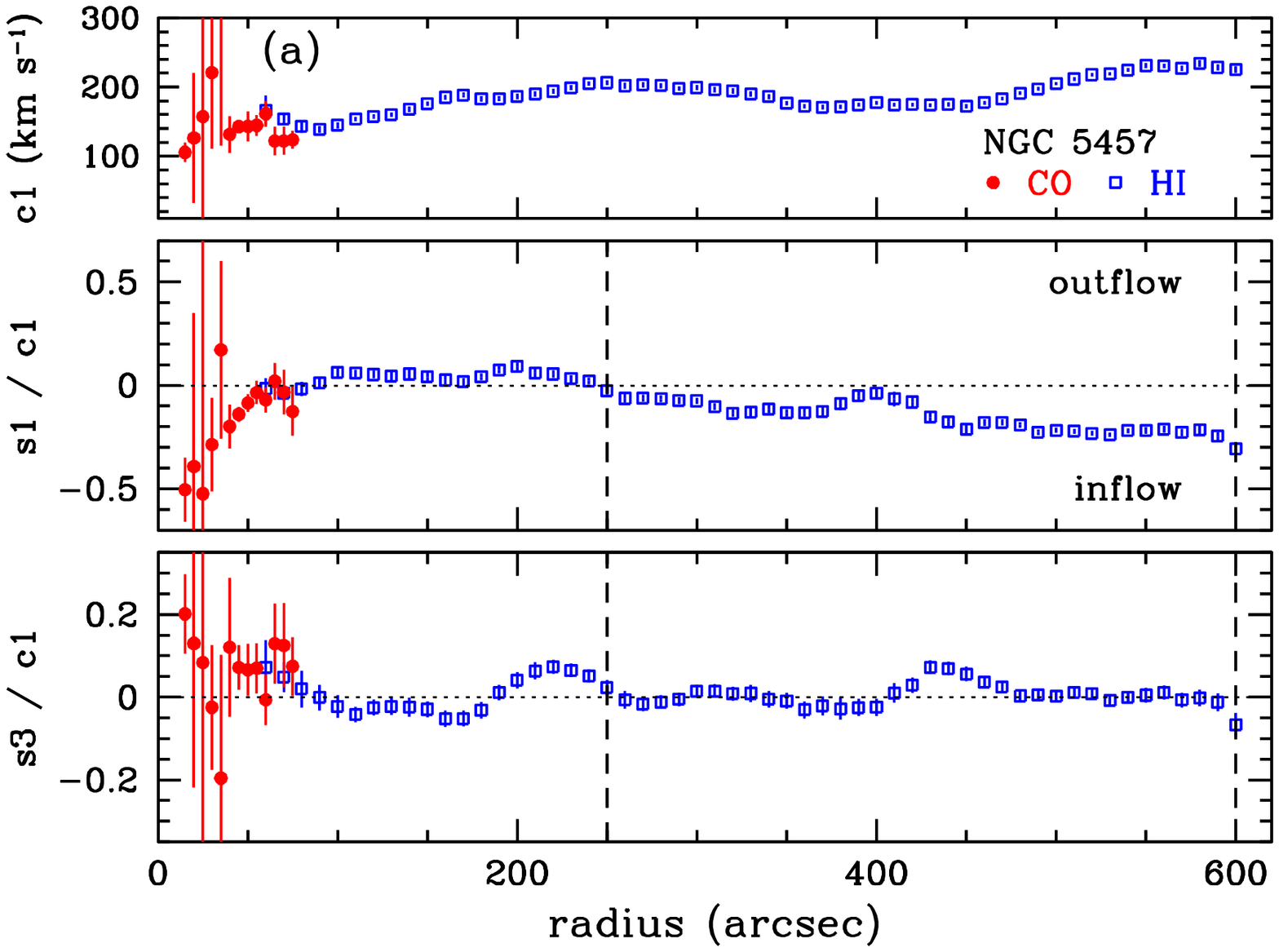}\\[2ex]
\includegraphics[scale=0.45]{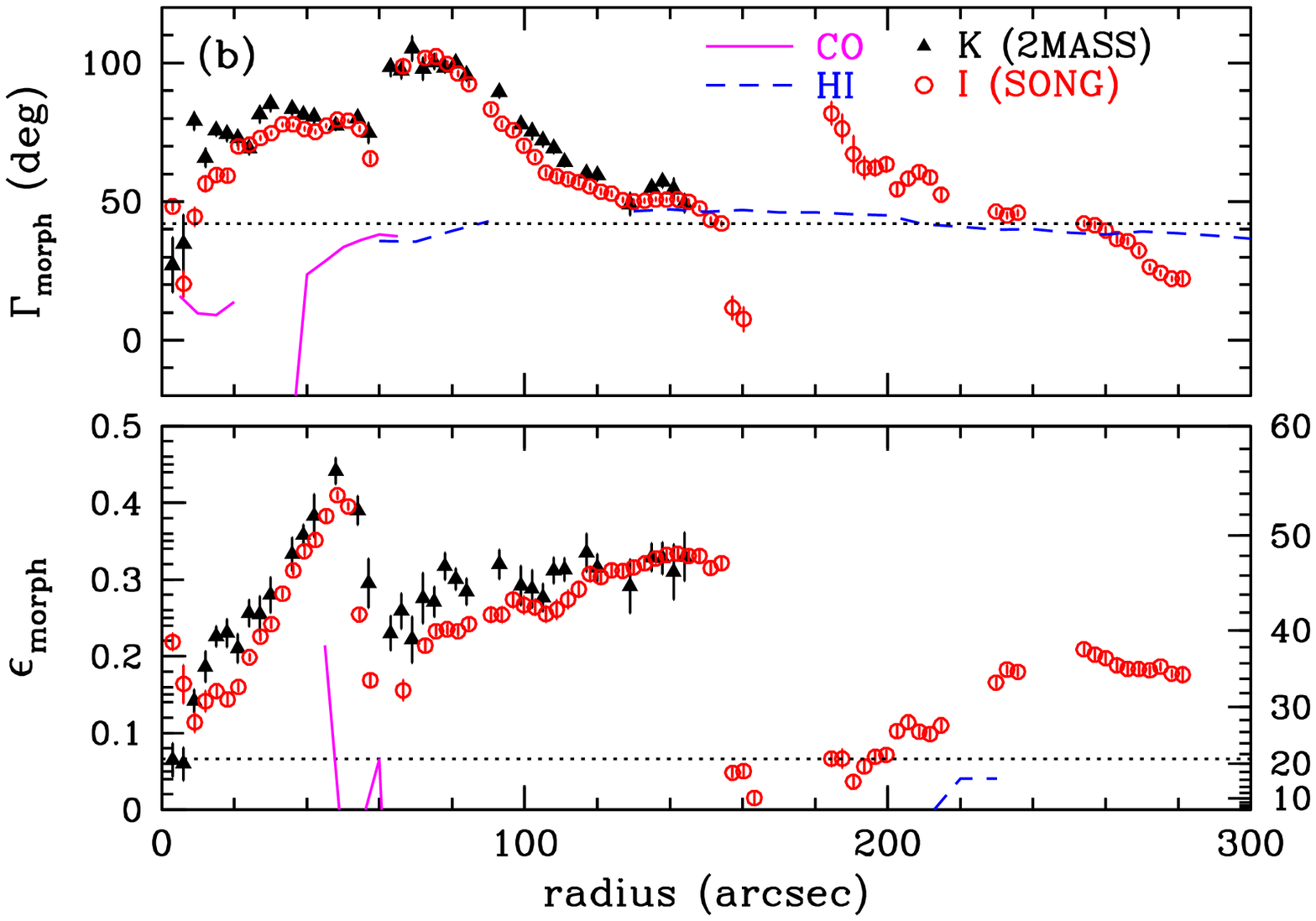}\\[2ex]
\includegraphics[scale=0.3]{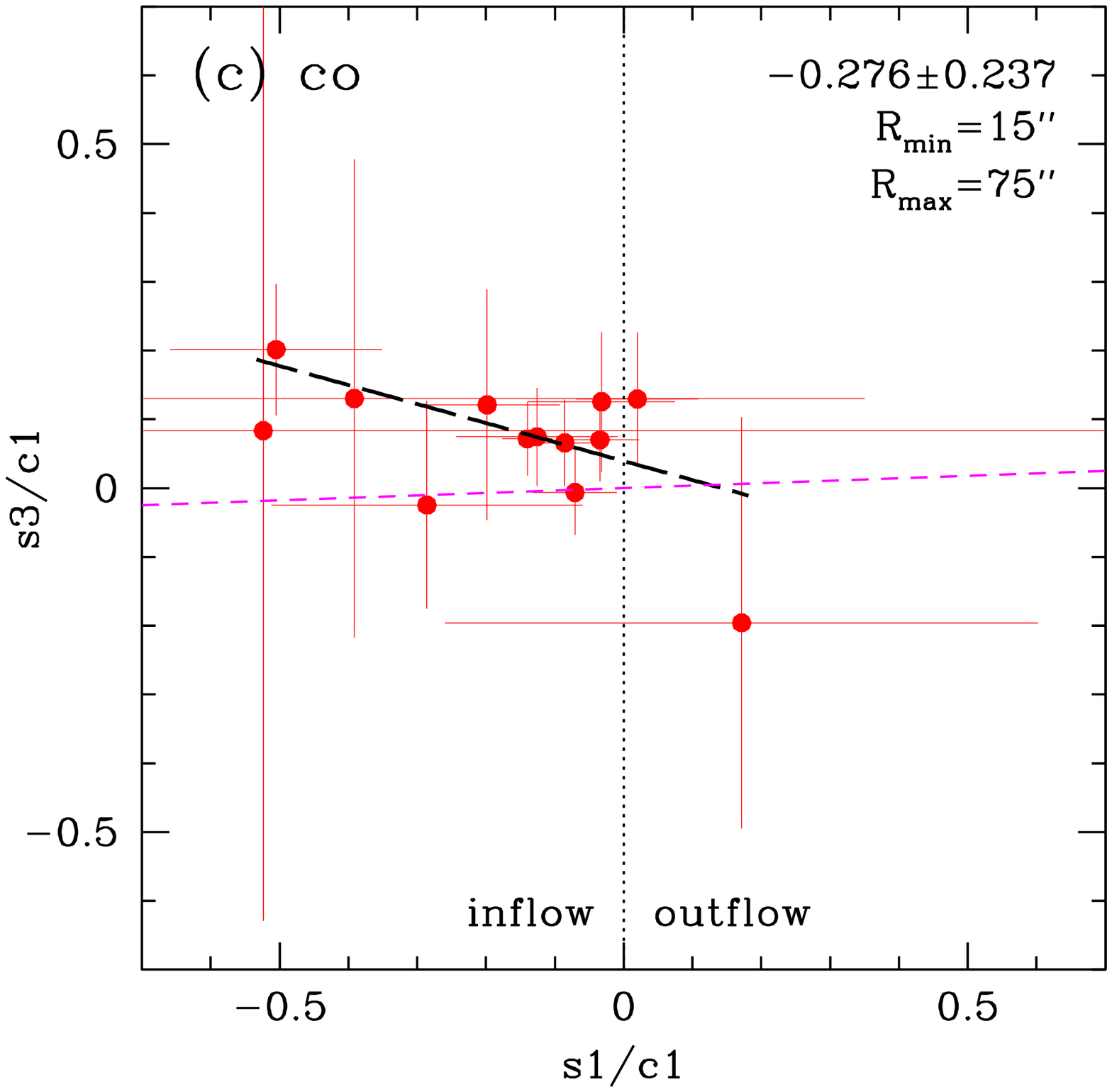}\\[3ex]
\includegraphics[scale=0.3]{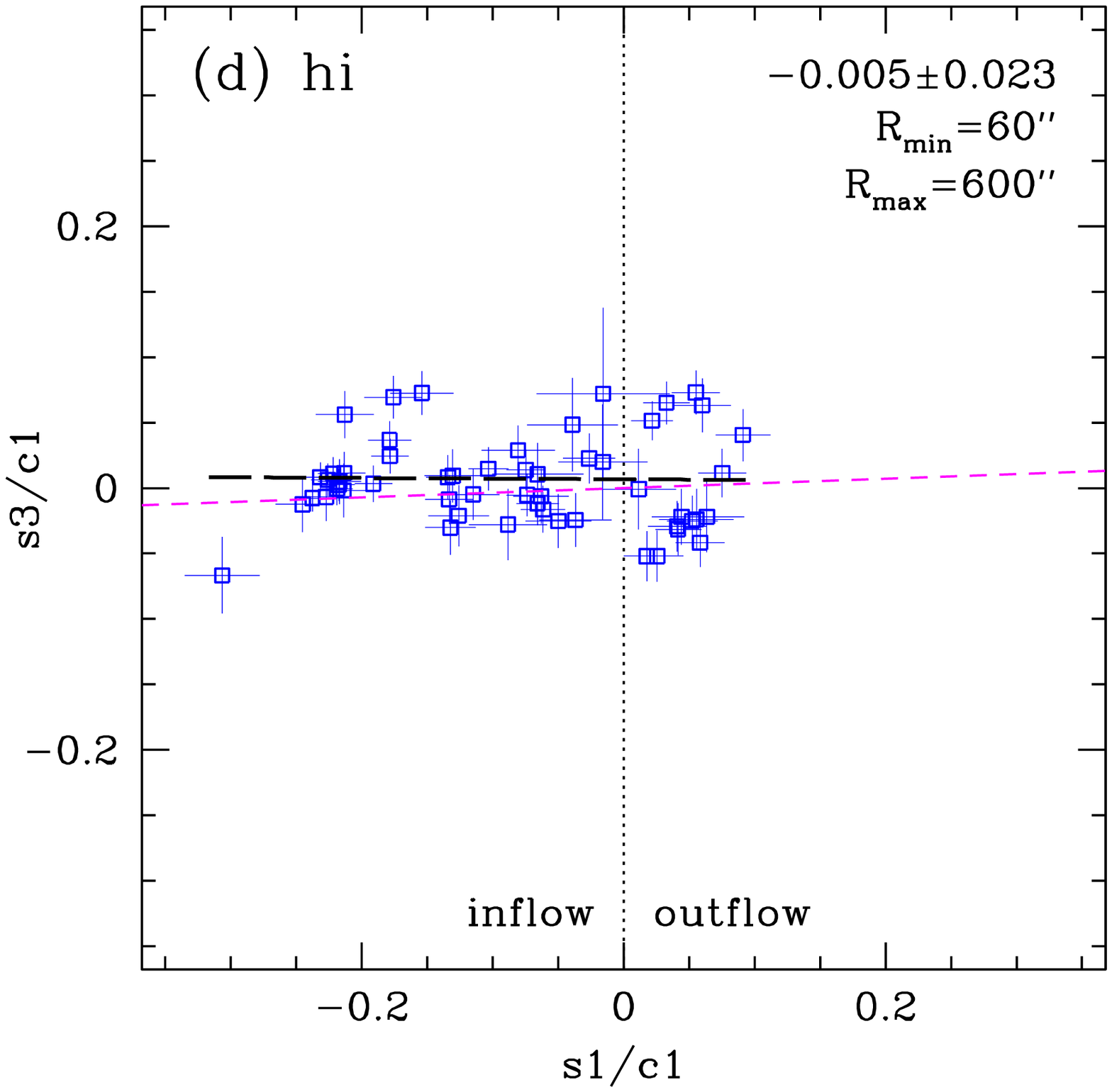}
\end{center}
\caption{
Same as Fig.~\ref{fig:s13:4321}, but for NGC 5457.
\label{fig:s13:5457}}
\end{figure}

\subsection{Harmonic Analysis of Velocity Fields}

Using the GIPSY task RESWRI \citep{Schoen:th}, each CO or \HI\
velocity field was subdivided into rings using the adopted orientation
parameters, and a third-order Fourier series was fitted to the
velocities in each ring (cf.\ Eq.~\ref{eqn:decomp} with $n$=3).  The
program employs a least-squares fitting procedure (singular value
decomposition) rather than a direct Fourier expansion since the points
may not be uniformly distributed in azimuth.  No extra weight is given
to points near the major axis, and in fact it is points near the {\it
minor} axis that provide the most information about non-circular
motions.

The $s1$ and $s3$ coefficients are plotted against radius for each
galaxy in Figures~\ref{fig:s13:4321}(a)--\ref{fig:s13:5457}(a),
after normalization by $c_1 \approx v_*$, which is also plotted.  The
error bars represent the formal least-squares errors (corrected for
the number of pixels per beam), using the sum of fractional errors for
the ratios.  While the $c_1$ and $c_3$ terms are also affected by
elliptical or spiral streaming, they are also quite susceptible to
beam smearing effects and, in the case of the $c_1$ term, dominated by
the overall circular rotation.  Fluctuations in the $c_0$, $c_2$, and
$s_2$ terms, on the other hand, reflect an error in the kinematic
center (in which case the deviations should fall as $R^{-1}$) or the
presence of an $m$=1 mode (lopsidedness) in the potential.  Although
lopsidedness is clearly revealed in several galaxies, it is usually at
a fairly low level ($|c_2| \lesssim 5$ \kms), except in the outer \HI\
disks of strongly warped galaxies.  A more thorough analysis of this
phenomenon is beyond the scope of this paper.

The lower panels of Figures~\ref{fig:s13:4321}--\ref{fig:s13:5457}
show, for each galaxy, the measured $s_1$ and $s_3$ terms plotted
against each other, normalized by $c_1 \approx v_c \sin i$.
Coefficients for the CO and \HI\ data are plotted separately as panels
(c) and (d), since
they are derived at different spatial resolutions.  The slope of a
linear least-squares fit (heavy long-dashed line) is given in the
upper right corner of each plot; we consider slopes of $\lesssim 0.1$
in absolute value, corresponding to $|s_1/s_3| \gtrsim 10$, to be good
candidates for radial flows.  The dashed line through $(0,0)$ is the
warp line---the locus of points along which an error in the disk
position angle ($\Gamma$) should fall, assuming a flat rotation curve.
A tendency for the data points fall close to this line would be
consistent with a warp.  In addition, a shift in the assumed
P.A. ($\Gamma_0$) would be equivalent to a shift of all data points in
a direction parallel to this line.

If the $s_1$ term is interpreted as radial flow, its sign corresponds
to inflow or outflow depending on the sense of rotation in the galaxy.
Assuming that spiral arms are always trailing, we infer that all
galaxies besides NGC~4736 and 5055 rotate counterclockwise in the
sky, so that except for these two galaxies the criterion $s_1>0$
corresponds to radial outflow, as labeled at the bottom of each
$s_1$-$s_3$ plot.  We discuss results for each galaxy individually in
\S\ref{sec:indiv}.

\subsection{Comparison with Isophotal Data}\label{sec:kinmorph}

Figures~\ref{fig:s13:4321}(b)--\ref{fig:s13:5457}(b)
show, as functions of radius, the morphological position angle and
inclination, as derived from the isophote fits, and the kinematic
minor axis position angle and inclination, as derived from the harmonic
decomposition of the CO and \HI\ velocity fields.  Horizontal
dotted lines are used to indicate the adopted values of the position
angle and inclination from the ROTCUR analysis (\S\ref{sec:rotcur}),
except for NGC~4736 where the adopted inclination is based on the
photometry of \citet*{Mollen:95}.  The inclination $i$ has been
expressed in terms of the ellipticity of the projected disk, $\epsilon
\equiv 1 - \cos i$.  Note that the morphological inclination,
$\epsilon_{\rm morph}$, is expected to decrease in the inner,
bulge-dominated region of the disk as the isophotes become more
circular.  This general behavior is seen most clearly in NGC~4414,
4501, 5033, and 5055, which are sufficiently inclined so that the
fitted values of $\Gamma_{\rm morph}$ and $\epsilon_{\rm morph}$ are
dominated by geometry rather than non-axisymmetric structure.

The kinematic values were derived as follows:

\begin{enumerate}

\item We determined the kinematic minor axis $\Gamma_{\rm min}$ from
the harmonic coefficients for each ring by setting $V_{\rm los}=V_{\rm
sys}$ in Eq.~\ref{eqn:decomp} and deriving the roots numerically using
Ridders' method \citep[][and references therein]{Press:92}.  There are
two roots spaced by $\sim\pi$, corresponding to the two halves of the
axis; we adopted the average of the two values (first offsetting one
of them by $\pi$) and took half their difference as the estimated
error $\sigma$.  Only points with $\sigma < 5\arcdeg$ are shown.

\item The kinematic inclination or ``ellipticity'' is defined by
\begin{equation}
\epsilon_{\rm kin} = 1 - q_{\rm kin} = 1 - \widehat{q}
	\left(1-\frac{4c_3}{c_1}\right)
\end{equation}
where $\widehat{q}$ is the cosine of the adopted inclination used in
the harmonic decomposition.  This approximation is only accurate to
first order in $\delta q/q$, and additional errors can arise if the
rings are sufficiently tilted with respect to the assumed inclination that
points in the velocity image get assigned to the wrong ring.  We
checked that our results were consistent with tilted-ring fits in
which the inclination is allowed to vary, although such fits generally
give poorer results because of correlations between fit parameters.
Only points with $\sigma < 0.05$ (with errors derived from the
harmonic coefficients) are shown.

\end{enumerate}

Based on the results of \S\ref{sec:phot}, we expect that elliptical
streaming should be accompanied by a change in the morphological
position angle $\Gamma_{\rm morph}$ in a direction {\it opposite} to
the change in $\Gamma_{\rm min}$, assuming the gas orbits are aligned
with the isophotes.  An anti-correlation of this type is seen in
several of the galaxies, most notably NGC 4321, 4501, and 4736, as
discussed below.  The presence of a bar, warp, or strong spiral
structure would be expected to affect the morphological inclination
(via the isophote ellipticity) and quite possibly the kinematic
inclination (via the $c_3$ term) as well.

\subsection{Results for Individual Galaxies}\label{sec:indiv}

\subsubsection{NGC 4321 (M100)}

The CO velocity field of this well-studied Virgo cluster galaxy shows
a strong $s_1$ term peaking between $R$=20\arcsec\ and 40\arcsec\ with
$s_1/c_1 \approx 0.35$ and a slope $|ds_3/ds_1| \approx -0.05$, making
it a clear candidate for radial flows.  Interpreted as radial flow,
the $s_1$ term implies an outflow at $\sim$60 \kms\ in the plane of
the galaxy.  However, elliptical streaming must be considered as well,
since NGC 4321 is known to possess a stellar bar with a radial extent
of $\sim$60\arcsec\ \citep{Knapen:95}.  Further evidence for
elliptical streaming comes from a comparison of the kinematic and
morphological position angles: the CO kinematics show a positive
offset in $\Gamma_{\rm min}$ over the range $R$=20\arcsec--40\arcsec\
[Fig.~\ref{fig:s13:4321}(b)], corresponding to a negative offset in
$\Gamma_{\rm morph}$ over the same region.  Moreover, the \HI\
velocity field from $R$=40\arcsec--180\arcsec\ shows a negative slope
in the $s_1$-$s_3$ plane as expected for a bar potential
[Fig.~\ref{fig:s13:4321}(d)].  The presence of a dominant $s_1$ term
in the CO kinematics would be consistent with our modeling of gas flow
in the ILR region of a barred galaxy [cf.\
Fig.~\ref{fig:s13models}(a)].

\subsubsection{NGC 4414}

The CO data show nearly circular kinematics, aside from a region near
$R \sim 30\arcsec$ where $ds_3/ds_1 \approx -0.1$
[Fig.~\ref{fig:s13:4414}(c)], a possible indication of radial inflow.
The peak inflow speed would be $\sim$8 \kms, less than 4\% of the
circular speed.  On the other hand, the errorbars are large and do not
rule out values of $ds_3/ds_1$ that would be characteristic of an
elliptical streaming model.  While there is little evidence for a bar
in the $K$-band image ($\Gamma_{\rm morph}$ varies by $<2\arcdeg$ over
the region $R$=20\arcsec--50\arcsec), low-amplitude spiral structure
had been noted in the $K$-band image by \citet{Thornley:97b} after
removing an axisymmetric component from the light distribution.  Thus
the apparent inflow signature may be related to weak spiral streaming.

While the \HI\ data show a strong decrease in the $s_1$ component,
especially for $R > 150\arcsec$, it is well correlated with changes in
$s_3$ [Fig.~\ref{fig:s13:4414}(d)], suggesting the presence of a
strong warp.  The slope of the least-squares fit is a good match to
the warp line, although there is a constant offset from that line
which may indicate the effects of other non-circular motions.  A
decreasing trend in the morphological inclination supports an
interpretation in which the warp extends as far in as $R \approx
30$\arcsec, well within the optical radius ($R_{25} \sim 100\arcsec$).
This would explain the slight mismatch between the CO and \HI\
rotation curves, since the average CO and \HI\ inclinations would
actually be different.  We note, however, that effects other than a
warp, such as finite thickness or flaring of the disk, might also
affect the morphological inclination.

\subsubsection{NGC 4501 (M88)}

The CO data show an outflow signature very similar to that seen in NGC
4321, with implied outflow velocities of up to $\sim$45 \kms\ in the
region $R<40\arcsec$ [Fig.~\ref{fig:s13:4501}(a)].  As in the case of
4321, however, this could well be due to elliptical streaming, a
possibility that is strongly favored by the remarkably clear
anti-correlation between $\Gamma_{\rm min}$ and $\Gamma_{\rm morph}$
in this region.  It is unclear whether such a signature could be
attributed to spiral arms---strong spiral features are seen in both CO
and H$\alpha$ maps (Paper I), but at an inclination of 64\arcdeg\ the
projection effects are severe.  The $s_1$-$s_3$ plot for the \HI\ data
[Fig.~\ref{fig:s13:4501}(d)] shows a marginally negative slope at
larger radii (with $s_1>0$, as would be the case for outflow), which
is also likely to be a consequence of the bar.  Note that although the
isophotes in this outer region ($R$=50\arcsec--100\arcsec) are nearly
circular, isophotes become less sensitive to non-axisymmetric
structures as the disk inclination increases.  We conclude that NGC
4501 is a likely example of a galaxy that is optically classified as
unbarred but whose gas kinematics and inner near-infrared isophotes
strongly suggest the presence of a bar.

\subsubsection{NGC 4736 (M94)}

This prototypical ringed galaxy contains a nuclear stellar bar and a
corresponding CO bar \citep{Wong:00}.  We attribute the large $s_1$
and $s_3$ terms in the CO data to the effects of the bar, although the
scatter in the $s_1$-$s_3$ plot is large [Fig.~\ref{fig:s13:4736}(c)],
partly due to the complicating effects of a strong spiral arc at $R
\sim 25\arcsec$ whose streaming motions are distinct from the bar's.
The \HI\ kinematics between $R$=50\arcsec\ and 150\arcsec\ show a
strong $s_1$ term with $|ds_3/ds_1| \approx 0.2$.  In an earlier paper
\citep{Wong:00}, we had argued that the relative weakness of the $s_3$
term appeared to rule out an elliptical streaming model.  The more
sophisticated analysis presented in \S\ref{sec:ellstr} suggests that
instead we may be viewing elliptical streaming {\it in the vicinity of
an ILR}; in fact, the prominent H$\alpha$ ring in this galaxy at
$R=45\arcsec$ has been modeled as the ILR of a large-scale oval
distortion \citep*{Gerin:91}.  A pronounced decrease in $\Gamma_{\rm
morph}$ occurs between $R$=50\arcsec--100\arcsec, also consistent with
such a distortion [Fig.~\ref{fig:s13:4736}(b)].  While a warp model is
not ruled out by the kinematics alone, due to the low inclination of
the galaxy, the observed anti-correlation between $\Gamma_{\rm morph}$
and $\Gamma_{\rm min}$ would then be harder to account for.

\subsubsection{NGC 5033}

The CO data show some evidence for a warp based on the $s_1$-$s_3$
plot [Fig.~\ref{fig:s13:5033}(c)], although the uncertainties are
large.  No indication of radial inflow at speeds of $>$5\% of the
circular speed ($v_c \sim 200$ \kms) is seen for $R<55\arcsec$.  For
the \HI\ data, the $s_1$ and $s_3$ terms are positively correlated at
radii beyond 90\arcsec, raising the possibility that the galaxy's
warp, which only becomes prominent around $R_{25}$=320\arcsec, is
pervasive throughout the galaxy, as suggested for NGC 4414.  Such a
disturbance may be related to the weak interaction of NGC 5033 with
Holmberg VIII, as evidenced by the tidal \HI\ plume on the southern
side of the galaxy \citep{Thean:97}.  However, the peak in $s_1$ and
$s_3$ near $R$=110\arcsec\ [Fig.~\ref{fig:s13:5033}(a)] might instead
be the result of spiral streaming.  The morphological inclination also
falls with radius beyond $R$=30\arcsec--100\arcsec, although it does
not closely follow the kinematic inclination
[Fig.~\ref{fig:s13:5033}(b)].  It is possible that extinction or
projection effects due to the galaxy's high inclination may dominate
over the warp, as far as the isophotes are concerned.  A deeper
$K$-band image would be useful for clarifying the situation.

\subsubsection{NGC 5055}

A strong negative slope in the $s_1$-$s_3$ plot for the CO data
[Fig.~\ref{fig:s13:5055}(c)] suggests the presence of low-level
streaming motions centered around $R$=30\arcsec.  Although inspection
of the $K$-band image does not reveal clear evidence of a bar, the
isophotes do become ``boxier'' around this radius, leading to a dip in
the ellipticity $\epsilon_{\rm morph}$.  In spite of this localized
deviation, the smallness of the $s_1$ term places an upper limit of
$v_R \lesssim 5$ \kms\ (3\% of the circular speed) on the magnitude of
any radial flows in the region $R$=15\arcsec--70\arcsec.

The \HI\ kinematics are clearly dominated by a warp for $R>200\arcsec$
(0.5$R_{25}$).  Inside of this radius, a possible region of gas inflow
is seen, highlighted by the small box in Fig.~\ref{fig:s13:5055}(d).
We interpret this weak feature (again amounting to radial motions of
no more than 3\% of the circular speed) to spiral streaming motions,
since one finds CO and \HI\ spiral arms crossing the galaxy's minor
axis at approximately these radii, where their effects are more likely
to have an impact on the observed kinematics.  Fluctuations in the
isophotal parameters $\Gamma_{\rm morph}$ and $\epsilon_{\rm morph}$
are also apparent around $R \approx 70\arcsec$
[Fig.~\ref{fig:s13:5055}(b)], and are likely associated with spiral
arms.

Note that for this galaxy, $\epsilon_{\rm morph}$ never rises to a
value comparable to $\epsilon_{\rm kin}$.  Aside from a few humps that
may be attributed to spiral structure, the morphological inclination
reaches a fairly constant value of around 57\arcdeg, while the
kinematics favor an inclination of 63\arcdeg, also with very little
scatter (this comparison is made well inside the region of the warp).
One possibility is that the morphological inclination is reduced by
the finite thickness of the stellar disk.  A radius of 200\arcsec\
corresponds to $\sim$10 kpc for this galaxy, so the projected minor
axis at this radius should have a length of $\sim$4.5 kpc if the true
inclination is 63\arcdeg.  If the minor axis is extended by 1 kpc due
to a thick disk, however, the morphological inclination is reduced to
56\arcdeg, consistent with the observed value.

\subsubsection{NGC 5457 (M101)}

The inclination of this galaxy is not well determined by either
morphology or kinematics; the adopted value is based on fitting just
two rings in the \HI\ velocity field.  However, our value of
21\arcdeg\ is close to the 18\arcdeg\ value of RC3 as well as the
22\arcdeg\ that results from the Tully-Fisher relation of
\citet{Pierce:92} using an \HI\ linewidth of 190 \kms.

Despite poor signal-to-noise in the inner arcminute, the CO velocity
field shows strong evidence for large $s_1$ and $s_3$ terms
[Fig.~\ref{fig:s13:5457}(c)], and indeed a substantial twist in the
kinematic minor axis is seen in the tapered CO velocity field
(Fig.~\ref{fig:velfields}).  The slope of the $s_1$-$s_3$ relation
appears to be negative and thus indicative of elliptical streaming.
Identifying the bar in the optical images is difficult due to the
strong spiral structure in this galaxy, but the bar-like CO and
H$\alpha$ morphology, as well as the rise in the ellipticity of the
isophotes across the region $R$=20\arcsec--60\arcsec\ to a value much
larger than in the rest of the disk, confirm the presence of a weak
bar or oval distortion in the inner regions.  The \HI\ kinematics show
a strongly decreasing $s_1$ term beyond $R$=200\arcsec\ (0.2$R_{25}$)
which is probably dominated by a large outer warp, with additional
fluctuations in $s_1$ and $s_3$ at certain radii due to the strong
spiral structure in this galaxy.  Unfortunately, the inclination is
too low to clearly distinguish the effects of the warp from possible
radial flows.


\begin{table}
\begin{center}
\caption{Candidate Regions For Radial Flows, Based On $s_1/s_3$
\label{tbl:infcan}}
\bigskip
\begin{tabular}{ccccl}
\tableline\tableline
Galaxy & Tracer & Radial Range & Maximum $v_R$ & Interpretation\\[0.5ex]
\tableline
NGC 4321 & CO & 20\arcsec--40\arcsec & $56 \pm 5$ & Bar streaming\\
NGC 4414 & CO & 15\arcsec--45\arcsec & $-8 \pm 2$\tablenotemark{a} 
	& Spiral streaming\\
NGC 4501 & CO & 15\arcsec--40\arcsec & $48 \pm 5$ & Bar streaming\\
NGC 4501 & HI & 40\arcsec--150\arcsec & $10 \pm 4$ & Bar streaming\\
NGC 5055 & \HI & 50\arcsec--150\arcsec & $-6 \pm 1$ & Spiral streaming\\
NGC 5457 & \HI & 250\arcsec--600\arcsec & $-56 \pm 5$ & Outer warp\\
\tableline
\end{tabular}
\end{center}
\tablenotetext{a}{Negative values denote inflow.}
\end{table}


\section{Discussion and Conclusions}\label{sec:disc}

We have carried out a search for radial flow signatures in the CO and
\HI\ velocity fields of seven nearby spirals by decomposing each
elliptical ring into a third-order Fourier series.  By requiring that
$|s_1/s_3| \gtrsim 10$ over a significant range in radius, where $s_1$ amd
$s_3$ are the coefficients of the $\sin\psi$ and $\sin 3\psi$ terms,
we identified candidate regions for radial inflow or outflow as
summarized in Table~\ref{tbl:infcan}.  Aside from NGC 5457, where an
inflow signature is degenerate with a warp, we find photometric
evidence for bars or spiral structure in all such regions, suggesting
that elliptical streaming in a bar or spiral potential is the dominant
contributor to non-circular motions.  While inflow may be superposed
on these motions, we find no unequivocal evidence for radial inflow
alone.  Three of the galaxies, NGC 4414, 5033 and 5055, show nearly
pure circular rotation in their inner few kiloparsecs ($R \lesssim
60$\arcsec), with an upper limit to any radial inflows of $\sim$5--10
\kms, or 3--5\% of the circular speed.  These are probably the
strictest limits that can be placed on radial gas flows in external
galaxies.

Although our radial inflow limits are a small fraction of the circular
speed, they cannot be used to rule out theoretical models that invoke
radial flows, since such models generally predict even smaller inflow
speeds.  On the higher end of estimates is that by \citet{Blitz:96},
who estimates that an inflow velocity of $-7$ \kms\ would be needed to
completely replenish gas consumed by star formation in the inner
Galaxy with \HI\ from twice the solar radius.  Given that estimates of
the gas consumption time in our sample of galaxies are substantially
longer than in the Milky Way (Paper I), such a high inflow rate
would not be expected.  \citet{Struck:91} discusses how radial flows
can serve to maintain hydrodynamic stability in gaseous disks, but
such flows are always subsonic ($\lesssim$5 \kms) and in fact reduce
to zero for an $R^{-1}$ gas surface density profile.
\citet{Struck:99} further develop this idea into a steady-state model
for a turblent multiphase ISM, predicting slow ($\sim$3 \kms) inflows
of cold gas in the galaxy midplane.  Pure chemical evolution models
invoke even smaller inflow speeds (e.g. $\sim$1 \kms\ in
\citealt{Portinari:00}) since they are concerned with steepening
metallicity gradients in galaxies over timescales of many Gyr.  Thus,
our observations and analysis techniques have yet to reach the regime
where such flows can be easily detected.

The results presented in this paper underscore the difficulty of
detecting radial inflow in disks where a warp or $m$=2 distortion in
the potential exists.  Once such effects become dominant, they cannot
be easily fitted and removed without performing a more sophisticated
analysis than that performed here.  For instance, by neglecting the
important role of shocks in the gas component, we are unable to model
the very processes (bars and spiral arms) that are most likely to lead
to radial inflows in the inner regions of galaxies.  Even with more
sophisticated modeling, however, a leap in sensitivity and angular
resolution will be needed to examine shock regions in detail.
Similarly, our lack of knowledge about warps limits our ability to
analyze the gas kinematics in the outskirts of galaxies.  The default
assumption, that warped orbits remain circular even as they move away
from the principal plane, must be only a crude approximation.  Since
warps may be intimately related to infalling gas \citep{Binney:92},
which in turn could induce radial flows, further modeling is clearly
desirable.

An intriguing result from our analysis has been the possible detection
of low-level kinematic warp signatures in NGC 4414 and 5033, at radii
well inside the radius at which the warp becomes prominent (which is
roughly the optical radius $R_{25}$).  While spiral structure may
confuse the picture in NGC 5033, it is unlikely to play a major role
in NGC 4414, where the spiral structure is much more flocculent and
the isophote fits seem to indicate an optical warp as well.  The
prominent kinematic warps in NGC 5055 and 5457 also appear well inside
$R_{25}$, contrary to usual expectations \citep[e.g.,][]{Briggs:90}.
Since warps are often straightforward to identify in the $s_1$-$s_3$
plane, the harmonic decomposition technique will be valuable for
assessing the radius at which warping begins, which in turn may shed
light on the physical mechanisms that maintain warps.

\acknowledgements

We thank our collaborators on BIMA SONG, particularly D. Bock,
T. Helfer, M. Regan, K. Sheth, M. Thornley, and S. Vogel, for their
efforts in assembling the primary and ancillary SONG data sets.
C. McKee provided useful comments to an earlier draft of this paper.
We are indebted to R. Braun, V. Cayatte, R. Kennicutt, J. Knapen,
C. Mundell, and A. Thean for providing us with data for this study,
and J. van Gorkom for preparing the observing file for the VLA
observations of NGC 4501.  Finally, we thank the anonymous referee for
detailed and constructive comments.  This research is based on the
Ph.D. thesis of T.W. and has been supported by National Science
Foundation grants AST 96-13998 and 99-81308 to the U.C. Berkeley Radio
Astronomy Laboratory, and a Bolton Fellowship from the Australia
Telescope National Facility.  This publication makes use of data
products from the Two Micron All Sky Survey, which is a joint project
of the University of Massachusetts and the Infrared Processing and
Analysis Center/California Institute of Technology, funded by the
National Aeronautics and Space Administration and the National Science
Foundation.

\appendix

\section{Effect of Dissipation on Gaseous Orbits}\label{sec:damping}

\begin{figure*}[t]
\begin{center}
\plottwo{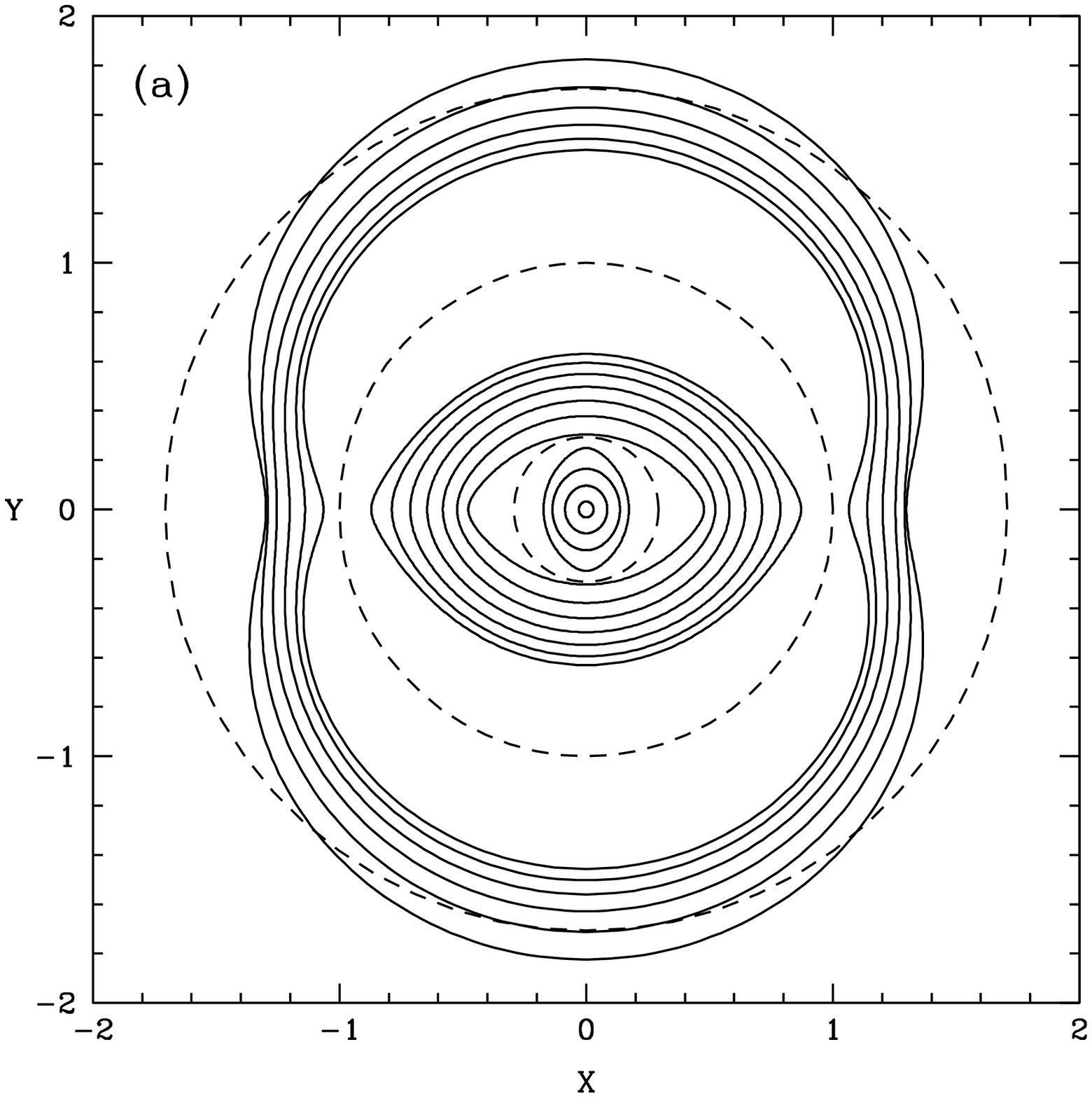}{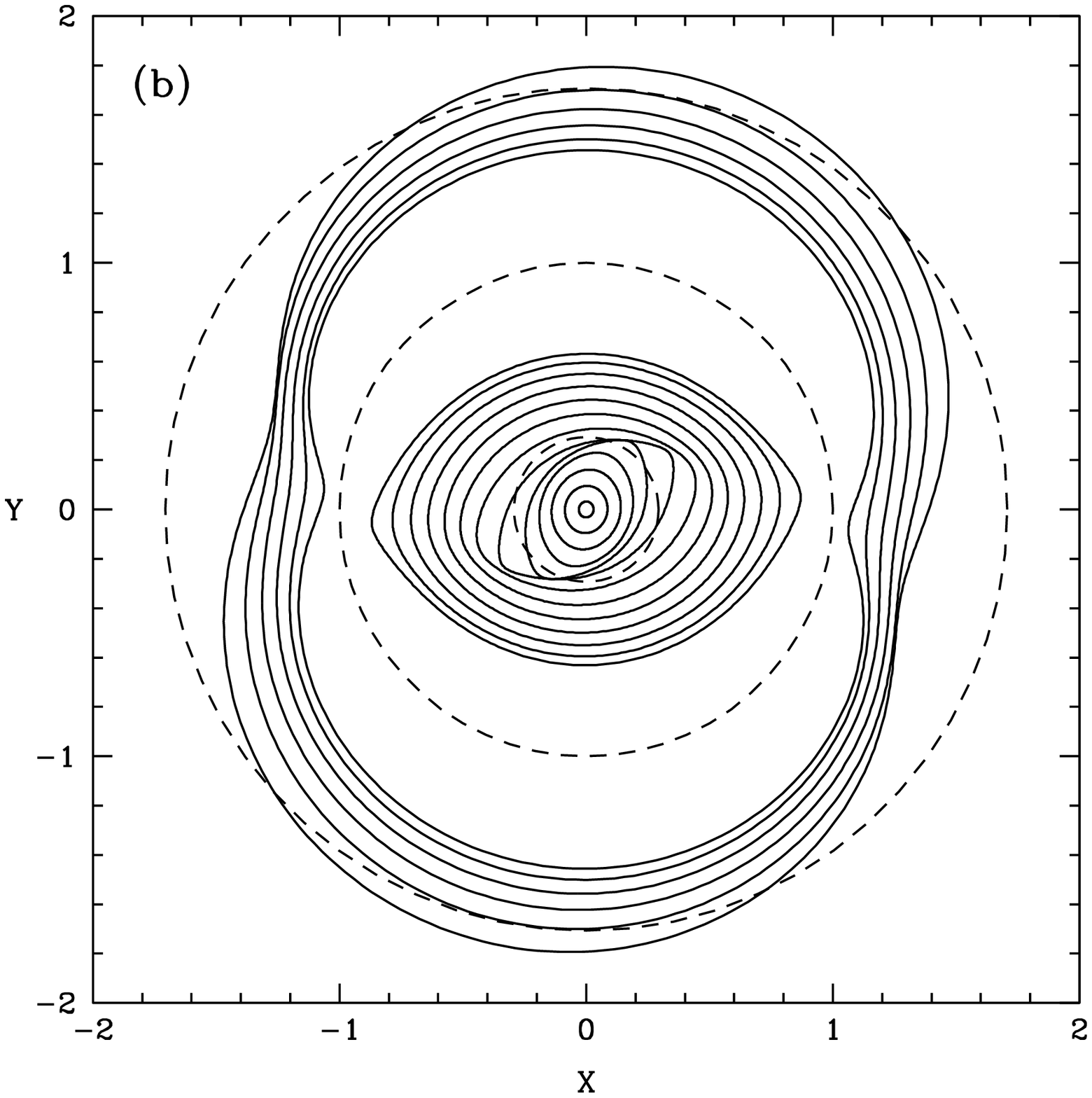}
\end{center}
\caption{
Orbits viewed in the rotating frame for a (a) dissipationless
bar model and (b) dissipative bar model with a flat rotation
curve.  The dashed lines at radii of $1-\sqrt{2}/2$, 1, and
$1+\sqrt{2}/2$ are the ILR, CR, and OLR respectively.
\label{fig:orbits}}
\end{figure*}

For a non-zero pattern speed, a collisionless weak bar model has the
property that the stable periodic orbits change orientation (from
parallel to perpendicular to the bar or vice versa) with each
resonance crossing [Figure~\ref{fig:orbits}(a)].  Many of the
resulting orbits intersect each other, and are hence not accessible to
the gas component, since only one velocity is permitted for a fluid at
a given spatial location.  Instead, the gaseous orbits are expected to
turn gradually between resonances, producing a spiral pattern [see
Figure~\ref{fig:orbits}(b)].  Since cloud collisions, shocks, and
energy dissipation lead to vastly more complex behavior than can be
described by a model such as the one given in \S\ref{sec:ellstr}, this
situation is generally examined with the help of numerical
simulations, employing either a hydrodynamical or ``sticky particle''
approach.

One can obtain an approximate {\it analytic} description of gaseous
orbits in a barred potential by introducing a damping term into the
equations of motion, as has been done by various authors
\citep[e.g.,][]{Lindblad:94,Wada:94,Byrd98}.  The additional term
simulates a frictional force that is proportional to the velocity
perturbation in the radial and/or azimuthal direction.  For simplicity
we consider damping in the radial direction only, following
\citet{Wada:94}.  (The more involved case of both radial and azimuthal
damping is treated by \citealt{Baker:th}).  The first-order equation
of motion (cf.\ \citealt{Binney:87}, p.~148) is then:
\begin{equation}
\ddot{R_1} + 2\Lambda \dot{R}_1 + \kappa_0^2R_1 =
	-\left[\frac{d\Phi_m}{dR} +
	\frac{2\Omega\Phi_m}{R(\Omega-\Omega_p)}\right]_{R_0} \cos
	m\phi_0
\end{equation}
where $\Lambda$ is the damping term and $\phi_0 \equiv
(\Omega_0-\Omega_p)t$.  The solutions for $(R,\phi,v_R,v_\phi)$ in the
rest frame of the bar are
given by \citet{Sakamoto:99}, although we have redefined some of the
constants here to facilitate comparison with \S\ref{sec:ellstr}:
\begin{eqnarray}
R & = & R_0 \left[1-\frac{A}{2}\cos (2\phi_0 + \delta_0)\right]\\ \phi
& = & \phi_0 + \frac{B}{4}\sin (2\phi_0 + \beta_0)\\ v_R & = &
v_c(1-\omega_p)[A\,\sin (2\phi_0+\delta_0)]\\ v_\phi & = &
v_c(1-\omega_p)\left[1 - \frac{A}{2} \cos(2\phi_0+\delta_0) +
\frac{B}{2}\cos(2\phi_0+\beta_0)\right]
\end{eqnarray}
The amplitudes of the epicyclic motion are given by
\begin{eqnarray}
A & = & \frac{2}{R_0\sqrt{\Delta_0^2 + 16\Lambda^2(\Omega_0-\Omega_p)^2}}
	\left[\frac{2\Phi_2(R_0)}{R_0(1-\omega_p)} +
	\Phi_2^\prime(R_0)\right]\\
B & = & \sqrt{(E+F)^2 + 2EF(\cos\delta_0 -1)}
\end{eqnarray}
where
\[E \equiv -\frac{2A}{1-\omega_p}\;,\quad
	F \equiv \frac{2\Phi_2(R_0)}{v_c^2(1-\omega_p)^2}\;. \]
The phase shifts of the orbit are given by
\begin{equation}
\tan\delta_0 = -\frac{4\Lambda(\Omega_0-\Omega_p)}{\Delta_0}\;,\quad
\tan\beta_0 = \frac{\sin\delta_0}{\cos\delta_0 + (F/E)}\;.
\end{equation}
Note that in the limit that $\Lambda \rightarrow 0$, we have $A
\rightarrow a_2$ and $B \rightarrow (a_2 + b_2)$, in terms of the
coefficients of the collisionless model.

Transforming to the galaxy rest frame, we find that for $t=0$:
\begin{eqnarray}
v_R & = & v_c\left[A(1-\omega_p) \sin (2\theta+\delta_0) \right]\\
v_\theta & = & v_c\left[1-\frac{A}{2}\cos
	(2\theta+\delta_0) + \frac{B}{2}(1-\omega_p)
	\cos (2\theta+\beta_0) \right]\;.
\end{eqnarray}
Recall that the harmonic coefficients are defined by:
\begin{eqnarray}
V_{\rm los} & = & V_{\rm sys} + (v_\theta \cos\psi + v_R \sin\psi) \sin i 
	\nonumber\\
	& = & c_0 + c_1 \cos\psi + s_1 \sin\psi + c_3 \cos3\psi + 
	s_3 \sin3\psi\;,
\label{eq_harmexp}
\end{eqnarray}
where $\psi = \theta - \theta_{\rm obs} - \pi/2$.  Thus we find:
\begin{eqnarray}
c_0 & = & V_{\rm sys}\\
\frac{c_1}{v_*} & = & 1-\frac{1}{4}[(-2\omega_p+1)
	A\cos(2\theta_{\rm obs}+\delta_0) + (1-\omega_p)
	B\cos(2\theta_{\rm obs}+\beta_0)] \nonumber\\
\frac{s_1}{v_*} & = & \frac{1}{4}[(-2\omega_p+1)
	A\sin(2\theta_{\rm obs}+\delta_0) + (1-\omega_p)
	B\sin(2\theta_{\rm obs}+\beta_0)] \nonumber\\
\frac{c_3}{v_*} & = & -\frac{1}{4}[(2\omega_p-3)
	A\cos(2\theta_{\rm obs}+\delta_0) + (1-\omega_p)
	B\cos(2\theta_{\rm obs}+\beta_0)] \nonumber\\
\frac{s_3}{v_*} & = & \frac{1}{4}[(2\omega_p-3)
	A\sin(2\theta_{\rm obs}+\delta_0) + (1-\omega_p)
	B\sin(2\theta_{\rm obs}+\beta_0)] \nonumber
\end{eqnarray}
where $v_* \equiv v_c \sin i$.  These relations give the observed
harmonic coefficients in terms of the orbit parameters 
$(A,B,\beta_0,\delta_0)$.


For the case of an elongated potential discussed in \S\ref{sec:ellstr}, the
orbit parameters ($A$, $B$, $\delta_0$, $\beta_0$) are given by:
\begin{eqnarray}
A & = & -\frac{\epot}{(1-\omega_p)\sqrt{
	(1-4\omega_p+2\omega_p^2)^2+4\lambda^2(1-\omega_p)^2}}\\
B & = & -\sqrt{\left(\frac{2A}{1-\omega_p}+\frac{\epot}{(1-\omega_p)^2}
	\right)^2 + \frac{4A\epot}{(1-\omega_p)^3}(\cos\delta_0 - 1)}\\
\tan \delta_0 & = & \frac{2\lambda(1-\omega_p)}{1-4\omega_p
	+2\omega_p^2}\\
\tan \beta_0 & = & \frac{\sin\delta_0}{\cos\delta_0 + \epot/[2A(1-
\omega_p)]}\;.
\end{eqnarray}
where $\lambda \equiv \Lambda/\Omega_0$.  
With our choice of signs,
$\delta_0 \rightarrow -\pi$ at $\omega_p$=0 for $\lambda=0$.
Model orbits calculated using these equations, using a flat rotation
curve, are shown in Figure~\ref{fig:orbits}(b).  Note that the damping
term leads to a gradual change in orbit orientation near the ILR
($\omega_p = 1-\sqrt{2}/2$), although the orbit oscillations still go
nonlinear near the CR ($\omega_p=1$). 

\clearpage

\bibliographystyle{apj}
\bibliography{thesis}

\end{document}